\documentclass{aa} 

\newcommand{\ltsim}{\protect\raisebox{-0.5ex}{$\:\stackrel{\textstyle <}
        {\sim}\:$}}

\usepackage{longtable}    
\usepackage{aalongtable}    
\usepackage{latexsym}
\usepackage{graphicx}
\usepackage{epsfig}
\usepackage{lscape}
\usepackage[authoryear]{natbib}
\bibpunct{(}{)}{;}{a}{}{,} 

\begin{document}

\title{The Chamaeleon II low-mass star-forming region: radial velocities, elemental abundances, and accretion properties\thanks{Based on 
FLAMES (GIRAFFE+UVES) observations collected at the Very Large Telescope (VLT; Paranal, Chile). Program 076.C-0385(A).} } 
   
\author{K. Biazzo \inst{1} \and J. M. Alcal\'a\inst{1} \and E. Covino\inst{1} \and A. Frasca\inst{2} \and F. Getman\inst{1} \and 
L. Spezzi \inst{3}}

\offprints{K. Biazzo}
\mail{katia.biazzo@oacn.inaf.it}

\institute{INAF - Capodimonte Astronomical Observatory, via Moiariello, 16, 80131 Naples, Italy
\and INAF - Catania Astrophysical Observatory, via S. Sofia, 78, 95123 Catania, Italy
\and ESO - European Southern Observatory, Karl-Schwarzschild-Str. 2, 85748 Garching bei M\"unchen, Germany
}

\date{Received / accepted }

\abstract
{Knowledge of radial velocities, elemental abundances, and accretion properties of members of star-forming regions is important for our 
understanding of stellar and planetary formation. While infrared observations reveal the evolutionary status of the disk, optical spectroscopy 
is fundamental to acquire information on the properties of the central star and on the accretion characteristics.}
{Existing {{\it 2MASS}} archive data and the {{\it Spitzer c2d}} survey of the Chamaeleon\,II dark cloud have provided disk properties of a 
large number of young stars. We complement these data with optical spectroscopy with the aim of providing physical stellar parameters 
and accretion properties.}
{We use FLAMES/UVES and FLAMES/GIRAFFE spectroscopic observations of 40 members of the Chamaeleon\,II star-forming region to 
measure radial velocities through cross-correlation technique, lithium abundances by means of curveso of growth, and for a suitable star 
elemental abundances of Fe, Al, Si, Ca, Ti, and Ni using the code MOOG. From the equivalent widths of the H$\alpha$, H$\beta$, and the \ion{He}{i} 
$\lambda$5876, $\lambda$6678, $\lambda$7065 \AA~emission lines, we estimate the mass accretion rates, $\dot M_{\rm acc}$, 
for all the objects.} 
{We derive a radial velocity distribution for the Chamaeleon\,II stars, which is peaked at $\langle V_{\rm rad}\rangle=11.4\pm2.0$~km\,s$^{-1}$. 
We find dependencies of $\dot M_{\rm acc} \propto M_\star^{1.3}$ and of $\dot M_{\rm acc} \propto Age^{-0.82}$ in the $\sim 0.1-1.0 M_\odot$ 
mass regime, as well as a mean mass accretion rate for Chamaeleon~II of $\dot M_{\rm acc} \sim 7^{+26}_{-5}\times10^{-10} M_\odot$\,yr$^{-1}$. 
We also establish a relationship between the \ion{He}{i} $\lambda$7065 \AA\ line emission and the accretion luminosity.}
{The radial velocity distributions of stars and gas in Chamaeleon\,II are consistent. 
The spread in $\dot M_{\rm acc}$ at a given stellar mass is about one order of magnitude and can not be ascribed entirely to 
short timescale variability. Analyzing the relation between $\dot M_{\rm acc}$ and the colors in {{\it Spitzer~c2d}} and {{\it 2MASS}} bands, 
we find indications that the inner disk changes from optically thick to optically thin at $\dot M_{\rm acc} \sim 10^{-10} M_\odot$\,yr$^{-1}$. 
Finally, the disk fraction is consistent with the age of Chamaeleon\,II. }
   
	   
\titlerunning{Radial velocity, abundances, and accretion of Chamaeleon II}
\authorrunning{K. Biazzo et al.}
\maketitle

\section{Introduction}

The study of accretion properties of members of star-forming regions (SFRs) is important for our understanding of stellar and planetary 
formation. While infrared observations provide information on the structure of the circumstellar disk, the accretion properties 
can be retrieved from photometry and spectroscopy using primary diagnostics, such as the UV excess emission (e.g., 
\citealt{gullbringetal1998, rigliacoetal2011a}), the Paschen/Balmer continuum and Balmer jump (e.g., 
\citealt{gullbringetal1998, herczeghillenbrand2008, rigliacoetal2011b}), or secondary tracers, like hydrogen recombination lines 
(H$\alpha$, H$\beta$, H$\gamma$, H9, Pa$\beta$, Pa$\gamma$, Br$\gamma$), and the \ion{He}{i}, \ion{Ca}{ii}, \ion{Na}{i} lines 
(e.g., \citealt{muzerolleetal1998a, muzerolleetal1998b, nattaetal2006, fangetal2009, rigliacoetal2011b, antoniuccietal2011}). 
The rate at which the central star accretes from disk material has been found to approximately scale with the square of the stellar 
mass and to decrease with age (see, e.g., \citealt{herczeghillenbrand2008}, and references therein). In addition, 
the accretion properties are also important to understand the planet-metallicity relation. In fact, the efficiency of dispersal of 
circumstellar disks is predicted to depend on stellar metallicity in the sense that the formation of planetesimals around stars is 
faster at higher metallicity (\citealt{ercolanoclarke2010}). Simultaneous measurements of accretion rates and elemental abundances in SFRs 
and young clusters are therefore crucial to shed light on the role of metallicity in disk dispersal and planetary formation.

The Chamaeleon II (hereafter Cha II) dark cloud, at a distance of $178\pm18$ pc (\citealt{whittetetal1997}), is one of the three main clouds 
of the Chamaeleon complex ($\alpha \sim 12^{\rm h}$, $\delta \sim -78^{\degr}$). It extends over $\sim 2\deg^2$ in the sky (see \citealt{luhman2008} 
for a recent review). The population of Cha II consists of some 20 classical T Tauri stars (CTTSs), $\sim$ten weak-lined T Tauri stars (WTTSs), 
an intermediate-mass Herbig Ae star (IRAS~12496$-$7650; see, e.g., \citealt{garcialopezetal2011}), a few Herbig-Haro objects 
(\citealt{alcalaetal2008}, and references therein), $\sim$three sub-stellar objects, and references therein)
and five very low-mass stars (\citealt{spezzietal2008} and references therein). Cha II is one of the five SFRs included in the 
{\it Spitzer Space Telescope} Legacy Program ``From Molecular Cores to Planet-forming Disks'' 
({\it c2d}; \citealt{evansetal2003, youngetal2005, porrasetal2007}). Through extensive work based on {\it c2d} IRAC and MIPS {\it Spitzer} fluxes 
and complementary data, a reliable census of the population in Cha II (down to $0.03M_{\sun}$) was achieved by \cite{alcalaetal2008}. They 
concluded that the cloud is dominated by objects with active accretion, with the Class II sources representing $\sim 60\%$. The same sample was 
investigated spectroscopically in the optical using GIRAFFE/UVES@VLT by \cite{spezzietal2008}, who derived stellar parameters and estimated 
a mean age of $4\pm2$ Myr for Cha II. However, studies of radial velocities, elemental abundances, and accretion properties of the 
cloud members were not addressed in these works. Recently, \cite{alcalaetal2011a} analyzed the star IRAS~12556$-$7731, 
concluding that it is indeed a background lithium-rich M-giant star unrelated to Cha~II.

As a continuation of the studies by \cite{alcalaetal2008} and \cite{spezzietal2008}, we derive here radial velocities, elemental abundances, 
and accretion properties for 40 pre-main sequence (PMS) stars in Cha~II. The outline of the paper is as follows. In Sect.~\ref{sec:obs}, 
we describe the spectroscopic observations and data reduction. In Sect.~\ref{sec:analysis}, we report determinations of 
radial velocities, elemental abundances, and accretion properties. The main results are discussed in Sect.~\ref{sec:discussion}, while our 
conclusions are presented in Sect.~\ref{sec:conclusions}. 

\section{Spectroscopic observations and data reduction}
\label{sec:obs}

The observations were conducted in February-March 2006 and February 2007 using FLAMES (GIRAFFE+UVES) at the VLT. A complete journal of 
the observations and instrumental setup is given in \cite{spezzietal2008}. The relevant information for this paper is summarized 
in Table~\ref{tab:logs}.

We observed 32 objects with GIRAFFE, 11 with UVES, and two with both spectrographs (see Table~\ref{tab:logs}). 
Despite the $\sim25'$ FLAMES field of view (FoV), it was not possible to assign a large number of fibers in each configuration to the PMS objects/candidates 
because of the large spatial scatter of the targets. 
The remaining fibers were allocated to young candidates and field stars\footnote{Hereafter, we refer to this sample as ``field stars''.} 
(numbers in parentheses in columns 4 and 5 of Table~\ref{tab:logs}; see \citealt{spezzietal2008} for more details).
Thirty-two objects were observed several (2--4) times within two days (see Table~\ref{tab:all_param}). 

While \cite{spezzietal2008} used a single spectrum per object to derive the spectral type and to confirm the presence of 
\ion{Li}{i} 6708\,\AA\ absorption, we use here the complete set of spectra to investigate accretion and short timescale variability. 
To this aim, we reprocessed the FLAMES/GIRAFFE and FLAMES/UVES observations. The GIRAFFE data were reduced using the GIRAFFE 
Base-Line Data Reduction Software 1.13.1 (girBLDRS; \citealt{blechaetal2000}): bias and flat subtraction, correction for the fiber 
transmission coefficient, wavelength calibration, and science frame extraction were performed. Then, a sky correction was applied to each 
stellar spectrum using the task {\sc sarith} in the IRAF\footnote{IRAF is distributed by the National Optical Astronomy Observatory, 
which is operated by the Association of the Universities for Research in Astronomy, inc. (AURA) under cooperative agreement with the 
National Science Foundation.} {\sc echelle} package and by subtracting the average of several sky spectra obtained 
simultaneously. The reduction of the UVES spectra was performed using the pipeline developed by \cite{modiglianietal2004}, which 
includes the following steps: subtraction of a master bias, \'echelle order definition, extraction of thorium-argon spectra, 
normalization of a master flat-field, frame extraction, wavelength calibration, and correction of the science frame 
for the normalized master flat-field. Sky subtraction was also performed with the IRAF task {\sc sarith} using the fibers allocated to the sky.

\setlength{\tabcolsep}{2.5pt}
\begin{table}
\caption{Summary of the observations.} 
\label{tab:logs}
\begin{center}
\begin{tabular}{lccccc}
\hline
\hline
Instrument &   Range & Resolution & \# stars & \# spectra\\ 
	   &   (\AA) & ($\lambda/\Delta\lambda$) &  & \\ 
\hline
GIRAFFE & 6438--7164 &  8\,600 & 32(+27) & 69(+45) \\
UVES	& 4764--6820 & 47\,000 & 11	 & 25 \\
\hline
\end{tabular}
\end{center}
\end{table}
\normalsize

\section{Data analysis and results}
\label{sec:analysis}

\subsection{Radial velocity distribution, membership, and binarity}
\label{sec:rad_vel}
We determined radial velocities (RVs) of each object, choosing Hn~23 and RX~J1303.1$-$7706 as UVES and GIRAFFE templates, respectively. 
These slowly rotating stars show no strong accretion signatures (see Table~\ref{tab:accret_param}). 
We measured the RV of each template (highest $S/N$) spectrum using the IRAF task {\sc rvidlines} inside the {\sc rv} package, which 
considers a line list. We used 50 and 10 lines for the UVES and GIRAFFE spectra, respectively, obtaining $V_{\rm rad}=12.5\pm0.4$ km s$^{-1}$ for 
RX~J1303.1$-$7706, and $V_{\rm rad}=15.2\pm0.3$ km s$^{-1}$ for Hn~23. The heliocentric RV of all targets was determined through the task 
{\sc fxcor} of the IRAF package {\sc rv}, which cross-correlates the target and template spectra, excluding regions affected 
by broad lines or prominent telluric features. The centroids of the cross-correlation function (CCF) peaks were determined by adopting 
Gaussian fits, and the RV errors were computed using a procedure that considers the fitted peak height and the antisymmetric noise 
(see \citealt{tonrydavis1979}). When more spectra were acquired, we computed the average RV for each object. 

In order to estimate the binary fraction, we considered as singles the stars with only one CCF peak and with night-to-night 
RV variations within 3$\sigma$. In the end, excluding seven objects for which we could not measure the RV, we find six spectroscopic 
binaries, which means a binary fraction of $18\%$. In the last column of Table~\ref{tab:all_param}, we list the most probable 
spectroscopic systems. Five stars are newly identified as spectroscopic binaries, while RXJ1301.0$-$7654a was already recognized 
as a double-lined (SB2) spectroscopic system (\citealt{covinoetal1997}).

\begin{figure}[h!]
\begin{center}
 \begin{tabular}{c}
\includegraphics[width=8.8cm]{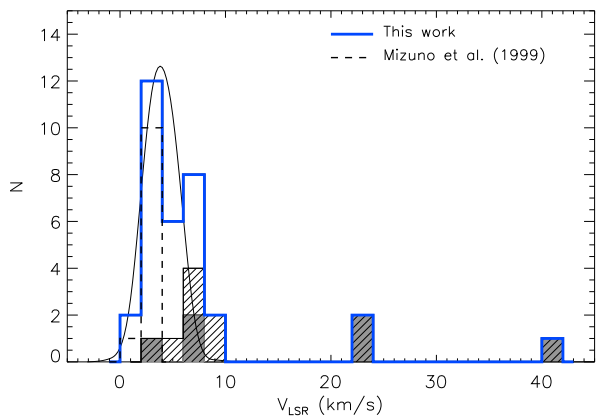}
\vspace{-.3cm}
 \end{tabular}
\caption{Average RV distribution in the LSR (solid thick line) of the Cha~II PMS stars. The distribution of the gas derived by 
\cite{mizunoetal1999} is overlaid (dashed line). The Gaussian fit to the PMS RV distribution is shown (thin line). The shaded histogram 
represents spectroscopic binary stars, while the hatched one marks the UVES observations. In the case of Hn~24 and Sz~54, where 
both UVES and GIRAFFE RVs were measured, we considered the UVES observations.}
\label{fig:vrad_distr}
 \end{center}
\end{figure}

In Fig.~\ref{fig:vrad_distr}, we show the Cha~II distribution of the RV measurements in the local standard of rest (LSR) obtained from 
both the UVES and GIRAFFE spectra, along with the RV distribution of the gas derived by \cite{mizunoetal1999} from the C$^{18}$O ($J=1-0$) 
transition in 11 dense molecular cores in Cha II. Excluding the spectroscopic binaries, the RV distribution of the Cha II members 
has a mean of $\langle V_{\rm rad}\rangle=11.4\pm2.0$ km s$^{-1}$, which translates to a value of 
$\langle V_{\rm LSR}\rangle =4.9\pm2.0$ km s$^{-1}$. The Gaussian fit of the distribution yields a mean value peaked at 
$\langle V_{\rm LSR}\rangle =3.9\pm1.6$ km s$^{-1}$, i.e. $\langle V_{\rm rad}\rangle=10.0\pm2.1$ km s$^{-1}$, which is in fairly 
good agreement with the average velocity of the gas ($\langle V_{\rm LSR}\rangle=3.0\pm0.7$ km s$^{-1}$, 
i.e. $\langle V_{\rm rad}\rangle =9.6\pm0.7$ km s$^{-1}$ (\citealt{mizunoetal1999}).

\subsection{Lithium equivalent width and radial velocity}
Lithium equivalent widths ($EW_{\rm Li}$) were measured by direct integration or by Gaussian fit using the IRAF task {\sc splot}. Errors in 
$EW_{\rm Li}$ were estimated in the following way: $i)$ when only one spectrum was available, the standard deviation of three $EW_{\rm Li}$ 
measurements was adopted; $ii)$ when more than one spectrum was gathered, the standard deviation of the measurements on the different 
spectra was adopted. Typical errors in $EW_{\rm Li}$ are of 0.001--0.087 \AA\ (higher values for GIRAFFE data); for the stars C62 and C66, 
$\sigma_{EW_{\rm Li}}\sim$0.15\,\AA. Our $EW_{\rm Li}$ measurements are consistent with the values of \cite{spezzietal2008} within 0.02\,\AA.

\begin{figure}[h!]
\begin{center}
 \begin{tabular}{c}
\includegraphics[width=8.8cm]{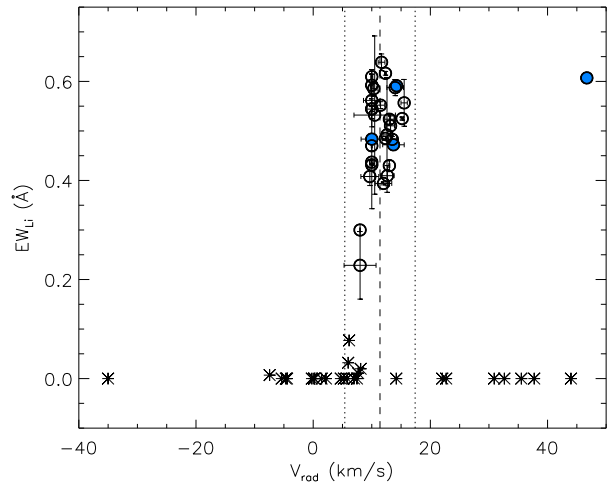}
\vspace{-.3cm}
 \end{tabular}
\caption{$EW_{\rm Li}$ versus RV for stars in the Cha~II FoV (see Table~\ref{tab:logs}). Open circles refer to 
most probable single stars, filled circles are multiple components, and asterisks refer to field stars. We excluded stars whose binarity 
was detected from RV variation at different phases (namely, Sz51 and Hn~24; see Table~\ref{tab:all_param}). In the case of Sz54, observed 
with both FLAMES configurations, we considered only the UVES RV values. Vertical lines represent the $\langle V_{\rm rad}\rangle\pm 3\sigma$ values, 
where $\sigma$=2.0 km s$^{-1}$ (see Section \ref{sec:rad_vel}).
}
\label{fig:vrad_ewli}
 \end{center}
\end{figure}

Figure~\ref{fig:vrad_ewli} shows the $EW_{\rm Li}$ versus RV for the 37 Cha~II members listed in Table~\ref{tab:all_param} 
(circles) and for the field stars (asterisks). The difference in RV distribution between the no-lithium or 
weak-lithium stars, and the strong-lithium stars is noticeable. The strong Li stars are confined to a narrow range of velocities 
(i.e., inside $\pm3\sigma$ from the peak of the RV distribution). The strong-lithium sample contains single stars and also the two most 
probable SB1 systems (C66 and IRAS~F13052$-$7653N) and the two SB2 systems (RXJ1301.0$-$7654a and Sz~54), while the rest can be considered 
as field stars. For the SB2 stars, the average of the RVs of their components falls inside the $\langle V_{\rm rad}\rangle\pm3\sigma$ distribution. 
The relatively narrow RV distribution of the strong-lithium sample confirms that these stars are all members of the same association.

\subsection{Elemental abundances}
\label{sec:abundances}
\subsubsection{Abundance measurements}
The FLAMES/UVES wide spectral coverage allows us to select several tens of \ion{Fe}{i}+\ion{Fe}{ii} lines and spectral features of 
other elements to measure abundances from line EWs. To this aim, as done in \cite{biazzoetal2011b}, we discarded stars 
with $T_{\rm eff} \ltsim 4000$ K (because of significant formation of molecules in the atmosphere), fast rotators (to avoid rotational blending), 
and strong accretors (for which accurate abundance analysis is hampered). In the end, only one star (Hn~23) fulfills the required criteria. 
Effective temperature and surface gravity ($\log g$) from the literature (Table~\ref{tab:liter_param}) were used as initial values, 
and initial microturbulence ($\xi$) was set to 1.5 km\,s$^{-1}$. Final values of the atmospheric parameters are listed in Table~\ref{tab:abundances}, 
together with abundance determinations, abundance internal errors, and number of lines considered (in parenthesis). The first source of internal 
error in abundance is due to uncertainties in line EWs, while another contribution comes from the uncertainties in stellar parameters. 
Systematic (external) errors, introduced by the code and/or model atmosphere, are negligible in comparison with the internal ones (see 
\citealt{biazzoetal2011a} for details on the treatment of errors). 

\begin{figure}	
\begin{center}
 \begin{tabular}{c}
\includegraphics[width=8.8cm]{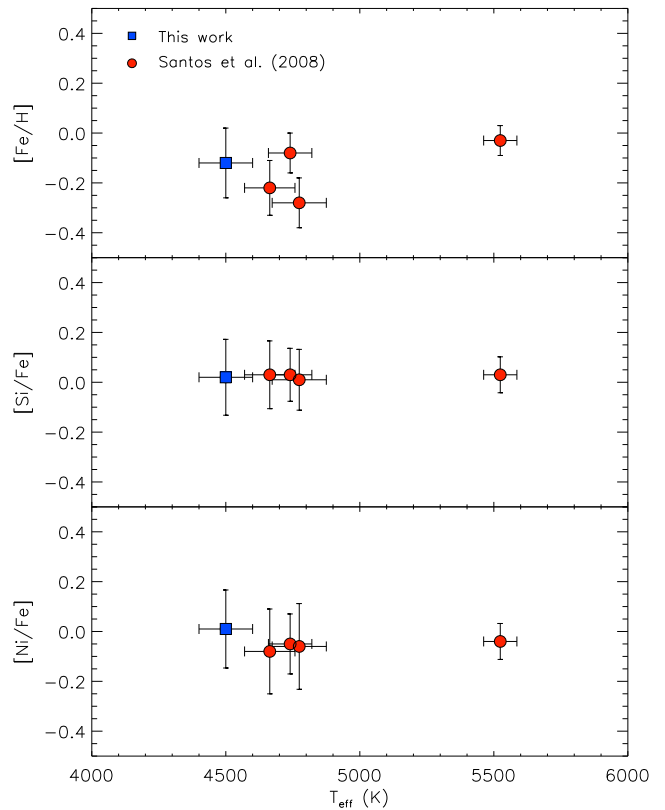}
\vspace{-.3cm}
 \end{tabular}
\caption{Iron, silicon, and nickel abundances versus spectroscopic temperature for Hn~23 and the stars analyzed 
by \cite{santosetal2008}.
}
\label{fig:abundances}
 \end{center}
\end{figure}

The [Fe/H] value of Hn~23 is slightly below the solar value and in agreement with the mean abundance of $<$[Fe/H]$>=-0.11\pm0.11$ found by 
\cite{santosetal2008} for the Chamaeleon Complex (see Fig.~\ref{fig:abundances}). This supports the suggestion that SFRs in the solar 
neighborhood are slightly more metal-poor than nearby young open clusters (\citealt{biazzoetal2011a}). This issue 
certainly deserves further study, using more stars in the region and homogeneous samples of PMS stars in as many SFRs as 
possible. Moreover, all other [X/Fe] abundances are close to the solar ones, with silicon and nickel close to the cluster mean value 
of $<$[Si/Fe]$>=0.03\pm0.01$ and $<$[Ni/Fe]$>=-0.05\pm0.02$ found by \cite{santosetal2008}. Titanium seems to be affected by NLTE effects 
(see Table~\ref{tab:abundances}), as previously found by other authors for stars with temperatures cooler than $\sim 5000$ K (see, e.g., 
\citealt{dorazirandich2009, biazzoetal2011a}, and references therein). However, detailed treatment of NLTE effects is beyond 
the scope of this paper. 

\begin{table}
\caption{Spectroscopic parameters and elemental abundances of Hn~23.} 
\label{tab:abundances}
\begin{center}
\begin{tabular}{ll}
\hline
\hline
\multicolumn{2}{c}{Spectroscopic parameters}\\
~\\
$T_{\rm eff}$ (K)  & $4500\pm100$   \\
$\log g$ (dex)     & $4.0\pm0.2$   \\
$\xi$ (km~s$^{-1}$)       & $1.9\pm0.2$   \\
\hline
\multicolumn{2}{c}{Elemental abundances}\\
~\\
$[$\ion{Fe}{i}/H$]$  & $-$0.12$\pm$0.14$\pm$0.06(41)  \\
$[$\ion{Fe}{ii}/H$]$ & $-$0.13$\pm$0.21$\pm$0.06(3)  \\
$[$Al/Fe$]$ & $-$0.02$\pm$0.14$\pm$0.05(2)   \\
$[$Si/Fe$]$ & $+$0.02$\pm$0.15$\pm$0.07(1)   \\
$[$Ca/Fe$]$ & $-$0.07$\pm$0.15$\pm$0.10(3)   \\
$[$\ion{Ti}{i}/Fe$]$ & $-$0.20$\pm$0.18$\pm$0.09(5)   \\
$[$\ion{Ti}{ii}/Fe$]$& $+$0.04$\pm$0.16$\pm$0.08(1)   \\
$[$Ni/Fe$]$ & $+$0.01$\pm$0.16$\pm$0.04(7)   \\
\hline
\end{tabular}
\end{center}
\end{table}

\subsubsection{Lithium abundance}
Mean lithium abundances were estimated from the average $EW_{\rm Li}$ listed in Table~\ref{tab:all_param} and $T_{\rm eff}$ values 
from \cite{spezzietal2008}, by using the LTE curves-of-growth reported by \cite{pavlenkomagazzu1996} for $T_{\rm eff}>3500$~K, 
and by \cite{pallaetal2007} for $T_{\rm eff}<3500$~K. The $\log g$ values were derived using the effective temperature, luminosity, and mean 
mass reported for each star in \cite{spezzietal2008}. The main source of error in $\log n{\rm (Li)}$ comes from the uncertainty in $T_{\rm eff}$, 
which is $\Delta T_{\rm eff}\sim100$~K (\citealt{spezzietal2008}). Taking this value and a mean error of 0.020\,\AA\ in $EW_{\rm Li}$ into account, 
we estimate a mean $\log n{\rm (Li)}$ error ranging from $\sim$0.07$-$0.10 dex for cooler stars ($T_{\rm eff}\sim3200$ K) down to 
$\sim$0.05$-$0.09 dex for warmer stars ($T_{\rm eff}\sim4200$ K), depending on the $EW_{\rm Li}$ value. Moreover, the $\log g$ value 
affects the lithium abundance, in the sense that the lower the surface gravity the higher the lithium abundance, and vice versa. In particular, 
the difference in $\log n{\rm (Li)}$ may rise to $\sim$$\pm0.05$ dex when considering stars with mean values of $EW_{\rm Li}=0.500$\,\AA\ and 
$T_{\rm eff}=4000$ K and assuming $\Delta \log g = \mp0.5$ dex.

\begin{figure}[t!]
\begin{center}
 \begin{tabular}{c}
\includegraphics[width=8.8cm]{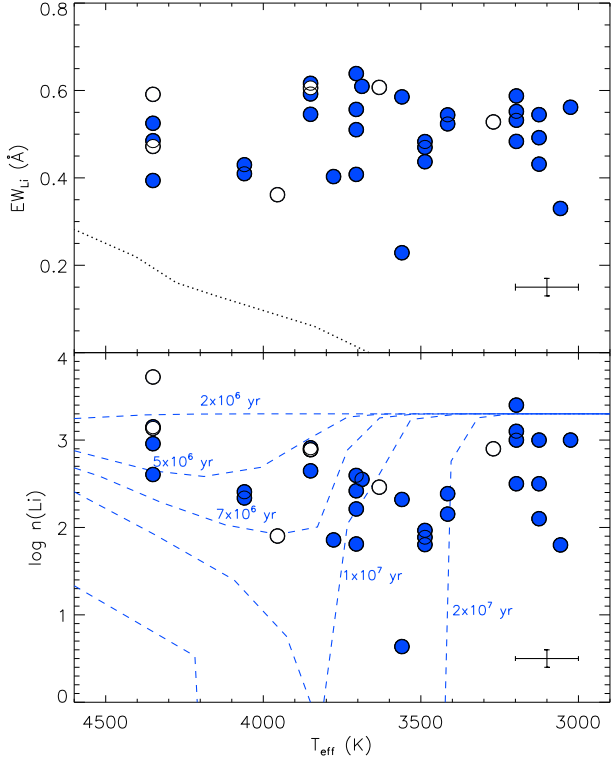}
\vspace{-1.3cm}
 \end{tabular}
\caption{{\it Upper panel:} $EW_{\rm Li}$ versus effective temperature. The upper envelope for the Pleiades, as 
adapted by \cite{soderblometal1993}, is overplotted as a dotted line. {\it Lower panel:} Lithium abundance versus effective temperature. 
The ``lithium isochrones'' by \cite{dantonamazzitelli1997} in the 2--20 Myr range are overlaid with dashed lines. 
In both panels, empty symbols represent spectroscopic binaries, while mean error bars are overplotted on the lower right corner.}
\label{fig:lognLi_Teff}
 \end{center}
\end{figure}

In Fig.~\ref{fig:lognLi_Teff} we show the mean lithium abundance as a function of the effective temperature (see Table~\ref{tab:all_param} 
for the $\log n{\rm (Li)}$ values). The average of $\log n{\rm (Li)}$ is about 2.5 dex with a dispersion of 0.6 dex. 
The lowest lithium abundance values are presumably due to spectral veiling, which affects the line EW.

\subsection{Accretion diagnostics and mass accretion rates}
\label{sec:accretion_rates}
The spectral coverage of our data allows us to select several lines (namely, H$\alpha$ $\lambda$6563 \AA, H$\beta$ $\lambda$4861 \AA, 
\ion{He}{i} $\lambda$5876 \AA, \ion{He}{i} $\lambda$6678 \AA, and \ion{He}{i} $\lambda$7065 \AA) that can be used to determine the accretion 
luminosity ($L_{\rm acc}^{\lambda}$). These emission lines are powered by processes related to accretion from the circumstellar disk 
(\citealt{herczeghillenbrand2008}). The use of these lines as secondary accretion diagnostics relies on empirical linear relationships between 
the observed line luminosity ($L^{\lambda}$) and the accretion luminosity (e.g., \citealt{herczeghillenbrand2008}). These relationships 
have been established through primary diagnostics, such as UV excess emission (\citealt{gullbringetal1998}). We used the 
$L^{\lambda}-L_{\rm acc}^{\lambda}$ empirical relations of \cite{herczeghillenbrand2008} to derive 
$L_{\rm acc}^{\lambda}$. The line luminosity was calculated as $L^{\lambda} = 4 \pi R_\star^2 F^{\lambda}$, where the stellar 
radius, $R_\star$, was taken from \cite{spezzietal2008} and the observed flux at the stellar radius, $F^{\lambda}$, was derived by multiplying 
the EW of each line ($EW_{\lambda}$) by the continuum flux at wavelengths adjacent to the line 
($F_{\rm continuum}^{\lambda\pm\Delta\lambda}$). 
The latter was gathered from the NextGen Model Atmospheres (\citealt{hauschildtetal1999}), assuming 
the corresponding stellar temperature and gravity (see Table~\ref{tab:liter_param}). The mass accretion rate, $\dot M_{\rm acc}^{\lambda}$, 
was then derived from $L_{\rm acc}^{\lambda}$ using the following relationship (\citealt{hartmann1998}):
\begin{equation}
\dot M_{\rm acc}^{\lambda} = \left(1 - \frac{R_\star}{R_{\rm in}}\right)^{-1} \frac{L_{\rm acc}^{\lambda} R_\star}{G M_\star}\,,
\end{equation}
\noindent{where the stellar radius $R_\star$ and mass $M_\star$ for each star were taken from \cite{spezzietal2008}, and the inner-disk 
radius $R_{\rm in}$, when available, from \cite{alcalaetal2008}. When no $R_{\rm in}$ was available, we assumed $R_{\rm in}=5R_\star$ 
(see \citealt{hartmann1998}), which is a good approximation for most accretors, as pointed out by \cite{alcalaetal2011b}. Contributions 
to the error budget on $\dot M_{\rm acc}$ include uncertainties on stellar mass, stellar radius, 
inner-disk radius, and $L_{\rm acc}^{\lambda}$. Assuming mean errors of $\sim0.15 M_\odot$ in $M_\star$ (\citealt{spezzietal2008}), 
$\sim 0.10 R_\odot$ in $R_\star$ (\citealt{spezzietal2008}), and $\sim 0.2$ AU in $R_{\rm in}$ (\citealt{alcalaetal2008}), $5-10\%$ as 
relative error in $EW_{\lambda}$, 10\% in $F_{\rm continuum}^{\lambda\pm\Delta\lambda}$, and the uncertainties in the relationships by 
\cite{herczeghillenbrand2008}, we estimate a typical error in $\log \dot M_{\rm acc}$ of $\sim 0.5$ dex. 
}

Apart from variability phenomena, which will be discussed in Section~\ref{sec:short_variability}, the mass accretion rates 
derived from the various diagnostics should be consistent with each other. In the following, we describe the results drawn from the hydrogen 
and helium emission lines.

\begin{figure}[h!]
\begin{center}
 \begin{tabular}{c}
\includegraphics[width=8.8cm]{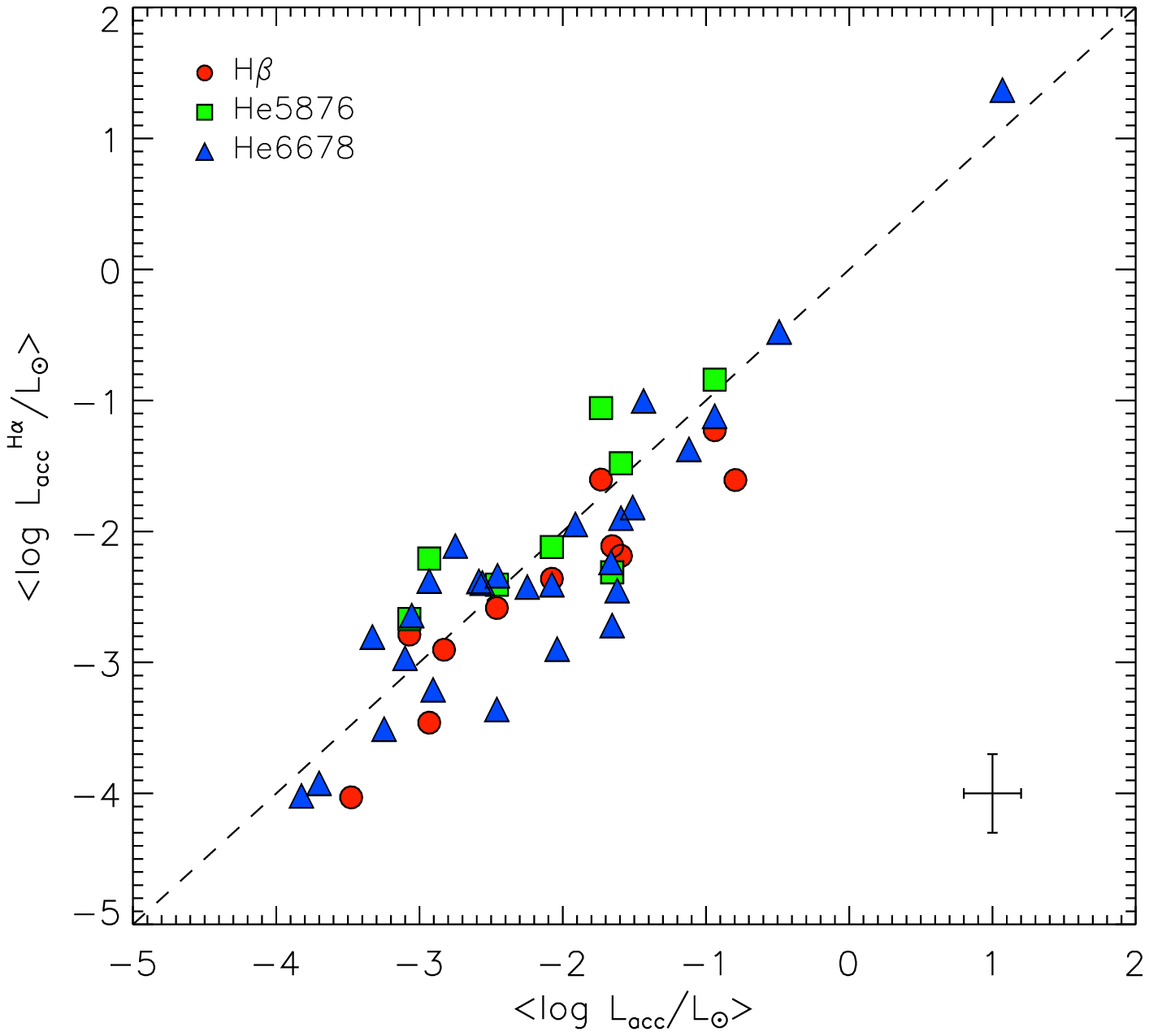}\\
\includegraphics[width=8.8cm]{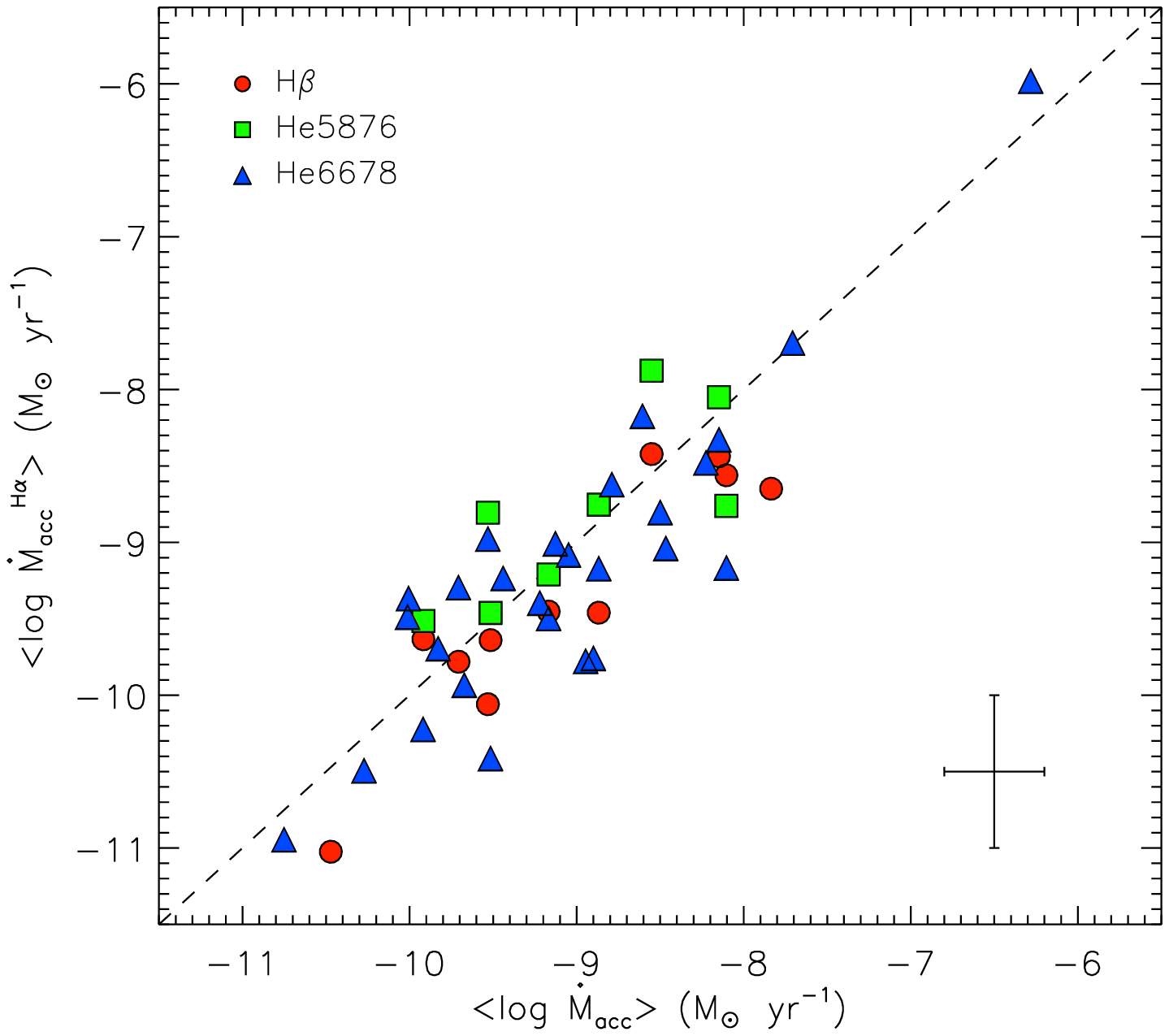}
\vspace{-.2cm}
 \end{tabular}
\caption{Average $L_{\rm acc}^{\rm H\alpha}$ ({\it top panel}) and $\dot M_{\rm acc}^{\rm H\alpha}$ ({\it bottom panel}) derived from 
the H$\alpha$ line as a function of $\langle L_{\rm acc}^{\rm \lambda}\rangle$ and $\langle\dot M_{\rm acc}^{\rm \lambda}\rangle$ obtained 
from the H$\beta$, \ion{He}{i} $\lambda$5876\,\AA, and \ion{He}{i} $\lambda$6678\,\AA\ lines. Mean error bars are overplotted on the 
lower right corner of each panel.}
\label{fig:mass_accr_rate_comparison}
 \end{center}
\end{figure}

\begin{figure}[h!]
\begin{center}
 \begin{tabular}{c}
\includegraphics[width=8.8cm]{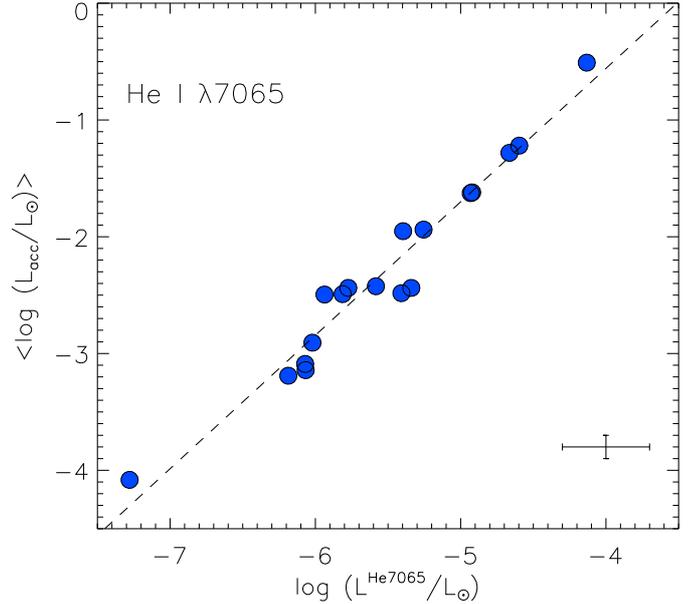}
\vspace{-.3cm}
 \end{tabular}
\caption{Average accretion luminosity versus line luminosity for the \ion{He}{i} $\lambda$7065\,\AA\ line. The linear fit given in the 
text (Equation~\ref{eq:lacc_l7065}) is represented by the dashed line. Mean error bars are overplotted on the lower right corner.
}
\label{fig:lum_accr_rate_he7065}
 \end{center}
\end{figure}

Figure~\ref{fig:mass_accr_rate_comparison} (upper panel) shows the mean accretion luminosity\footnote{This is the average of different 
observations for a given star.}, derived from the H$\alpha$ emission line, versus the mean accretion luminosities obtained from other hydrogen 
and helium emission lines. A fairly good agreement between these diagnostics is present. Comparing the mean mass accretion rate from the 
H$\alpha$ line with the mean $\dot M_{\rm acc}$ as obtained through the other diagnostics, the agreement is well reproduced (see lower panel 
in Fig.~\ref{fig:mass_accr_rate_comparison} and column 16 in Table~\ref{tab:accret_param}). This justifies the use of all the diagnostics 
to compute an average $\langle L_{\rm acc}\rangle$\footnote{This is the average of different diagnostics.} (and, hence, also an average 
$\langle\dot M_{\rm acc}\rangle$) for each star. A weighted average of $\langle L_{\rm acc}\rangle$ derived from the H$\alpha$, H$\beta$, 
\ion{He}{i} $\lambda$5876\,\AA, and \ion{He}{i} $\lambda$6678\,\AA\ emission lines allows us to analyze the relationship between 
$\langle L_{\rm acc}\rangle$ and the luminosity in the \ion{He}{i} $\lambda$7065\,\AA\ line, which is shown in Fig.~\ref{fig:lum_accr_rate_he7065}. 
A linear fit to the relationship gives
\begin{equation}
\langle\log L_{\rm acc}^{\rm He\lambda7065}\rangle = (4.01\pm0.35) + (1.14\pm0.06) \langle\log L^{\rm He\lambda7065}\rangle\,.
\label{eq:lacc_l7065}
\end{equation}

The good correlation justifies the use of the \ion{He}{i} $\lambda$7065\,\AA\ line as the reliable diagnostic of $L_{\rm acc}$. 

As already pointed out, the different line diagnostics yield consistent mass accretion rates (see 
Fig.~\ref{fig:mass_accr_rate_comparison} and Table \ref{tab:accret_param}). 
Considering, for instance, the H$\alpha$ line, which is observed in all targets, the mean difference in 
$\log \dot M_{\rm acc}^{\rm H\alpha}$ as compared to $\log \dot M_{\rm acc}^{\rm H\beta}$ is of $0.3\pm0.3$ $M_\odot$\,yr$^{-1}$
(with a maximum of 0.8~dex observed for Sz~54), while it is $-0.2\pm0.4$~dex with respect 
to $\log \dot M_{\rm acc}^{\rm He\lambda5876}$~(with a maximum of $-0.2$~dex observed for Sz~56), and $0.1\pm0.5$~dex with respect 
to $\log \dot M_{\rm acc}^{\rm He\lambda6678}$ (with a maximum of 1.1~dex for Sz~50). The mass accretion rate for the sample is in the 
range $10^{-11} \div 2\times10^{-8}$ $M_\odot$\,yr$^{-1}$, which is typical of Class II low-mass YSOs (see Fig.~2 in \citealt{sicilia-aguilaretal2010}). 
Excluding multiple systems, stars with $EW_{\rm H\alpha}\le10$ \AA, and the early-type star DK~Cha, 
we find an average mass accretion rate for Cha~II of $\langle\dot M_{\rm acc}\rangle\sim 7\times10^{-10}$ $M_\odot$\,yr$^{-1}$. 

\subsubsection{Comparison with previous studies}
Fourteen of our targets were also analyzed by \cite{antoniuccietal2011} as part of the POISSON (Protostellar Objects IR-optical Spectral 
Survey On NTT) project aimed at deriving the mass accretion rates of young stars in star-forming regions through low-resolution optical/near-IR 
spectroscopy. \cite{antoniuccietal2011} used the Br$\gamma$ line as an accretion tracer. They argue that this is the best diagnostic in their spectra 
when compared with other tracers (i.e., Pa$\gamma$, \ion{Ca}{ii}, H$\alpha$, and [\ion{O}{i}]). Comparing their $\log \dot M_{\rm acc}$ with 
our mean $\log \dot M_{\rm acc}$, their values tend to be systematically higher than ours. 
In particular, their values strongly diverge from ours at $\log \dot M_{\rm acc}<-8$, with a mean difference of $1.0\pm0.8$~dex. An analogous 
trend was found by the same authors when comparing $L_{\rm acc}^{\rm Pa\beta, [\ion{O}{i}], H\alpha, \ion{Ca}{ii}}$ with 
$L_{\rm acc}^{\rm Br\gamma}$ (see their Fig.~4), which is more evident in their $L_{\rm acc}^{\rm H\alpha}-L_{\rm acc}^{\rm Br\gamma}$ 
diagram. They ascribe this behavior to enhanced chromospheric emission or absorption from outflowing material, direct photoionization 
at higher luminosities, or flux losses due to winds.

\begin{figure*}	
\begin{center}
 \begin{tabular}{c}
\includegraphics[width=18.5cm]{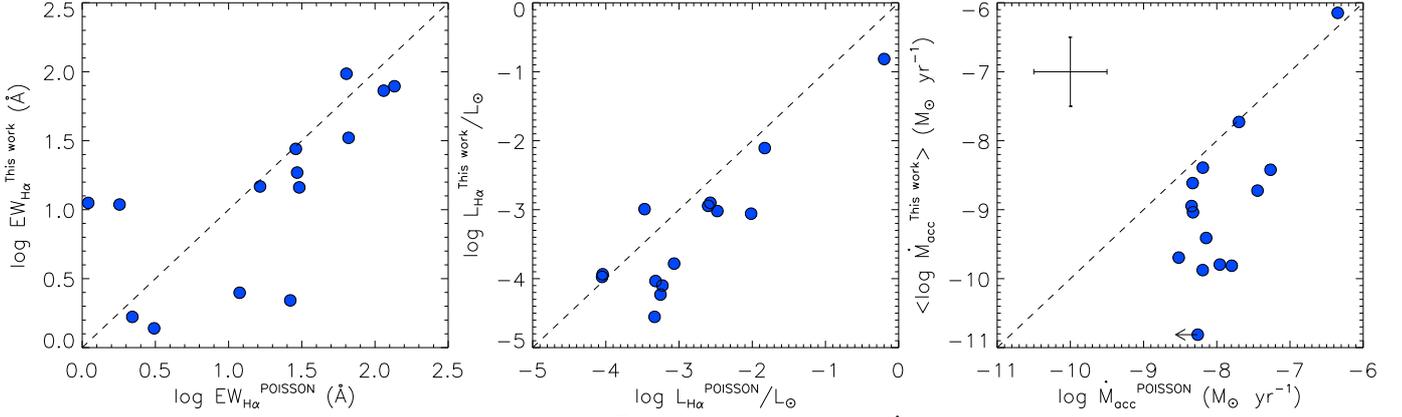}
\vspace{-.4cm}
 \end{tabular}
\caption{Comparison between our $EW_{\rm H\alpha}$ ({\it left panel}), $L^{\rm H\alpha}$ ({\it middle panel}), and 
$\dot M_{\rm acc}$ ({\it right panel}) values and those obtained by \cite{antoniuccietal2011}. In the right panel, 
mean error bars are overplotted on the upper left corner.
}
\label{fig:our_Antoniucci_comp}
 \end{center}
\end{figure*}

When comparing the POISSON's $EW_{\rm H\alpha}$ with the values derived by us (left-hand panel of Fig.~\ref{fig:our_Antoniucci_comp}), the 
difference is $\Delta EW_{\rm H\alpha}=10\pm23$ \AA. Considering that the observations were performed at different epochs (our run was 
in 2006, while their run was in 2009) and that their $EW_{\rm H\alpha}$ were measured on low-resolution spectra ($R\sim700$), the agreement 
is fairly good. In fact, a good correlation is found between our estimated H$\alpha$ line luminosities and the POISSON's values, with a 
difference of $0.04\pm0.13 L_{\odot}$ (see Fig.~\ref{fig:our_Antoniucci_comp}, middle panel). Thus, the differences in mass accretion rate arise 
when deriving $L_{\rm acc}$. A similar behavior as the one shown in Fig.~\ref{fig:our_Antoniucci_comp} for the mass accretion rate is found when 
our $L_{\rm acc}$ values are compared with those of POISSON, in agreement with the results by \cite{antoniuccietal2011}.
This means that the differences between POISSON and our determinations are mainly due to the tracers used to derive $L_{\rm acc}$. 
As shown in Fig.~\ref{fig:mass_accr_rate_comparison}, very good correlations of $L_{\rm acc}$ as derived from the \ion{He}{i}~and H$\alpha$ 
lines are found. Should there be an important contribution to the line diagnostics by winds and/or chromospheric activity, our estimates of 
$L_{\rm acc}$ would be in excess with respect to those derived from the Br$\gamma$ line. Instead, the opposite is observed. We thus exclude 
the possibility that the differences are due to the influence of winds and chromospheric activity. A possible explanation for the difference 
between our average $L_{\rm acc}$ values and the POISSON's $L_{\rm acc}^{\rm Br\gamma}$ may be that the various accretion diagnostics originate 
in different regions at different physical conditions. In order to investigate this issue, measurements of primary accretion tracers, 
and Br$\gamma$ measurements are needed. Such analysis cannot be conducted with the data available here, but will 
be addressed in our future studies exploiting X-Shooter@VLT data (cf. \citealt{alcalaetal2011b}).

\subsubsection{Other lines in the optical}
A number of optical lines are seen in emission in the spectra of several stars in our sample. These are the forbidden lines of 
[\ion{O}{i}] $\lambda$6300.3\,\AA, and $\lambda$6363.8\,\AA, [\ion{S}{ii}] $\lambda$6715.8\,\AA, and $\lambda$6729.8\,\AA, 
[\ion{N}{ii}] $\lambda$6548.4\,\AA, and $\lambda$6583.4\,\AA, as well as \ion{Fe}{ii} multiplets. These lines trace mainly 
stellar/disk winds, jets, disk surfaces, and outflow activity (\citealt{cabritetal1990, hartiganetal1995}), and their detailed 
treatment is beyond the scope of this paper. Here, we only note that the [\ion{O}{i}] $\lambda$6300.3\,\AA\ line is detected in 6/11 objects, 
the [\ion{O}{i}] $\lambda$6363.8\,\AA\ line in 1/11 objects, the [\ion{S}{ii}] 6715.8\,\AA\ and 6729.8\,\AA\ emission is observed in 7/40 and 
9/40 sources, respectively, and the [\ion{N}{ii}] 6548.4\,\AA\ and 6583.4\,\AA\ lines are present in 5/40 and 5/40 stars (see 
Table~\ref{tab:all_param}). In the case of Sz\,51, the star showing the strongest H$\alpha$ in our FLAMES/UVES sample, there is 
evidence of other emission lines (such as the \ion{Mg}{i} triplet at $\lambda$5167.3, 5172.7, 5183.6\,\AA, the multiplets 
42 and 49 of \ion{Fe}{ii}, etc.) indicating mass loss.

\section{Discussion}
\label{sec:discussion}

\subsection{Accretion versus stellar age and mass}
\label{sec:accretion_parameters}
\begin{figure}	
\begin{center}
 \begin{tabular}{c}
\includegraphics[width=9cm]{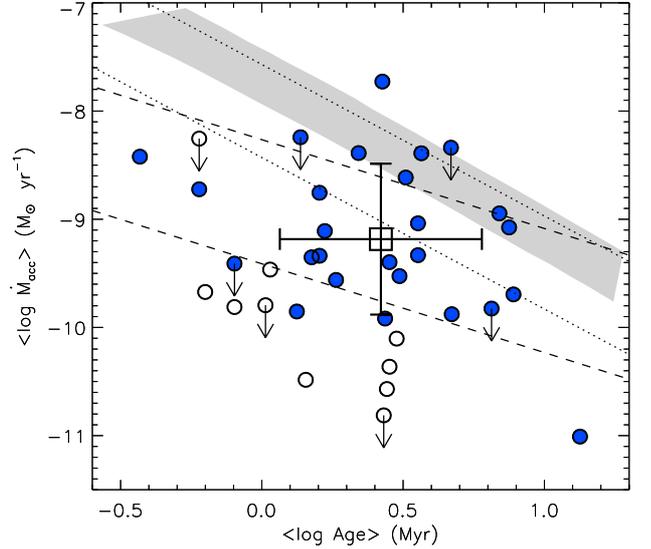}
\vspace{-.4cm}
 \end{tabular}
\caption{Mean mass accretion rate versus age. Open symbols represent the targets with mean $EW_{\rm H\alpha}\le10$\,\AA, 
while binaries are evidenced as upper limits. The big square represents the mean position of Cha II, 
taking into account the single stars with mean $EW_{\rm H\alpha}>10$\,\AA\ (vertical 
and horizontal error bars correspond to the standard deviations from the mean $\dot M_{\rm acc}$ and age, respectively).
The dashed lines mark minimum and maximum limits of Eq.~\ref{eq:macc_age}, while dotted lines represent the analogous 
limits of the relationship derived for Cha I by \cite{hartmannetal1998}. The collection of viscous disk evolutionary models 
for solar-type stars with initial disk masses of $0.1-0.2 M_\odot$, constant viscosity $\alpha=10^{-2}$, and viscosity exponent 
$\gamma=1$ reported by \cite{sicilia-aguilaretal2010} is also displayed (filled region).}
\label{fig:mass_accr_rate_age}
 \end{center}
\end{figure}

Figure~\ref{fig:mass_accr_rate_age} shows the mean mass accretion rate versus stellar age (see Table~\ref{tab:liter_param}) 
for all the targets except the early-type star DK~Cha, which has a massive disk and high $\dot M_{\rm acc}$ (\citealt{sicilia-aguilaretal2010}). 
A slightly decreasing trend with age may be present, though over a narrow age interval ($\sim 0.4-13.4$ Myr) 
and with a large scatter in $\dot M_{\rm acc}$. This would be consistent with the evolution of a viscous disk (see, e.g., 
\citealt{hartmannetal1998, sicilia-aguilaretal2010}, and references therein), although the mean $\dot M_{\rm acc}$ is slightly lower 
than that expected from the model at the Cha~II age. In order to quantify the degree of anti-correlation between $\log \dot M_{\rm acc}$ 
and $\log Age$, we calculated the Spearman's rank correlation coefficient using the IDL procedure R\_CORRELATE (\citealt{pressetal1986}). 
We find a correlation coefficient $\rho=-0.38$, with a probability of obtaining such $\rho$ from randomly distributed data of $p=0.08$. 
This seems to confirm a moderate anti-correlation between $\dot M_{\rm acc}$ and age, with an average mass accretion rate of 
$7^{+26}_{-5} \times 10^{-10} M_\odot$\,yr$^{-1}$ at a mean age of $3^{+3}_{-1}$ Myr. The linear relation we obtain for the likely single 
stars with mean $EW_{\rm H\alpha}$ higher than 10\,\AA\ is
\begin{equation}
\langle\log\dot M_{\rm acc}~{(M_\odot\,{\rm yr}^{-1})}\rangle=-8.84\pm0.22-0.82\pm0.40\log Age~{\rm (Myr)\,.}
\label{eq:macc_age}
\end{equation} 
The slope is higher than, yet consistent within the errors with, that obtained by \cite{hartmannetal1998} in Cha~I (i.e., 
$\log \dot M_{\rm acc}~{(M_\odot\,{\rm yr}^{-1})}=-8.00\pm0.10-1.40\pm0.29\log Age~{\rm (Myr)}$). We warn, however, about 
the large uncertainties in absolute ages derived from theoretical models for stars younger than $\sim 10$ Myr (see, e.g., 
\citealt{spezzietal2008}). We note also that a strong constraint on the apparent trend in $\log \dot M_{\rm acc} $ versus age is set 
by only one object in the sample (C41, age $\sim 13$ Myr). The trend disappears if this object is not considered.

\begin{figure}	
\begin{center}
 \begin{tabular}{c}
\includegraphics[width=9cm]{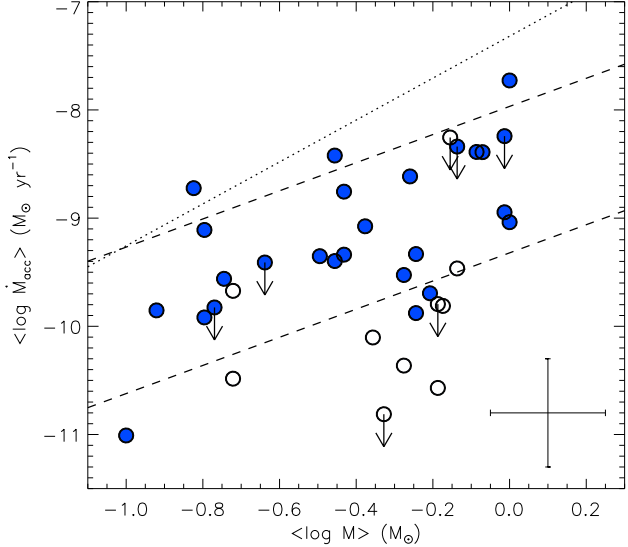}
\vspace{-.4cm}
 \end{tabular}
\caption{Mean mass accretion rate versus mean stellar mass. Symbols as in Fig.~\ref{fig:mass_accr_rate_age}. 
The dashed lines mark minimum and maximum limits of Eq.~\ref{eq:macc_mass}, while the dotted line represents the mean 
relationship derived for Cha I by \cite{antoniuccietal2011}. Mean error bars are overplotted on the lower right corner.}
\label{fig:mean_mass_accr_rate_mass} 
 \end{center}
\end{figure}

Figure~\ref{fig:mean_mass_accr_rate_mass} shows the mean mass accretion rate versus mean stellar mass (see Table~\ref{tab:liter_param}). 
Taking into account single stars with $EW_{\rm H\alpha}>10$\,\AA, we find the following linear relationship:
\begin{equation}
\label{eq:macc_mass}
\langle\log\dot M_{\rm acc}\,{(M_\odot{\rm yr}^{-1})}\rangle = -8.64\pm0.21+1.30\pm0.41\log M_\star\,{(M_\odot)\,,}
\end{equation}
where the slope is slightly lower than that found for Cha~I ($\sim 2$; see \citealt{antoniuccietal2011}, and references therein). 
A correlation of $\dot M_{\rm acc}$ with stellar mass is evident. Spearman's rank correlation coefficient is $\rho=0.51$ with a probability 
$p=0.02$. The exponent of the $\dot M_{\rm acc}-M_\star$ power law is consistent with the range of values $\sim 1.0-2.8$ found for low-mass 
stars in other SFRs (see, e.g., \citealt{herczeghillenbrand2008}, \citealt{fangetal2009}; \citealt{antoniuccietal2011}, and references therein).

\subsection{Short timescale variability}
\label{sec:short_variability}

Young PMS stars are known to be variable, due to the combination of different processes (\citealt{herbstetal1994}) occurring on 
different timescales: on short timescale (days), variability can be induced by rotation of cool or hot spots (type I 
variability), and on long timescales (months--years), accretion rate changes (type II variability) or obscuration by circumstellar dust 
might occur (type III variability, e.g., \citealt{schisano2009}). The time span of our observations is $\sim 48$ hours. 
Emission line variability on timescales shorter than two days is observed for several objects and for all the analyzed lines (see 
Fig.~\ref{fig:mass_accr_rate_mass}). In particular, Figs.~\ref{fig:halpha_profiles_giraffe}, \ref{fig:halpha_profiles_uves}, and 
\ref{fig:hbeta_profiles_uves} show the H$\alpha$ and H$\beta$ profiles of the stars displaying the strongest line variations, while 
Fig.~\ref{fig:mass_accr_rate_mass} shows the mass accretion rates derived from the H$\alpha$, H$\beta$, \ion{He}{i} $\lambda$5876\,\AA, 
and \ion{He}{i} $\lambda$6678\,\AA\ lines versus stellar mass. The vertical bars in these plots represent the range of $\dot M_{\rm acc}$ 
as due to the two-day variability of the corresponding diagnostics.

In general, on the timescale of only two days, the H$\alpha$ equivalent width of some stars changes up to a factor of $2-3$ and, 
{\it assuming} that this is due to accretion, variation in $\log \dot M_{\rm acc}$ would be $0.2-0.6$ dex, i.e., a factor of $1.6-4.0$. We 
conclude that this $\log \dot M_{\rm acc}$ variability, even if large, cannot explain the $\log \dot M_{\rm acc}$ spread at a given 
mass. This means that other stellar properties besides mass must also affect the variations. 

\subsection{Accretion and Metallicity}
\label{sec:accretion_metallicity}
Investigating the dependency of the mass accretion rate upon iron abundance in SFRs is important for two reasons. First, while the correlation 
between stellar metallicity and presence of giant planets around solar-type stars is well established (see, e.g., 
\citealt{johnsonetal2010} and references therein), the metallicity-planet connection in the early stages of planetary formation is 
still a matter of debate. The evolution of $\dot M_{\rm acc}$ is affected by possible planetary formation in the disk, and hence it might 
provide important clues on the planet-metallicity correlation. Second, the efficiency of the dispersal of circumstellar (or protoplanetary) 
disks and hence the dispersal timescale are predicted to depend on metallicity in the sense that planetary formation is faster in disks with 
higher metallicity (\citealt{ercolanoclarke2010}). \cite{yasuietal2010} find that the disk fraction in significantly low-metallicity clusters 
([O/H]$\sim -0.7$) declines rapidly in $<1$ Myr, which is much faster than the value of $\sim 5-7$ Myr observed in solar-metallicity clusters. 
Since the shorter disk lifetime reduces the time available for planetary formation, they suggest that this could be one of the reasons for the 
strong planet-metallicity correlation.

Recent studies by \cite{demarchietal2011} and \cite{spezzietal2012} in the Large and Small Magellanic Clouds show that metal-poor stars 
accrete at higher rates compared with solar-metallicity stars in galactic SFRs. Summarizing the mean $\dot M_{\rm acc}$ of 
low-mass ($0.1-1.0 M_\odot$) Class II stars members of $\sim 3-4$ Myr old nearby SFRs for which iron abundance has been recently measured 
(see Table~\ref{tab:fe_macc}), it is only possible to point out that for [Fe/H]$\sim0$ the mass accretion rate is $\sim 10^{-10}$ $M_\odot$\,yr$^{-1}$.

\begin{table*}
\caption{Iron abundances and mass accretion rates in nearby SFRs.} 
\label{tab:fe_macc}
\begin{center}
\begin{tabular}{lrl|ll}
\hline
Star-Forming Region & [Fe/H] & Reference & $\dot M_{\rm acc}$ & Reference\\
    & (dex)  &  &   ($M_\odot$\,yr$^{-1}$)   & \\
\hline
Orion Nebula Cluster & $-0.13\pm0.02$ & \cite{biazzoetal2011a}       & $3\times10^{-9}$ & \cite{robbertoetal2004}\\
$\sigma$~Orionis     & $-0.02\pm0.09$ & \cite{gonzalez-hernandez2008}& $3\times10^{-10}$& \cite{rigliacoetal2011a}\\
Taurus               & $0.00\pm0.07$  & \cite{dorazietal2011}        & $3\times10^{-9}$ & \cite{gudeletal2007}\\
Chamaeleon II        & $-0.12\pm0.14$ & This work                    & $7\times10^{-10}$& This work\\
\hline
\end{tabular}
\end{center}
\end{table*}

\subsection{Fraction of accretors versus age}
Excluding the six spectroscopic binaries, 27 of the studied stars result in having mean $EW_{\rm H\alpha}$ higher than 10\,\AA, which would 
imply a percentage of accretors of about 26/34$=$79\% (34 being the total number of single stars in the sample). However, the 
majority of PMS stars in Cha~II have a spectral type later than K7, with most of them later than M3. Therefore, according 
to the criterion by \cite{whitebasri2003}, a more adequate dividing line between most probable accretors and non-accretors in Cha~II is 
$EW_{\rm H\alpha}$=20\,\AA. Using this criterion, 19 stars can be classified as true accretors, leading to a percentage of $\sim55\pm5$\%. 
This fraction of accretors is consistent with the average age of the cloud members. In fact, following the mass accretion rate evolution 
with time shown in Fig.~3 of \cite{fedeleetal2010}, the fraction of stars with ongoing mass accretion decreases fast with age, going 
from $\sim 60$\% at $1.5-2.0$~Myr down to $\sim 2$\% at 10~Myr.

\subsection{Color-$\dot M_{\rm acc}$ diagrams}
Near-infrared colors can be used to probe the inner disk region. \cite{hartiganetal1995}, studying a sample of 42 T~Tauri stars and using 
the $K-L$ color excesses, pointed out that the disk dispersion is mainly due to the formation of micron-sized dust particles, which 
combine to create planetesimals and protoplanets at the end of the CTTS phase. Protoplanets may clear the inner disk of gas and dust, causing 
the disk to lose its near-infrared color excess and at the same time opening a gap in the disk (\citealt{linpapaloizou1993}), thereby 
terminating accretion from the disk onto the star.

\begin{figure*}	
\begin{center}
 \begin{tabular}{c}
\includegraphics[width=18cm]{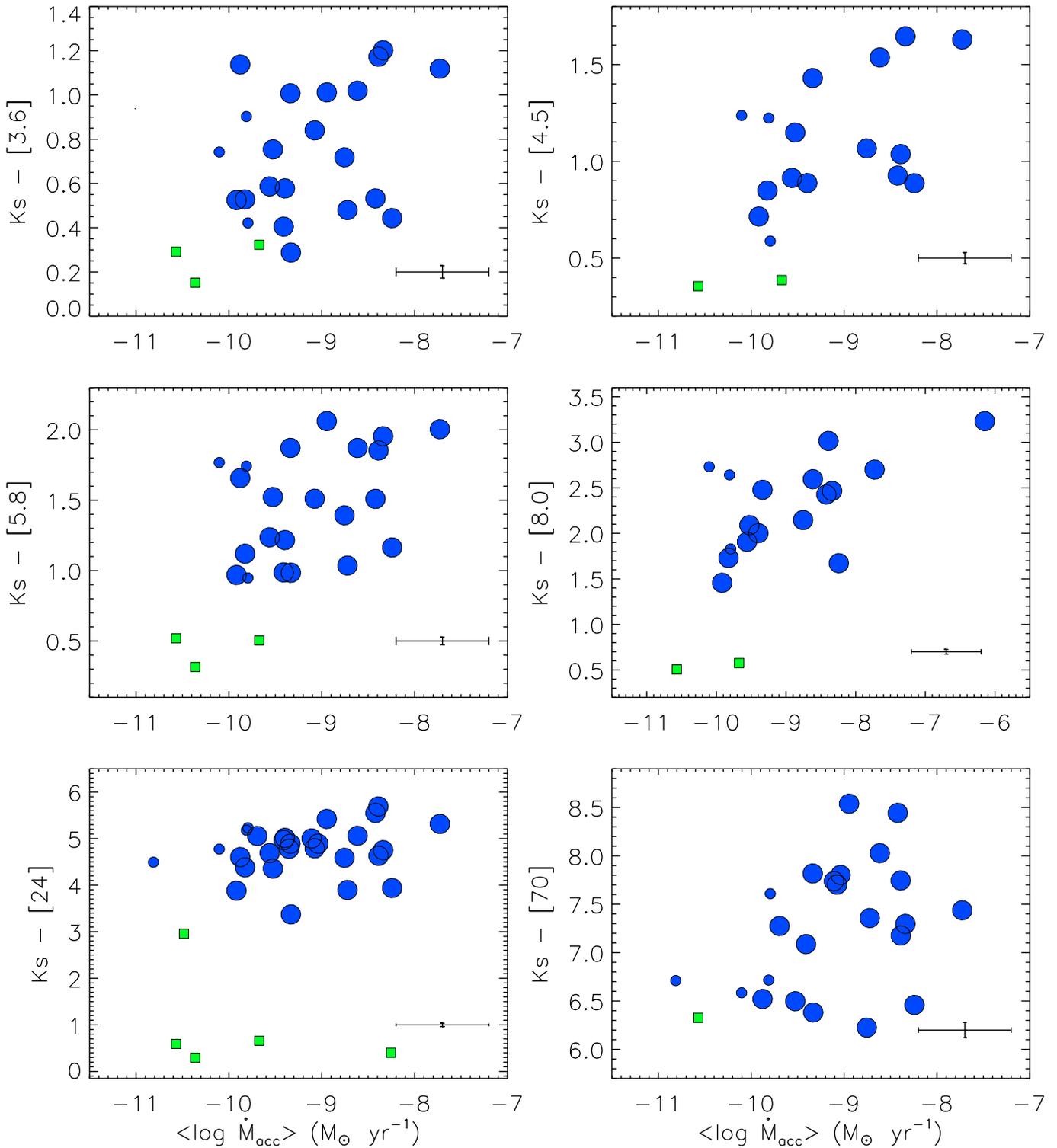}
\vspace{-.4cm}
 \end{tabular}
\caption{Infrared {\it 2MASS}-{\it Spitzer} colors versus mean mass accretion rates. Squares and dots correspond to Class III and Class II 
objects, respectively (see Table~\ref{tab:liter_param}). Symbol sizes represent stars with $EW_{\rm H\alpha}<10$\,\AA\ (small squares and dots) 
and with $EW_{\rm H\alpha}>10$\,\AA\ (big dots). Mean error bars are overplotted on the lower right corner of each panel.}
\label{fig:macc_2MASS_Spitzer_colors}
 \end{center}
\end{figure*}

\begin{figure*}	
\begin{center}
 \begin{tabular}{c}
\includegraphics[width=18cm]{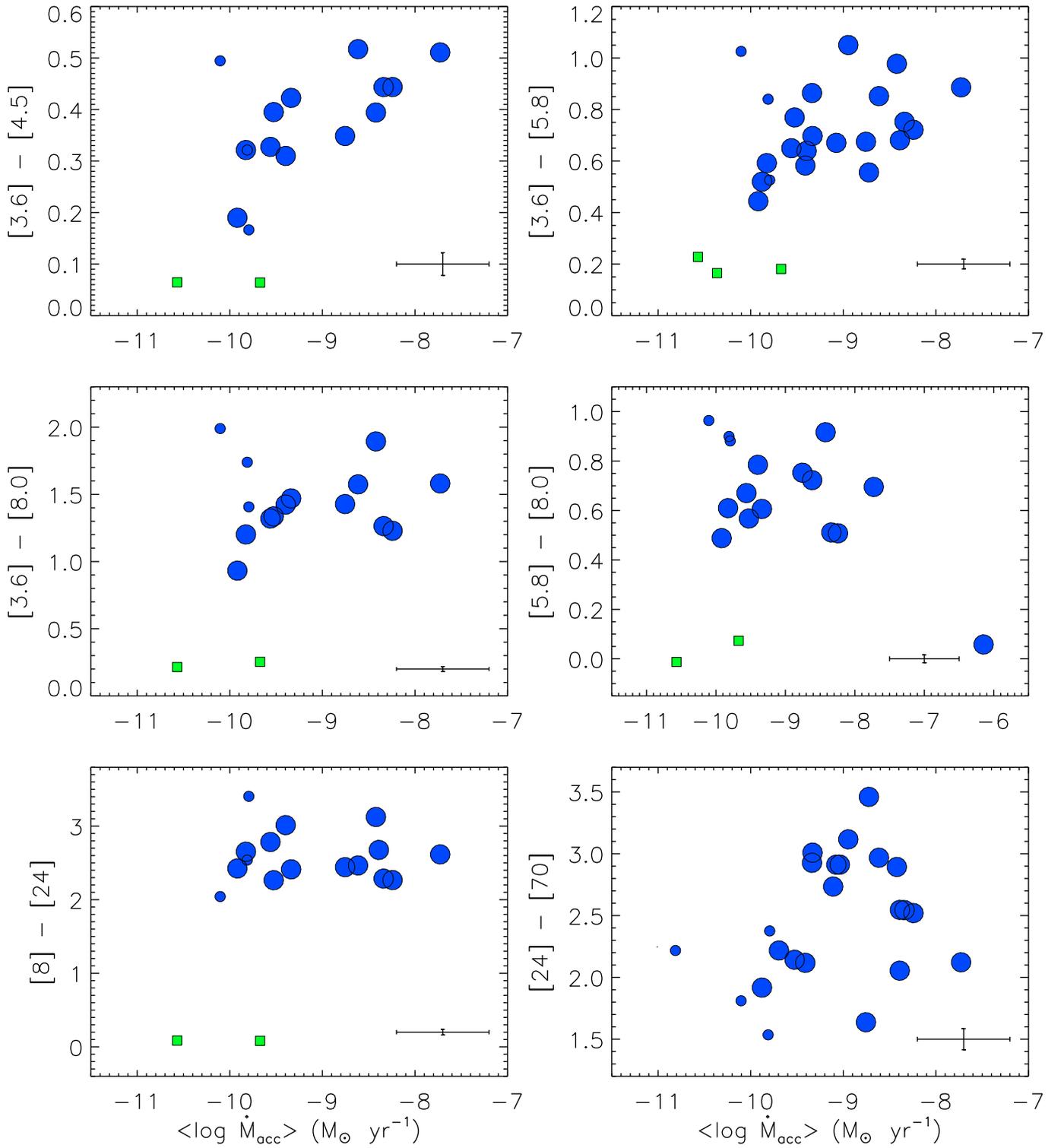}
\vspace{-.4cm}
 \end{tabular}
\caption{{\it Spitzer}-{\it Spitzer} colors versus mean mass accretion rates. Symbols as in Fig.~\ref{fig:macc_2MASS_Spitzer_colors}. 
Mean error bars are overplotted on the lower right corner of each panel.
}
\label{fig:macc_Spitzer_colors}
 \end{center}
\end{figure*}

With the aim of investigating possible relationships between near-infrared colors and accretion properties, we used $2MASS$ and {\it Spitzer}
\footnote{We considered the IRAC@{\it Spitzer} fluxes at 3.6, 4.5, 5.8, and 8.0 $\mu$m and the MIPS@{\it Spitzer} 
fluxes at 24 and 70 $\mu$m.} colors (see Figs.~\ref{fig:macc_2MASS_Spitzer_colors} and \ref{fig:macc_Spitzer_colors}). 
We considered these colors as disk tracers from the inner to the outer zone because they may estimate the magnitude of the near- and 
mid- infrared excesses above the photospheric level. In order to quantify the degree of correlation between these diagnostics, we calculated 
the Spearman's rank correlation coefficient, as we did in Section~\ref{sec:accretion_parameters}, and considered Class II stars with 
$EW_{\rm H\alpha}>10$\,\AA. The correlation coefficients, together with the probabilities, are listed in Table~\ref{tab:color_macc} and show 
that the best agreements are obtained for the $K$s$-$[8.0] and [3.6]$-$[4.5] colors versus $\langle\log \dot M_{\rm acc}\rangle$ parameters. 
Also, $K$s$-$[4.5], $K$s$-$[5.8], [3.6]$-$[5.8], and [3.6]$-$[8.0] colors versus $\langle\log \dot M_{\rm acc}\rangle$ show good agreement. 
This implies that objects with detectable accretion have optically thick inner disks. In particular, we can define the regions where 
$\dot M_{\rm acc} > 1.0\times10^{-10} M_\odot$\,yr$^{-1}$ and $K$s$-$[8.0]$>$1.5 or [3.6]$-$[4.5]$>$0.2, or [3.6]$-$[5.8]$>$0.5, or 
[3.6]$-$[8.0]$>$0.9 as those where accreting objects with infrared excess are found in Cha~II. The value 
$\dot M_{\rm acc}\sim10^{-10} M_\odot$\,yr$^{-1}$ is a reasonable threshold for the transition from optically thick to optically thin 
(inner) disk (\citealt{dalessioetal2006}), as also found by \cite{rigliacoetal2011a} in the $\sigma$~Orionis SFR. The rough trend we 
tentatively observe among optically thin and optically thick disks, which needs to be confirmed on larger samples, might indicate 
a link between the mass accretion rate and the grain properties. This link, in turn controls the disk geometry, a connection that is worth 
exploring further.

\begin{table}
\caption{Spearman's correlation coefficients ($\rho$) and probabilities ($p$) for different color-$\dot M_{\rm acc}$ relations.}
\label{tab:color_macc}
\begin{center}
\begin{tabular}{lrr}
\hline
Color & $\rho$ & $p$ \\
\hline
$K$s$-$[3.6]	& 0.28  & 24\% \\
$K$s$-$[4.5]	& 0.56  & 5\% \\
$K$s$-$[5.8]	& 0.52  & 2\% \\
$K$s$-$[8.0]	& 0.68  & 1\% \\
$K$s$-$[24]	& 0.29  & 16\% \\
$K$s$-$[70]	& 0.30  & 19\% \\
$[$3.6$]-$[$4.5]$ & 0.80  & 1\% \\
$[$3.6$]-$[$5.8]$ & 0.61  & 1\% \\
$[$3.6$]-$[$8.0]$ & 0.52  & 8\% \\
$[$5.8$]-$[$8.0]$ &$-$0.07& 80\% \\
$[$8.0$]-$[$24]$  &$-$0.09& 75\% \\
$[$24$]-$[$70]$ & 0.05  & 85\% \\
\hline
\end{tabular}
\end{center}
\end{table}

\section{Conclusions}
\label{sec:conclusions}
In this paper, we determined radial velocities, lithium abundances, and accretion properties of 40 members of the Chamaeleon~II star-forming region 
from FLAMES@VLT optical spectroscopy. Elemental abundances of Fe, Al, Si, Ca, Ti, and Ni for a suitable pre-main sequence star of the region were also 
measured. Our main results can be summarized as follows:
\begin{enumerate}
\item The average radial velocity of the stars is consistent with that of the gas ($\sim 10-12$ km s$^{-1}$). 
The dispersion of the radial velocity distributions of the stars and gas ($\sim 1-2$ km s$^{-1}$) are also in agreement. 
Similar results were found by \cite{dubathetal1996} in Chamaeleon~I (where $\langle V_{\rm rad}\rangle_{\rm Cha~I}\sim15.0\pm0.5$ km s$^{-1}$). 
\item A binary fraction of 18\% is found in Chamaeleon~II, which is similar to that in other T associations with comparable star density 
(see, e.g., the Taurus-Auriga association; \citealt{torresetal2002}).
\item The metallicity of the suitable member is slightly below the solar value, as also found by \cite{santosetal2008} for the Chamaeleon Complex.
\item We find an average lithium abundance of the star-forming region of $2.5\pm0.6$ dex.
\item Mass accretion rates derived through several secondary diagnostics (e.g., H$\alpha$, H$\beta$, \ion{He}{i} $\lambda$5876\,\AA, and 
\ion{He}{i} $\lambda$6678\,\AA) are consistent with each other, justifying the use of all of them to compute an average mass accretion 
rate for each star. 
\item We provide a relationship between accretion luminosity, $L_{\rm acc}$, and the luminosity of the \ion{He}{i} $\lambda$7065 \AA\ line.
\item The relationship between mass accretion rate ($\dot M_{\rm acc}$) and stellar mass ($M_\star$) in Chamaeleon~II is consistent 
with that found by previous studies in other T associations. 
\item Although slightly lower, the average mass accretion rate in Chamaeleon~II fits well the relationship $\dot M_{\rm acc}$ versus $Age$ reported 
by \cite{sicilia-aguilaretal2010}.
\item We cannot exclude that significant variability on timescales longer than the time span of our observations and possibly due to 
episodes of variable mass accretion may produce the vertical scatter observed in the $\dot M_{\rm acc}$ versus $M_\star$ plot. 
However, we suggest that such scatter of about two orders of magnitude in $\dot M_{\rm acc}$ at a given mass is also affected by other stellar properties. 
\item The color-$\dot M_{\rm acc}$ relationships suggest that the circumstellar disks in Cha~II become optically thin at $\sim 10^{-10} M_\odot$\,yr$^{-1}$. 
\item The fraction of accretors in Chamaeleon~II is $\sim 50$\%, which, according to the evolution of mass accretion rate in star-forming regions 
by \cite{fedeleetal2010}, is consistent with the estimated age for the region ($\sim 3$ Myr). 

\end{enumerate}
 
\begin{acknowledgements}
The authors are very grateful to the referee Ralph Neuh\"auser for carefully reading the paper and for his useful remarks. 
This research made use of the SIMBAD database, operated at the CDS (Strasbourg, France). KB acknowledges the funding support from 
the INAF Postdoctoral fellowship. We thank S. Antoniucci for discussions on accretion luminosity from the Br$\gamma$ line. 
We thank G. Capasso and F. Cioffi for their support with the OAC computers.
We also thank G. Attusino for his warm assistence during the preparation of the manuscript.
\end{acknowledgements}

\bibliographystyle{aa}

\pagestyle{empty}
\setlength{\tabcolsep}{3pt}

\tiny
\begin{longtable}{llclcccrccl}
\caption[ ]{\label{tab:all_param} Observing log, radial velocities, and lithium content. }\\
\hline\hline
$2MASS$& Other ID & $JD$ & Instr. & \# & $V_{\rm rad}^{\rm a}$& $EW_{\rm Li}$ &$\langle\log n({\rm Li})>$& Comment$^{\rm b, c}$\\
Name   & 	   & ($+2\,450\,000$) & & obs. & (km~s$^{-1}$)  & (\AA)  & (dex) & \\
\hline
\endfirsthead
\caption{continued.}\\
\hline\hline
$2MASS$& Other ID & $JD$ & Instr. & \# & $V_{\rm rad}^{\rm a}$& $EW_{\rm Li}$ &$\langle\log n({\rm Li})>$& Comment$^{\rm b, c}$\\
Name   & 	   & ($+2\,450\,000$) & & obs. & (km~s$^{-1}$)  & (\AA)  & (dex) & \\
\hline
\endhead
\hline
\endfoot
\hline
12531722$-$7707106&IRAS 12496$-$7650/DK Cha	 &3795.7806&GIRAFFE            &2&		...&0	&   & C/HV; S1, S2   \\
                &				 &3797.7351&	"	       & &		...&0	&   & C/HV; S1, S2 \\
12563364$-$7645453&Sz 46N 			 &3796.7383&UVES      	       &3&     14.9$\pm$1.0&0.61&2.4& O1\\
                &				 &3797.8682&	"  	       & &     15.9$\pm$0.5&0.54&   & O1\\
                &				 &3797.8963&	"  	       & &     15.8$\pm$2.0&0.52&   & O1    \\
12565864$-$7647067&Sz 47  			 &3796.7382&GIRAFFE            &3&		...&0	&   & C/HV   \\ 
                &				 &3797.8681&	"	       & &		...&0	&   & C/HV  \\ 
                &				 &3797.8963&	"	       & &		...&0	&   & C/HV  \\     
12571172$-$7640111&IRAS 12535$-$7623/CHIIXR 2	 &3796.7383&UVES      	       &3&     10.2$\pm$1.3&0.58&2.9&\\
                &				 &3797.8682&	"	       & &	9.9$\pm$1.5&0.58&   &	    \\
                &				 &3797.8963&	"	       & &	9.9$\pm$1.8&0.62&   &	    \\
12585614$-$7630104&WFI J12585611$-$7630105	 &3796.7382&GIRAFFE            &3&  11$^{\rm \ast}$&0.53&3.0&	\\
                &				 &3797.8681&	"	       & &   9$^{\rm \ast}$&0.54&     &   \\
                &				 &3797.8963&	"	       & &  11$^{\rm \ast}$&0.62&     &   \\
12590984$-$7651037&C41				 &3795.3476&GIRAFFE            &1&		...&0.24&1.8& BC; S1, S2; N1, N2\\
12591013$-$7712139&Iso-Cha II 29  		 &4151.1828&GIRAFFE            &2&		...&0.57&2.6& BC; S1, S2; N1, N2\\
                &				 &4152.2866&	"	       & &		...&0.52&   & BC    \\
13005346$-$7709086&Sz 48SW/CHIIXR 7		 &3795.7394&GIRAFFE            &3&  12$^{\rm \ast}$&0.66&2.6& \\
                &				 &3796.8740&	"	       & &  12$^{\rm \ast}$&0.63&   & \\
                &				 &3797.6925&	"	       & &  11$^{\rm \ast}$&0.63&   &  \\
13005532$-$7710222&Sz 50/Iso-Cha II 52/CHIIXR 8	 &3795.7394&UVES      	       &3&     12.5$\pm$1.3&0.52&2.4& O1  \\
                &				 &3796.8740&	"	       & &     13.9$\pm$1.2&0.54&   & O1\\
                &				 &3797.6926&	"	       & &     12.6$\pm$1.1&0.52&   & O1\\
13005534$-$7708296&WFI J13005531$-$7708295	 &3795.7394&GIRAFFE            &1&  10$^{\rm \ast}$&0.61&2.6&  \\
13005622$-$7654021&RX J1301.0$-$7654a		 &3795.8653&UVES      	       &1&   $-$4.2$\pm$2.4&0.59&3.7&SB2 \\
                &				 &	"  &	"  	       & &     32.6$\pm$2.7&	&   &  \\ 
13005927$-$7714027&IRAS F12571$-$7657/Iso-Cha II 54&3797.6925&GIRAFFE          &1&   8$^{\rm \ast}$&0.30&   & LSN\\
13015891$-$7751218&Sz 51/BC~Cha			 &3795.8943&UVES      	       &2&     45.3$\pm$0.7&0.42&1.9&SB1; O1, O2; F  \\
                &				 &3797.7786&	"  	       & &     14.3$\pm$0.7&0.30&   &	O1, O2; F\\
13021351$-$7637577&CM Cha/IRAS 12584$-$7621	 &3795.8653&GIRAFFE            &1&  13$^{\rm \ast}$&0.43&2.4&  \\
13022287$-$7734494&C50				 &3795.8936&GIRAFFE            &2&  12$^{\rm \ast}$&0.57&2.5&  \\
                &				 &3797.7785&	"  	       & &  13$^{\rm \ast}$&0.41&   &  \\
13030444$-$7707027&RX J1303.1$-$7706		 &3795.7394&GIRAFFE            &3&     12.5$\pm$0.4&0.61&2.9&  \\
                &				 &3796.8740&	"	       & &     12.2$\pm$0.4&0.62&   &	 \\
                &				 &3797.6925&	"	       & &     12.3$\pm$0.4&0.62&   &	 \\
13030905$-$7755596&C51				 &3795.8936&GIRAFFE            &2&  12$^{\rm \ast}$&0.55&3.1&	  \\
                &				 &3797.7785&	"  	       & &  11$^{\rm \ast}$&0.56&   & \\
13042410$-$7650012&Hn 23  			 &3795.8653&UVES      	       &1&     15.2$\pm$0.2&0.53&3.2& O1  \\
13042489$-$7752303&Sz 52  			 &3795.8936&GIRAFFE            &2&  14$^{\rm \ast}$&0.49&1.9& S1, S2; N1, N2\\
                &				 &3797.7785&	"  	       & &  13$^{\rm \ast}$&0.48&   & S1, S2; N1, N2  \\
13045571$-$7739495&Hn 24  			 &3795.6959&GIRAFFE            &3&         12$\pm$3&0.60&2.9&	\\
                &				 &3796.8263&	"	       & &         11$\pm$3&0.61&   &	\\
                &				 &3797.6463&	"	       & &         11$\pm$3&0.61&   &  \\
                &				 &3795.8943&UVES               &2&     46.6$\pm$0.7&0.61&2.9&SB1  \\
                &				 &3797.7786&	"	       & &     14.1$\pm$0.8&0.61&   &  \\    
13050855$-$7733425&Hn 25  			 &3795.6959&GIRAFFE            &1&  10$^{\rm \ast}$&0.47&2.0&	\\   
13051269$-$7730525&Sz 53  			 &3795.6959&GIRAFFE            &3&   8$^{\rm \ast}$&0.40&1.8& S1, S2; N1, N2  \\ 
                &				 &3796.8263&	"	       & &  10$^{\rm \ast}$&0.40&   & S1, S2; N1, N2	\\ 
                &				 &3797.6464&	"	       & &  11$^{\rm \ast}$&0.43&   & S1, S2; N1, N2 \\      
13052072$-$7739015&Sz 54/BF~Cha			 &3795.8936&GIRAFFE            &2&  14$^{\rm \ast}$&0.46&3.1& \\     
                &				 &3797.7785&	"	       & &  10$^{\rm \ast}$&0.47&   & \\     
                &				 &3795.6960&UVES               &3&   $-$1.0$\pm$1.4&0.47&3.1& SB2; O1	\\
                &				 &	"  &	"	       & &     30.9$\pm$1.6&	&   &  \\ 
                &				 &3796.8263&	"	       & &  $-$10.8$\pm$2.0&0.47&   & O1\\
                &				 &	"  &	"	       & &     34.0$\pm$1.6&	&   &  \\ 
                &				 &3797.6464&	"	       & &   $-$6.3$\pm$2.3&0.47&   & O1\\
                &				 &	"  &	"	       & &     35.4$\pm$2.8&	&   &  \\ 
13052169$-$7738102&SSTc2d J130521.7$-$773810	 &4151.2721&GIRAFFE            &1&		...&0.33&   & BC; C; S1, S2; N1, N2\\
13063053$-$7734001&Sz 55  			 &3795.6959&GIRAFFE            &3&   8$^{\rm \ast}$&0.15&0.6& S1, S2; N1, N2  \\
                &				 &3796.8263&	"	       & &  13$^{\rm \ast}$&0.26&   & S1, S2; N2\\
                &				 &3797.6464&	"	       & &  13$^{\rm \ast}$&0.27&   & S1, S2; N1, N2 \\
13063882$-$7730352&Sz 56  			 &3795.6960&UVES$^{\rm \circ}$  &3&        ...      &0.56&2.9& very LSN; O1 \\
                &				 &3796.8263&	"	       & &        ...      &0.50&   & very LSN; O1\\
                &				 &3797.6464&	"	       & &	  ...	   &0.53&   & very LSN; O1\\
13065656$-$7723094&Sz 57/C60			 &3795.6960&GIRAFFE            &3&  10$^{\rm \ast}$&0.54&3.0&	  \\
                &				 &3796.8263&	"	       & &  10$^{\rm \ast}$&0.57&    &    \\
                &				 &3797.6464&	"	       & &  10$^{\rm \ast}$&0.53&    &    \\
13065744$-$7723415&Sz 58/IRAS 13030$-$7707/C61	 &3795.6960&GIRAFFE            &3&         13$\pm$4&0.49&3.0& S1, S2\\	  
                &				 &3796.8263&	"	       & &         12$\pm$5&0.49&   & S1, S2 \\   
                &				 &3797.6464&	"	       & &         12$\pm$4&0.48&   & S1, S2   \\ 
13070922$-$7730304&Sz 59/BK~Cha			 &3795.6959&GIRAFFE            &3&         12$\pm$3&0.39&2.3& S2    \\	  
                &				 &3796.8263&	"	       & &         14$\pm$3&0.43&   & S2     \\   
                &				 &3797.6464&	"	       & &         12$\pm$2&0.41&   & S2     \\   
13071806$-$7740529&C62				 &3795.6959&GIRAFFE            &2&  13$^{\rm \ast}$&0.65&3.0&	   \\
                &				 &3797.6463&	"  	       & &   8$^{\rm \ast}$&0.42&   &  \\
13072241$-$7737225&Sz 60W 			 &3795.6960&UVES      	       &3&  13.9$\pm$0.8&0.49&2.2&  \\
                &				 &3796.8263&	"	       & &     13.2$\pm$1.0&0.51&   & \\
                &				 &3797.6464&	"	       & &     12.5$\pm$0.9&0.53&   & \\
13074851$-$7741214&Hn 26  			 &3795.6959&GIRAFFE            &3&  11$^{\rm \ast}$&0.59&2.3& \\
                &				 &3796.8263&	"	       & &  10$^{\rm \ast}$&0.57&   &	     \\
                &				 &3797.6463&	"	       & &  10$^{\rm \ast}$&0.60&   &	   \\
13080628$-$7755051&Sz 61/BM Cha  		 &3796.6956&GIRAFFE            &2&         13$\pm$3&0.40&2.6&	 \\
                &				 &3797.8201&	"  	       & &         11$\pm$3&0.39&   &  \\
13082714$-$7743232&C66				 &3795.6959&GIRAFFE            &4&   9$^{\rm \ast}$&0.55&2.5&SB1?    \\
                &				 &3796.6956&	"	       & &   8$^{\rm \ast}$&0.63&   &		 \\
                &				 &3797.6463&	"	       & &  11$^{\rm \ast}$&0.45&   &	      \\
                &				 &3797.8201&	"	       & &  12$^{\rm \ast}$&0.30&   &	    \\
13090987$-$7709437&IRAS F13052$-$7653NW/CHIIXR 60&4151.2219&GIRAFFE	       &1&	    ...    &0.40&1.9& BC; S2\\
13091071$-$7709443&IRAS F13052$-$7653N/CHIIXR 60 &3795.8228&UVES	       &1&46.7$\pm$0.8$^{\rm \ast\ast}$&0.61&2.5&SB1?   \\
13095036$-$7757240&Sz 62  			 &3796.6956&GIRAFFE            &2&   10$^{\rm \ast}$&0.44&1.8&  \\
                &				 &3797.8201&	"  	       & &   10$^{\rm \ast}$&0.44&   &  \\
13100415$-$7710447&Sz 63  			 &3795.8227&GIRAFFE            &1&   10$^{\rm \ast}$&0.54&2.2&  \\
13125238$-$7739182&2MASS J13125238$-$7739182	 &3796.6956&GIRAFFE            &2&   15$^{\rm \ast}$&0.58&3.4&   \\
                &				 &3797.8201&	"  	       & &   13$^{\rm \ast}$&0.60&     &   \\
13140369$-$7753076&Sz 64  			 &3796.6956&GIRAFFE            &2&   10$^{\rm \ast}$&0.43&2.1&       \\
                &				 &3797.8201&	"  	       & &   10$^{\rm \ast}$&0.43&     &     \\  
\hline							  
\end{longtable}
\footnotesize{Notes:
\begin{itemize}
\item[] $^{\rm \ast}$ Due to low $S/N$ ratio, few lines, late spectral type, and/or short wavelength coverage, the radial velocity error may be up to 60\%.
\item[] $^{\rm \ast\ast}$ This can be classified as a {\it bona-fide} PMS star based on other criteria (see \citealt{alcalaetal2008, spezzietal2008}); its high radial velocity suggests it may be a SB1.
\item[] $^{\rm a}$ For spectroscopic binaries, the different radial velocity measurements of the corresponding CCF peaks are given.
\item[] $^{\rm b}$ SB1: single-lined spectroscopic binary; SB2: double-lined spectroscopic binary; SB1?: suspected SB1; 
BC: bad CCF; C: continuum-type spectrum; HV: heavily veiled star; LSN: low $S/N$ spectrum.
\item[] $^{\rm c}$ O1=[\ion{O}{i}] $\lambda$6300.8\,\AA, O2=[\ion{O}{i}] $\lambda$6363.8\,\AA, S1=[\ion{S}{ii}] $\lambda$6715.8\,\AA, 
S2=[\ion{S}{ii}] $\lambda$6729.8\,\AA, N1=[\ion{N}{ii}] $\lambda$6548.4\,\AA, N2=[\ion{N}{ii}] $\lambda$6583.4\,\AA, F=[\ion{Fe}{ii}] multiplets.
\item[] $^{\rm \circ}$ In \cite{spezzietal2008}, it was mistakenly reported as GIRAFFE observations.
\end{itemize}
}

\newpage
\topmargin 6 cm
\pagestyle{empty}

\setlength{\tabcolsep}{2.3pt}
\begin{landscape}
\scriptsize
\begin{longtable}{lccrccrccrcrccrr}
\caption[ ]{\label{tab:accret_param} Star name (column1); equivalent width, observed flux, and mass accretion rate of the 
H$\alpha$, H$\beta$, \ion{He}{i} $\lambda$5876 \AA, and $\lambda$6678 \AA\ lines (columns 2--13); \ion{He}{i} $\lambda$7065 \AA\ line 
equivalent width and observed line luminosity (columns 14--15); average $\dot M_{\rm acc}$ as derived from the H$\alpha$, 
H$\beta$, \ion{He}{i} $\lambda$5876 \AA, and $\lambda$6678 \AA\ lines (column 16).}\\
\hline\hline
ID name & $EW_{\rm H\alpha}$&$\log F^{\rm H\alpha}$&$\log \dot M_{\rm acc}^{\rm H\alpha}$&$EW_{\rm H\beta}$&$\log F^{\rm H\beta}$&$\log \dot M_{\rm acc}^{\rm H\beta}$& $EW_{\rm He5876}$ &$\log F^{\rm He5876}$&$\log \dot M_{\rm acc}^{\rm He5876}$&$EW_{\rm He6678}$&$\log F^{\rm He6678}$&$\log \dot M_{\rm acc}^{\rm He6678}$& $EW_{\rm He7065}$ &$\log L^{\rm He7065}$&$\langle\log \dot M_{\rm acc}\rangle$\\
	& (\AA)             & (erg\,s$^{-1}$\,cm$^2$)       & ($M_\odot$\,yr$^{-1}$                     )& (\AA)           & (erg\,s$^{-1}$\,cm$^2$)       & ($M_\odot$\,yr$^{-1}$)                       & (\AA)	   & (erg\,s$^{-1}$\,cm$^2$)    &    ($M_\odot$\,yr$^{-1}$)		      & (\AA)		&  (erg\,s$^{-1}$\,cm$^2$)    & ($M_\odot$\,yr$^{-1}$)			   & (\AA)	      & ($L_\odot$)	     & ($M_\odot$\,yr$^{-1}$)	       \\
\hline
\endfirsthead
\caption{continued.}\\
\hline\hline
ID name  & $EW_{\rm H\alpha}$&$\log F^{\rm H\alpha}$&$\log \dot M_{\rm acc}^{\rm H\alpha}$&$EW_{\rm H\beta}$&$\log F^{\rm H\beta}$&$\log \dot M_{\rm acc}^{\rm H\beta}$& $EW_{\rm He5876}$ &$\log F^{\rm He5876}$&$\log \dot M_{\rm acc}^{\rm He5876}$&$EW_{\rm He6678}$&$\log F^{\rm He6678}$&$\log \dot M_{\rm acc}^{\rm He6678}$& $EW_{\rm He7065}$ &$\log L^{\rm He7065}$&$\langle\log \dot M_{\rm acc}\rangle$\\
	 & (\AA)             & (erg\,s$^{-1}$\,cm$^2$)       & ($M_\odot$\,yr$^{-1}$			 )& (\AA)	    & (erg\,s$^{-1}$\,cm$^2$)	   & ($M_\odot$\,yr$^{-1}$)			  & (\AA)	    & (erg\,s$^{-1}$\,cm$^2$)	 &    ($M_\odot$\,yr$^{-1}$)			& (\AA) 	  &  (erg\,s$^{-1}$\,cm$^2$)	& ($M_\odot$\,yr$^{-1}$)		      & (\AA)		 & ($L_\odot$)  	& ($M_\odot$\,yr$^{-1}$)	  \\
\hline
\endhead
\hline
\endfoot
\hline
IRAS 12496$-$7650/DK Cha	& 78.5  	 &9.1& $-$6.3&  	    &	&	&    &    &	 &0.65&7.0& $-$6.1&    &      &$-$6.1	 \\ 
				& 91.5  	 &9.2& $-$6.2&  	    &	&	&    &    &	 &1.07&7.2& $-$5.8&    &      &       \\
Sz 46N  			& 10.5$^\bullet$ &6.7& $-$9.7& 4.6$^\bullet$&6.1& $-$9.8&0.50&5.2 &$-$9.6&0.06&4.2&$-$10.5&    &      &$-$9.5  \\
				& 15.6$^\bullet$ &6.8& $-$9.5& 8.5$^\bullet$&6.4& $-$9.5&0.65&5.3 &$-$9.4&0.07&4.3&$-$10.4&    &      &      \\
				& 17.4$^\bullet$ &6.9& $-$9.4& 5.9$^\bullet$&6.2& $-$9.7&0.69&5.3 &$-$9.4&    &   &	  &    &      & 	 \\
Sz 47				& 32.5  	 &   &       &  	    &	&	&    &    &	 &    &   &	  &    &      &...	\\ 
				& 32.5  	 &   &       &  	    &	&	&    &    &	 &    &   &	  &    &      &       \\       
				& 32.3  	 &   &       &  	    &	&	&    &    &	 &    &   &	  &    &      &       \\     
IRAS 12535$-$7623/CHIIXR 2	& 2.2$^\bullet$  &6.0& $-$9.6& 0.8$^\bullet$&5.6& $-$9.6&    &    &	 &    &   &	  &    &      &$-$9.8 \\
				& 1.5$^\bullet$  &5.9& $-$9.8& 0.6$^\bullet$&5.5& $-$9.7&    &    &	 &    &   &	  &    &      & 	  \\
				& 1.3$^\bullet$  &5.8& $-$9.8& 0.3$^\bullet$&5.2&$-$10.1&    &    &	 &    &   &	  &    &      & 	  \\
WFI J12585611$-$7630105 	& 29.2  	 &6.3& $-$9.7&  	    &	&	&    &    &	 &0.35&4.2&$-$10.1&    &      &$-$9.9	\\
				& 31.2  	 &6.3& $-$9.7&  	    &	&	&    &    &	 &0.55&4.4& $-$9.8&    &      &       \\
				& 31.5  	 &6.3& $-$9.7&  	    &	&	&    &    &	 &0.52&4.4& $-$9.9&    &      &       \\
C41				& 46.7  	 &6.8&$-$10.8&  	    &	&	&    &    &	 &0.95&4.9&$-$10.9&0.18&$-$7.3&$-$11.0 \\
Iso-Cha II 29			& 0.9		 &5.6&$-$10.6&  	    &	&	&    &    &	 &    &   &	  &    &      &$-$10.6 \\
				& 1.0		 &5.7&$-$10.5&  	    &	&	&    &    &	 &    &   &	  &    &      &        \\
Sz 48SW/CHIIXR 7		&  9.9$^\bullet$ &6.6& $-$9.9&  	    &	&	&    &    &	 &    &   &	  &    &      &$-$9.9  \\
				& 14.7$^\bullet$ &6.8& $-$9.7&  	    &	&	&    &    &	 &    &   &	  &    &      &      \\
				&  8.1$^\bullet$ &6.6&$-$10.0&  	    &	&	&    &    &	 &    &   &	  &    &      &       \\
Sz 50/Iso-Cha II 52/CHIIXR 8	& 33.8$^\bullet$ &6.8& $-$8.1& 9.7$^\bullet$&5.9& $-$8.6&0.43&4.6 &$-$8.7&0.09&4.1& $-$9.1&    &      &$-$8.4	 \\
				& 35.4$^\bullet$ &6.8& $-$8.1&11.8$^\bullet$&6.0& $-$8.5&0.46&4.7 &$-$8.7&0.08&4.0& $-$9.2&    &      &      \\
				& 30.4$^\bullet$ &6.7& $-$8.1& 8.5$^\bullet$&5.9& $-$8.6&0.32&4.5 &$-$8.9&0.07&4.0& $-$9.2&    &      &      \\
WFI J13005531$-$7708295 	& 2.1		 &6.0&$-$10.4&  	    &	&	&    &    &	 &    &   &	  &    &      &$-$10.4  \\
RX J1301.0$-$7654a		& 3.1		 &6.8& $-$8.6& 1.4	    &6.5& $-$8.4&0.16&5.6 &$-$7.9&    &   &	  &    &      &$-$8.3  \\
IRAS F12571$-$7657/Iso-Cha II 54& 16.0  	 &   &       &  	    &	&	&    &    &	 &    &   &	  &    &      &...   \\
Sz 51				& 132.5$^\bullet$&8.1& $-$8.1&39.0$^\bullet$&7.5& $-$8.3&2.92&6.4 &$-$8.0&1.11&6.0& $-$8.3&    &      &$-$8.3	 \\
				& 110.0$^\bullet$&8.0& $-$8.2&26.0$^\bullet$&7.3& $-$8.5&2.29&6.3 &$-$8.1&0.91&5.9& $-$8.4&    &      &        \\
CM Cha/IRAS 12584$-$7621	& 27.6  	 &7.4& $-$8.6&  	    &	&	&    &    &	 &0.72&5.8& $-$8.2&0.50&$-$4.6&$-$8.4	\\
C50				& 37.5  	 &6.6& $-$9.8&  	    &	&	&    &    &	 &1.34&5.1& $-$9.5&0.39&$-$6.2&$-$9.9	\\
				& 38.3  	 &6.7& $-$9.8&  	    &	&	&    &    &	 &0.78&4.8& $-$9.9&0.48&$-$6.1&       \\
RX J1303.1$-$7706		& 2.6		 &6.1& $-$9.5&  	    &	&	&    &    &	 &    &   &	  &    &      &$-$9.5	\\
				& 2.7		 &6.1& $-$9.5&  	    &	&	&    &    &	 &    &   &	  &    &      & 	\\
				& 2.6		 &6.1& $-$9.5&  	    &	&	&    &    &	 &    &   &	  &    &      & 	\\
C51				& 9.5		 &6.0& $-$9.6&  	    &	&	&    &    &	 &    &   &	  &    &      &$-$9.7	 \\
				& 6.3		 &5.8& $-$9.8&  	    &	&	&    &    &	 &    &   &	  &    &      &        \\
Hn 23				& 11.9  	 &7.4& $-$8.9& 1.5	    &6.5& $-$9.5&0.26&5.7 &$-$8.8&0.08&5.2& $-$9.2&    &      &$-$9.0	 \\
Sz 52				& 49.4  	 &7.1& $-$9.2&  	    &	&	&    &    &	 &0.70&5.1& $-$9.4&0.45&$-$5.6&$-$9.4 \\
				& 49.9  	 &7.1& $-$9.2&  	    &	&	&    &    &	 &0.62&5.1& $-$9.4&0.43&$-$5.6&        \\
Hn 24				& 1.3$^\bullet$  &5.8&$-$10.0&  	    &	&	&    &    &	 &    &   &	  &    &      &$-$9.8	 \\
				& 1.2$^\bullet$  &5.8&$-$10.0&  	    &	&	&    &    &	 &    &   &	  &    &      &        \\
				& 1.9$^\bullet$  &6.0& $-$9.7&  	    &	&	&    &    &	 &    &   &	  &    &      &       \\
				& 1.1		 &5.7&$-$10.0& 0.9	    &5.7& $-$9.6&0.12&4.6 &$-$9.5&    &   &	  &    &      &       \\
				& 1.4		 &5.8& $-$9.9& 0.8	    &5.6& $-$9.7&    &    &	 &    &   &	  &    &      &       \\
Hn 25				& 22.5  	 &6.6& $-$9.4&  	    &	&	&    &    &	 &0.67&4.9& $-$9.2&0.54&$-$5.4&$-$9.3	 \\
Sz 53				&145.0$^\bullet$ &7.8& $-$8.2&  	    &	&	&    &    &	 &1.05&5.5& $-$8.7&1.37&$-$4.7&$-$8.6	 \\
				& 46.9$^\bullet$ &7.3& $-$8.8&  	    &	&	&    &    &	 &0.70&5.3& $-$8.9&0.64&$-$5.1& 	 \\
				& 98.3$^\bullet$ &7.6& $-$8.4&  	    &	&	&    &    &	 &0.91&5.4& $-$8.8&0.79&$-$5.0&       \\
Sz 54				& 23.8  	 &7.7& $-$7.8&  	    &	&	&    &    &	 &    &   &	  &    &      &$-$8.2  \\
				& 22.7  	 &7.7& $-$7.8&  	    &	&	&    &    &	 &    &   &	  &    &      &      \\
				& 21.2  	 &7.7& $-$7.9&1.9	    &6.6& $-$8.6&    &    &	 &    &   &	  &    &      & 	\\
				& 22.4  	 &7.7& $-$7.8&2.2	    &6.7& $-$8.5&    &    &	 &    &   &	  &    &      &      \\
				& 22.6  	 &7.7& $-$7.8&1.2	    &6.4& $-$8.8&    &    &	 &    &   &	  &    &      &      \\
SSTc2d J130521.7$-$773810	& 26.1  	 &   &       &  	    &	&	&    &    &	 &0.21&   &	  &    &      &...   \\
Sz 55				& 101.1$^\bullet$&7.6& $-$9.1&  	    &	&	&    &    &	 &2.04&5.7& $-$9.1&1.39&$-$5.3&$-$9.1	 \\
				& 127.2$^\bullet$&7.7& $-$9.0&  	    &	&	&    &    &	 &1.93&5.7& $-$9.1&1.41&$-$5.2&      \\
				& 130.8$^\bullet$&7.7& $-$9.0&  	    &	&	&    &    &	 &1.97&5.7& $-$9.1&1.14&$-$5.3&       \\
Sz 56				& 11.1$^\bullet$ &6.1& $-$9.7& 3.5$^\bullet$&5.1&$-$10.3&1.90&5.0 &$-$9.0&0.60&4.7& $-$9.1&    &      &$-$9.4	 \\
				& 14.3$^\bullet$ &6.2& $-$9.5&12.8$^\bullet$&5.7& $-$9.7&3.70&5.3 &$-$8.6&0.78&4.8& $-$8.9&    &      &      \\
				& 18.8$^\bullet$ &6.3& $-$9.4& 4.8$^\bullet$&5.2&$-$10.2&2.97&5.2 &$-$8.8&0.88&4.8& $-$8.9&    &      &      \\
Sz 57/C60			& 23.0$^\bullet$ &6.4& $-$8.7&  	    &	&	&    &    &	 &0.50&4.6& $-$8.5&0.12&$-$5.9&$-$8.7	 \\
				& 16.5$^\bullet$ &6.3& $-$8.8&  	    &	&	&    &    &	 &0.39&4.5& $-$8.7&0.11&$-$5.9& 	\\
				& 16.2$^\bullet$ &6.3& $-$8.9&  	    &	&	&    &    &	 &0.41&4.5& $-$8.7&0.10&$-$6.0& 	\\
Sz 58/IRAS 13030$-$7707/C61	& 18.2$^\bullet$ &7.6& $-$8.8&  	    &	&	&    &    &	 &    &   &	  &0.15&$-$5.0&$-$8.9  \\
				& 14.8$^\bullet$ &7.5& $-$8.9&  	    &	&	&    &    &	 &0.02&4.6&$-$10.2&0.21&$-$4.9&       \\
				& 10.5$^\bullet$ &7.3& $-$9.1&  	    &	&	&    &    &	 &0.08&5.2& $-$9.4&0.20&$-$4.9& 	\\
Sz 59				& 52.8$^\bullet$ &7.7& $-$8.1&  	    &	&	&    &    &	 &0.44&5.6& $-$8.3&0.48&$-$4.5&$-$8.4	 \\
				& 35.3$^\bullet$ &7.5& $-$8.3&  	    &	&	&    &    &	 &0.30&5.4& $-$8.5&0.30&$-$4.7& 	  \\
				& 39.1$^\bullet$ &7.6& $-$8.3&  	    &	&	&    &    &	 &0.25&5.3& $-$8.6&0.30&$-$4.7& 	  \\
C62				& 37.2  	 &6.6& $-$9.7&  	    &	&	&    &    &	 &1.40&5.0& $-$9.4&0.45&$-$6.1&$-$9.6	  \\
				& 39.6  	 &6.6& $-$9.7&  	    &	&	&    &    &	 &1.85&5.2& $-$9.2&0.60&$-$6.0&       \\
Sz 60W  			& 25.6$^\bullet$ &7.1& $-$9.3& 7.5$^\bullet$&6.4& $-$9.6&1.22&5.5 &$-$9.1&0.38&5.0& $-$9.5&    &      &$-$9.3	\\
				& 37.6$^\bullet$ &7.2& $-$9.1&11.8$^\bullet$&6.5& $-$9.4&0.93&5.4 &$-$9.3&0.31&4.9& $-$9.6&    &      &      \\
				& 36.8$^\bullet$ &7.2& $-$9.1&12.4$^\bullet$&6.6& $-$9.4&1.06&5.5 &$-$9.2&0.39&5.0& $-$9.5&    &      &      \\
Hn 26				& 6.6$^\bullet$  &6.3&$-$10.1&  	    &	&	&    &    &	 &0.08&4.3&$-$10.5&0.09&$-$6.1&$-$10.1 \\
				& 8.1$^\bullet$  &6.5&$-$10.0&  	    &	&	&    &    &	 &0.18&4.6&$-$10.0&0.08&$-$6.1& 	  \\
				& 13.4$^\bullet$ &6.7& $-$9.7&  	    &	&	&    &    &	 &0.11&4.4&$-$10.2&0.11&$-$6.0& 	 \\
Sz 61/BM Cha			& 81.1$^\bullet$ &8.2& $-$7.6&  	    &	&	&    &    &	 &0.68&6.1& $-$7.7&0.80&$-$4.0&$-$7.7	\\
				& 64.7$^\bullet$ &8.1& $-$7.8&  	    &	&	&    &    &	 &0.60&6.1& $-$7.7&0.53&$-$4.2&       \\
C66				& 20.8  	 &6.5&$-$10.2&  	    &	&	&    &    &	 &0.67&4.9&$-$10.0&0.50&$-$6.2&$-$9.8	\\
				& 29.1  	 &6.7&$-$10.0&  	    &	&	&    &    &	 &1.87&5.3& $-$9.4&0.40&$-$6.3& 	   \\
				& 35.9  	 &6.8& $-$9.9&  	    &	&	&    &    &	 &2.03&5.4& $-$9.3&0.99&$-$5.9& 	   \\
				& 25.7  	 &6.6&$-$10.0&  	    &	&	&    &    &	 &2.10&5.4& $-$9.3&0.90&$-$5.9& 	   \\
IRAS F13052$-$7653NW/CHIIXR 60  & 11.2  	 &6.8&$-$10.0&  	    &	&	&    &    &	 &0.73&5.4& $-$9.4&0.50&$-$5.3&$-$9.7	\\
IRAS F13052$-$7653N/CHIIXR 60	& 2.2		 &5.9&$-$10.5&  0.5	    &5.0&$-$11.0&    &    &	 &    &   &	  &    &      &$-$10.8   \\
Sz 62				& 115.8 	 &7.3& $-$8.4&  	    &	&	&    &    &	 &0.74&4.9& $-$9.1&0.35&$-$5.6&$-$8.4	\\
				& 150.8 	 &7.4& $-$8.4&  	    &	&	&    &    &	 &0.98&5.0& $-$9.0&0.88&$-$5.2&       \\
Sz 63				& 93.6  	 &7.2& $-$8.9&  	    &	&	&    &    &	 &0.41&4.6& $-$9.8&0.35&$-$5.8&$-$9.4	\\
2MASS J13125238$-$7739182	& 5.8		 &5.8&$-$10.3&  	    &	&	&    &    &	 &0.11&3.9&$-$10.5&    &      &$-$10.5   \\
				& 7.0		 &5.9&$-$10.2&  	    &	&	&    &    &	 &0.11&3.9&$-$10.5&    &      &        \\
Sz 64				& 132.1 	 &7.1& $-$9.1&  	    &	&	&    &    &	 &2.82&5.3& $-$9.0&0.79&$-$5.9&$-$9.1	 \\
				& 121.4 	 &7.0& $-$9.1&  	    &	&	&    &    &	 &3.00&5.3& $-$9.0&1.22&$-$5.7&     \\  	   
\hline							  
\end{longtable}
\footnotesize{Notes:
\begin{itemize}
\item $^\bullet$ Variable line profile.
\end{itemize}
}
\end{landscape}

\topmargin -1 cm
\begin{longtable}{llccrrcrrc}
\caption[ ]{\label{tab:liter_param} Parameters taken from the literature. }\\
\hline\hline
& Sp.T.$^{\rm a}$& $T_{\rm eff}^{\rm a}$ & $<M_{\rm \star}>^{\rm a}$ & $\log L_{\rm \star}^{\rm a}$ & $R_{\rm \star}^{\rm a}$ & $\langle\log g>^{\rm a}$ & $R_{\rm in}^{\rm a}$ & $<Age>^{\rm a}$ & Class$^{\rm b}$\\
& & (K) & ($M_{\odot}$) & ($L_{\odot}$) & ($R_{\odot}$) &(dex)& (AU) & (Myr) & \\
\hline
\endfirsthead
\caption{continued.}\\
\hline\hline
& Sp.T.$^{\rm a}$& $T_{\rm eff}^{\rm a}$ & $<M_{\rm \star}>^{\rm a}$ & $\log L_{\rm \star}^{\rm a}$ & $R_{\rm \star}^{\rm a}$ & $\langle\log g>^{\rm a}$ & $R_{\rm in}^{\rm a}$ & $<Age>^{\rm a}$ & Class$^{\rm b}$\\
& & (K) & ($M_{\odot}$) & ($L_{\odot}$) & ($R_{\odot}$) &(dex)& (AU) & (Myr) & \\
\hline
\endhead
\hline
\endfoot
\hline
IRAS 12496$-$7650/DK Cha        & F0 &7200& 2.00 &    1.27 &  2.77 & 3.9  & 1.79      & 4.75 &II    \\
Sz 46N  		        & M1 &3705& 0.53 & $-$0.48 &  1.39 & 3.9  & 0.13      & 3.07 &II   \\
Sz 47			        & ...&... & ...  &    ...  &  ...  & ...  & 0.2$^\ast$& ...  &III	     \\ 
IRAS 12535$-$7623/CHIIXR 2      & M0 &3850& 0.67 &    0.14 &  2.71 & 3.4  & 0.49      & 0.80 &II  \\
WFI J12585611$-$7630105 	& M5 &3025& 0.12 & $-$1.03 &  1.13 & 3.4  & ...	      & 1.33 &III  \\
C41				&M5.5&3057& 0.10 & $-$1.95 &  0.37 & 4.3  & 0.98      &13.37 &Flat  \\
Iso-Cha II 29			& M0 &3850& 0.65 & $-$0.19 &  1.85 & 3.7  &70.86      & 2.77 &III   \\
Sz 48SW/CHIIXR 7		& M1 &3705& 0.57 & $-$0.58 &  1.25 & 4.0  & 2.06      & 4.70 &II   \\
Sz 50/Iso-Cha II 52/CHIIXR 8	& M3 &3415& 0.35 &    0.06 &  3.10 & 3.0  & 0.07      & 0.37 &II   \\
WFI J13005531$-$7708295         &M2.5&3687& 0.53 & $-$0.45 &  1.46 & 3.8  & 0.2$^\ast$& 2.83 &III   \\
RX J1301.0$-$7654a		& K5 &4350& 0.70 &    0.38 &  2.67 & 3.4  & 0.2$^\ast$& 0.60 &III  \\
IRAS F12571$-$7657/Iso-Cha II 54& ...&... & ...  &    ...  &  ...  & ...  & 0.09      & ...  &II   \\
Sz 51				&K8.5&3955& 0.73 & $-$0.36 &  1.37 & 4.0  & 0.20      & 4.67 &II      \\
CM Cha/IRAS 12584$-$7621  	& K7 &4060& 0.85 & $-$0.14 &  1.78 & 3.9  & 0.78      & 3.67 &II    \\
C50				& M5 &3125& 0.16 & $-$1.19 &  0.89 & 3.8  & 0.08      & 2.73 &II    \\
RX J1303.1$-$7706 		& M0 &3850& 0.73 &    0.10 &  2.61 & 3.5  & 0.2$^\ast$& 1.07 &III   \\
C51				&M4.5&3197& 0.19 & $-$0.53 &  1.77 & 3.2  & 0.2$^\ast$& 0.63 &III    \\
Hn 23				& K5 &4350& 1.00 & $-$0.06 &  1.60 & 4.0  & 0.18      & 3.57 &II   \\
Sz 52				&M2.5&3487& 0.35 & $-$0.75 &  1.15 & 3.8  & 0.35      & 2.83 &II  \\
Hn 24				& M0 &3850& 0.65 &    0.02 &  2.37 & 3.5  & 0.06      & 1.03 &II   \\
Hn 25				&M2.5&3487& 0.37 & $-$0.48 &  1.56 & 3.6  & 0.19      & 1.60 &II   \\
Sz 53				& M1 &3705& 0.55 & $-$0.49 &  1.39 & 3.9  & 0.03      & 3.23 &II   \\
Sz 54				& K5 &4350& 0.97 &    0.29 &  2.42 & 3.6  & 0.09      & 1.37 &II   \\
SSTc2d J130521.7$-$773810 	& ...&... & ...  &    ...  &  ...  & ...  & 0.65      & ...  &Flat \\
Sz 55				& M2 &3560& 0.42 & $-$0.90 &  0.91 & 4.1  & 0.08      & 7.50 &II   \\
Sz 56				& M4 &3270& 0.23 & $-$0.47 &  1.78 & 3.3  & 0.43      & 0.80 &II   \\
Sz 57/C60			& M5 &3125& 0.15 & $-$0.39 &  2.21 & 2.9  & 0.05      & 0.60 &II   \\
Sz 58/IRAS 13030$-$7707/C61	& K5 &4350& 0.97 & $-$0.16 &  1.43 & 4.1  & 1.75      & 6.93 &II   \\
Sz 59				& K7 &4060& 0.82 & $-$0.05 &  1.96 & 3.8  & 0.37      & 2.20 &II   \\
C62				&M4.5&3197& 0.18 & $-$1.05 &  0.97 & 3.7  & 0.02      & 1.83 &II    \\
Sz 60W  			& M1 &3705& 0.57 & $-$0.54 &  1.32 & 4.0  & 0.07      & 3.57 &III$^\ddagger$ \\
Hn 26				& M2 &3560& 0.44 & $-$0.59 &  1.30 & 3.8  & 0.23      & 3.00 &II   \\
Sz 61/BM Cha			& K5 &4350& 1.00 &    0.07 &  1.87 & 3.8  & 0.66      & 2.67 &II    \\
C66				&M4.5&3197& 0.17 & $-$1.30 &  0.73 & 3.9  & 0.01      & 6.50 &II      \\
IRAS F13052$-$7653NW/CHIIXR 60	&M0.5&3777& 0.62 & $-$0.70 &  1.03 & 4.2  & 0.11      & 7.77 &II  \\
IRAS F13052$-$7653N/CHIIXR 60	&M1.5&3632& 0.47 & $-$0.47 &  1.49 & 3.8  & 3.54      & 2.70 &II     \\
Sz 62				&M2.5&3487& 0.37 & $-$0.48 &  1.56 & 3.6  & 0.05      & 1.60 &II  \\
Sz 63				& M3 &3415& 0.32 & $-$0.64 &  1.38 & 3.7  & 3.02      & 1.50 &II  \\
2MASS J13125238$-$7739182	&M4.5&3197& 0.19 & $-$0.81 &  1.28 & 3.5  & 0.03      & 1.43 &III     \\
Sz 64				& M5 &3125& 0.16 & $-$1.04 &  1.04 & 3.6  & 0.18      & 1.67 &II     \\
\hline				   			    
\end{longtable}
\footnotesize{Notes:
\begin{itemize}
\item[] $^{\rm a}$ Spectral types, effective temperatures, luminosities, and radii are from \cite{spezzietal2008}. Masses and ages are 
the average of three values reported in \cite{spezzietal2008} and derived using three different sets of evolutionary models, with mean 
standard deviations of $\sim 0.15M_\odot$ and $\sim 1.8$ Myr, respectively. Mean surface gravities were computed through the relation: 
$\langle\log g\rangle = 4.4377 + \langle\log M_{\rm \star}\rangle + 4 \log T_{\rm eff}/5777 - \log L_{\rm \star}$. $R_{\rm in}$ was adopted from \cite{alcalaetal2008}. 
\item[] $^{\rm b}$ Lada Class as derived in \cite{alcalaetal2008}.
\item[] $^\ddagger$ In \cite{alcalaetal2008}, it appears as Class III, but it is a Class II object.
\item[] $^\ast$ $R_{\rm in}=5 R_{\rm \star}$ (see Section~\ref{sec:accretion_rates}).
\end{itemize}
}
\normalsize

\Online

{\bf To be published in electronic form only}\\

\begin{appendix}
\section{Examples of H$\alpha$/H$\beta$ line profiles}
\begin{figure}[h!]
\begin{center}
\includegraphics[width=4.5cm]{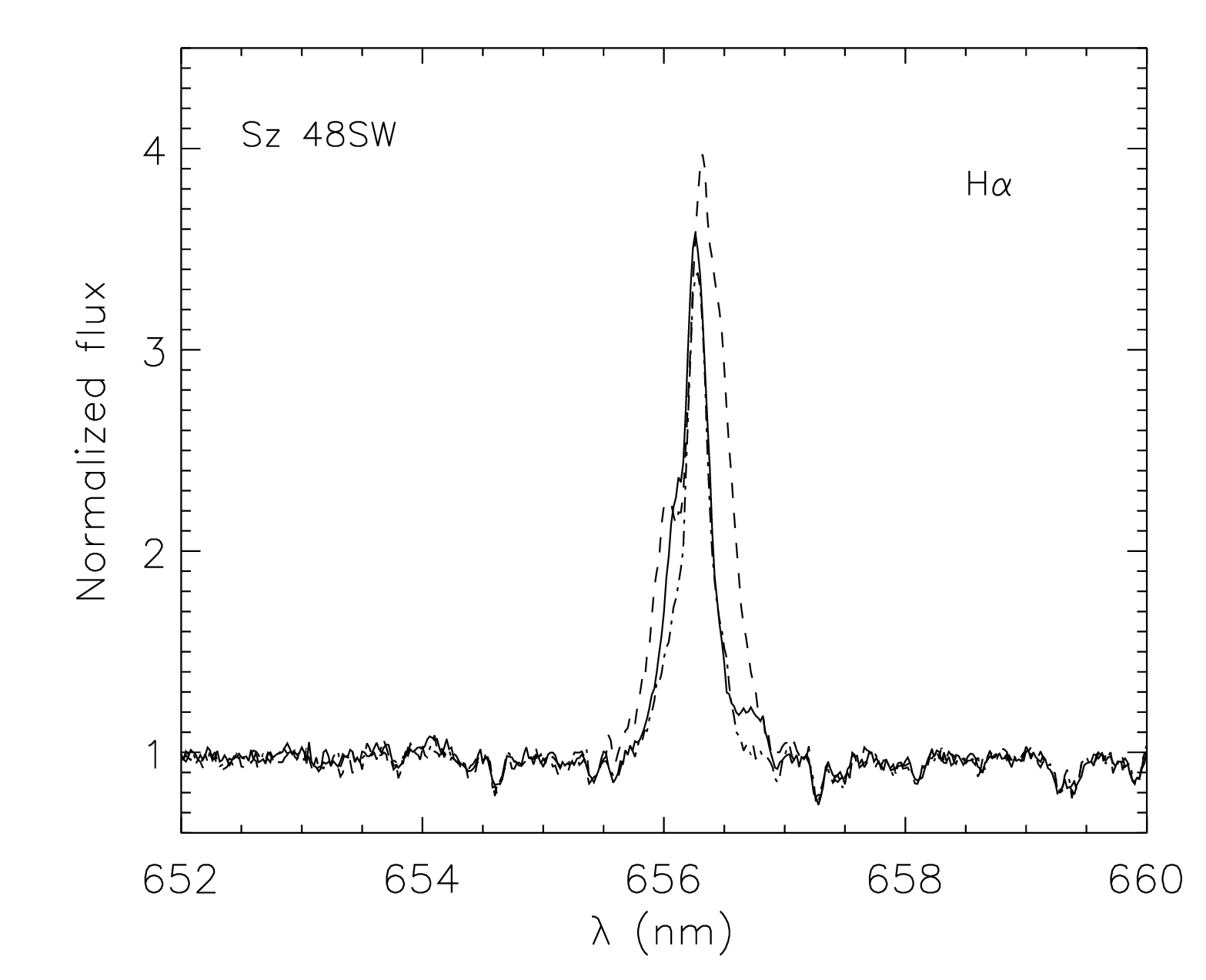}
\includegraphics[width=4.5cm]{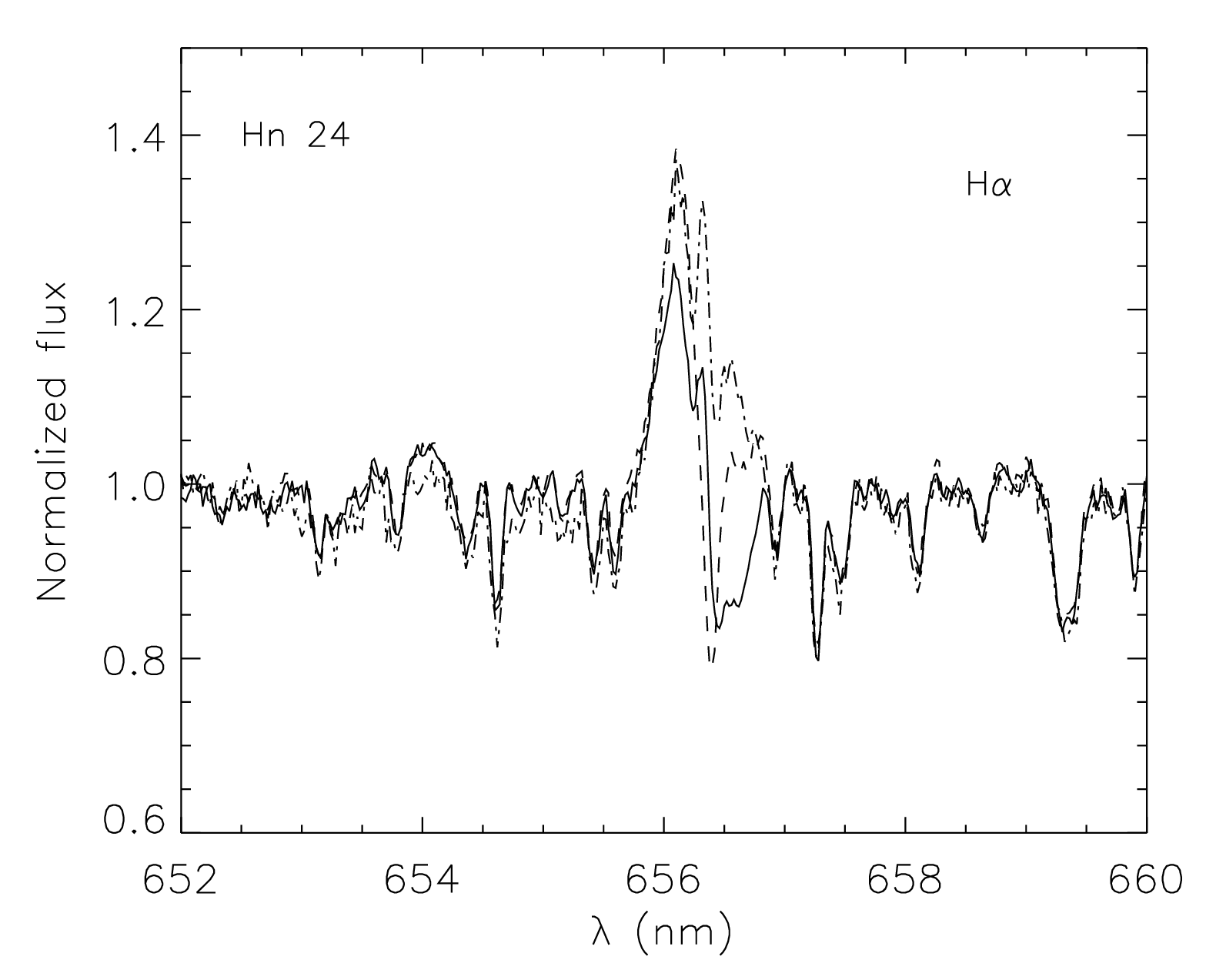}
\includegraphics[width=4.5cm]{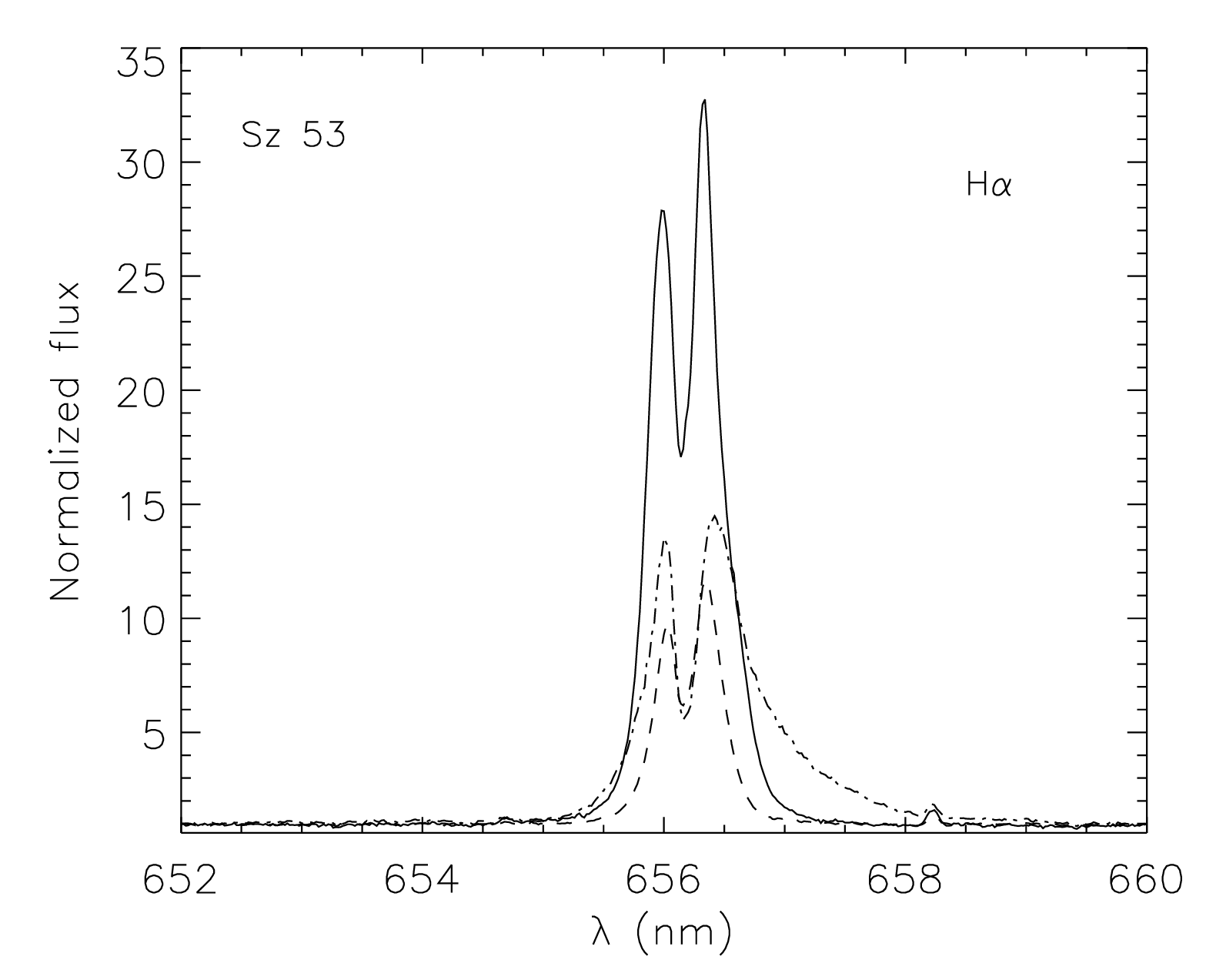}\\
\includegraphics[width=4.5cm]{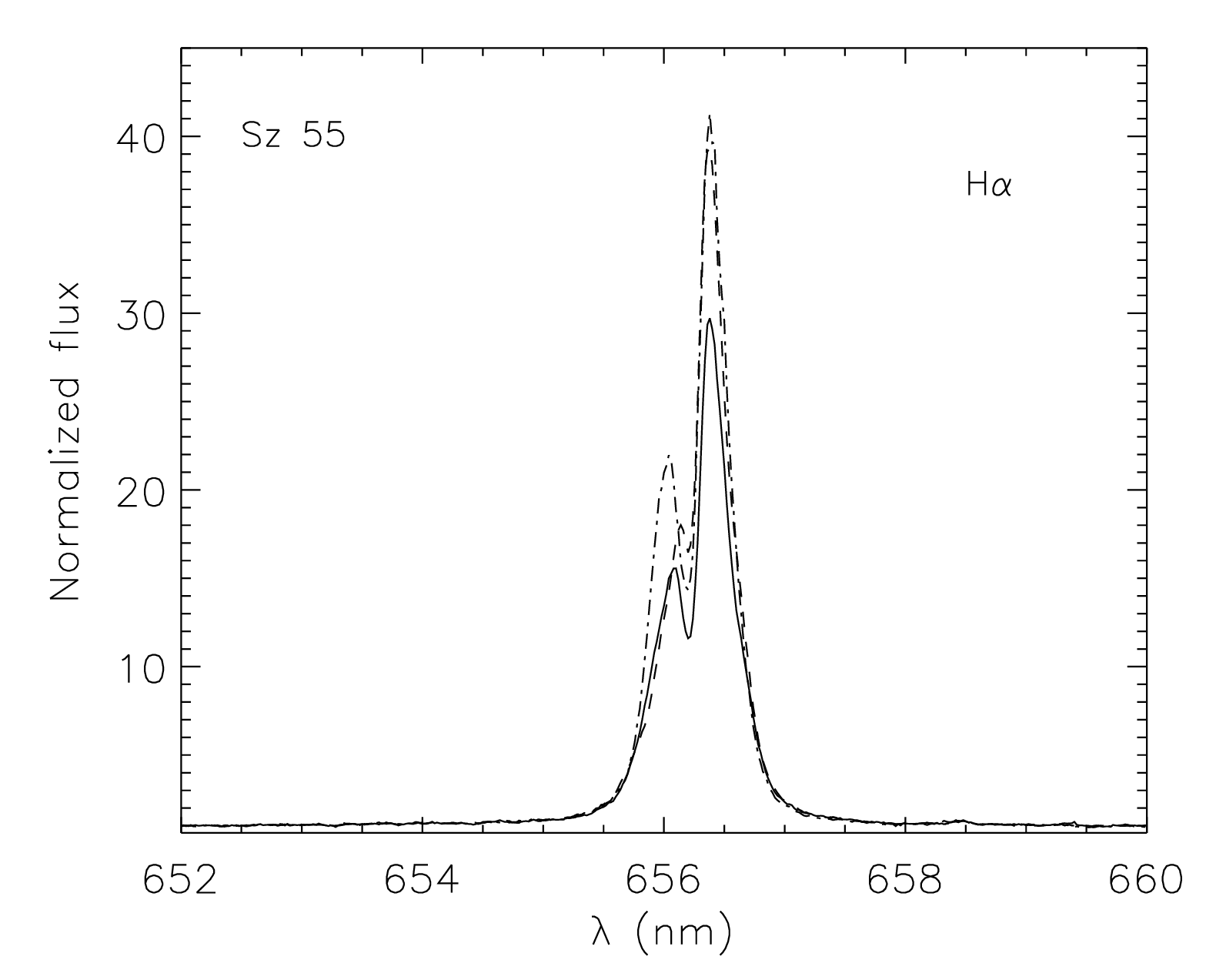}
\includegraphics[width=4.5cm]{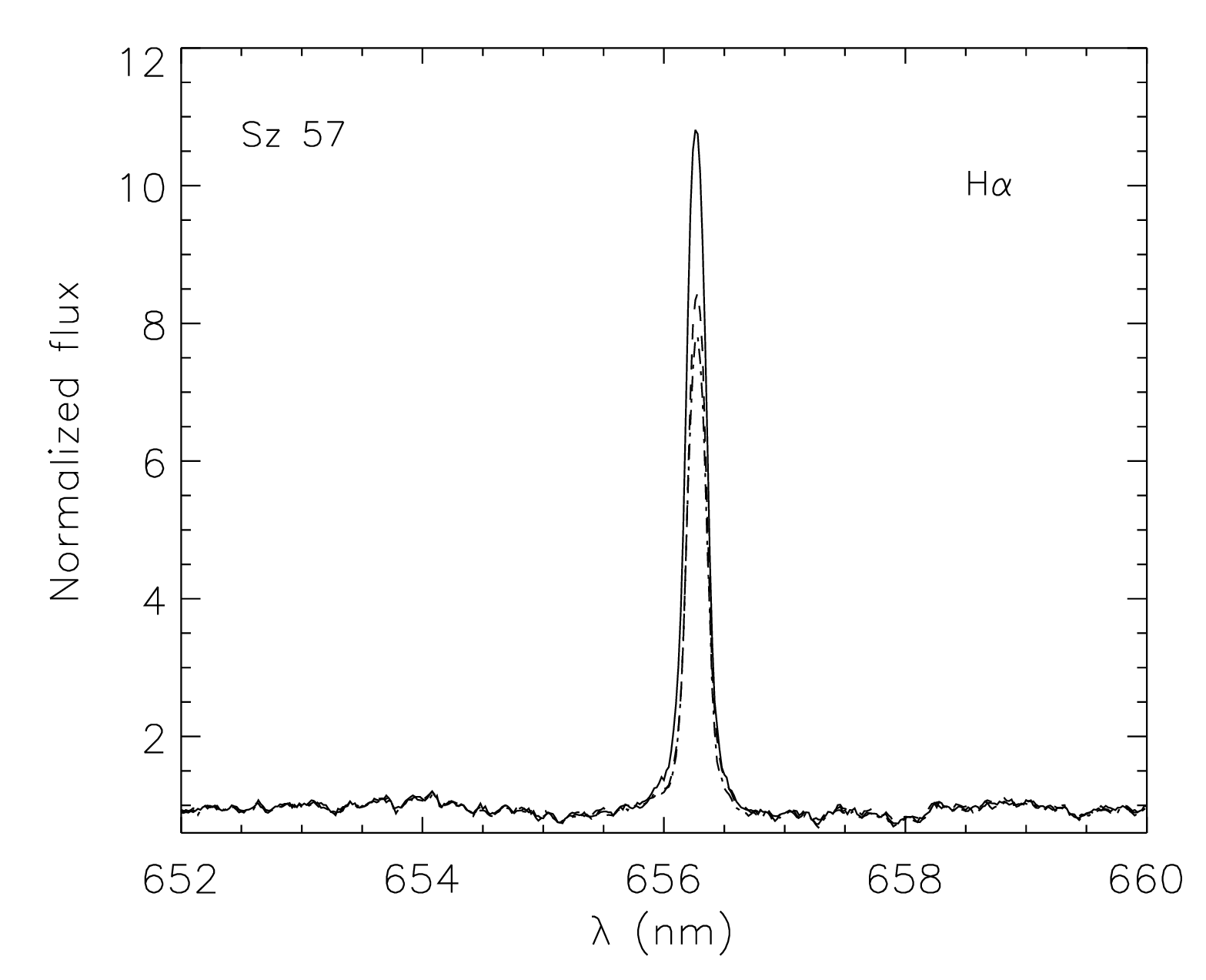}
\includegraphics[width=4.5cm]{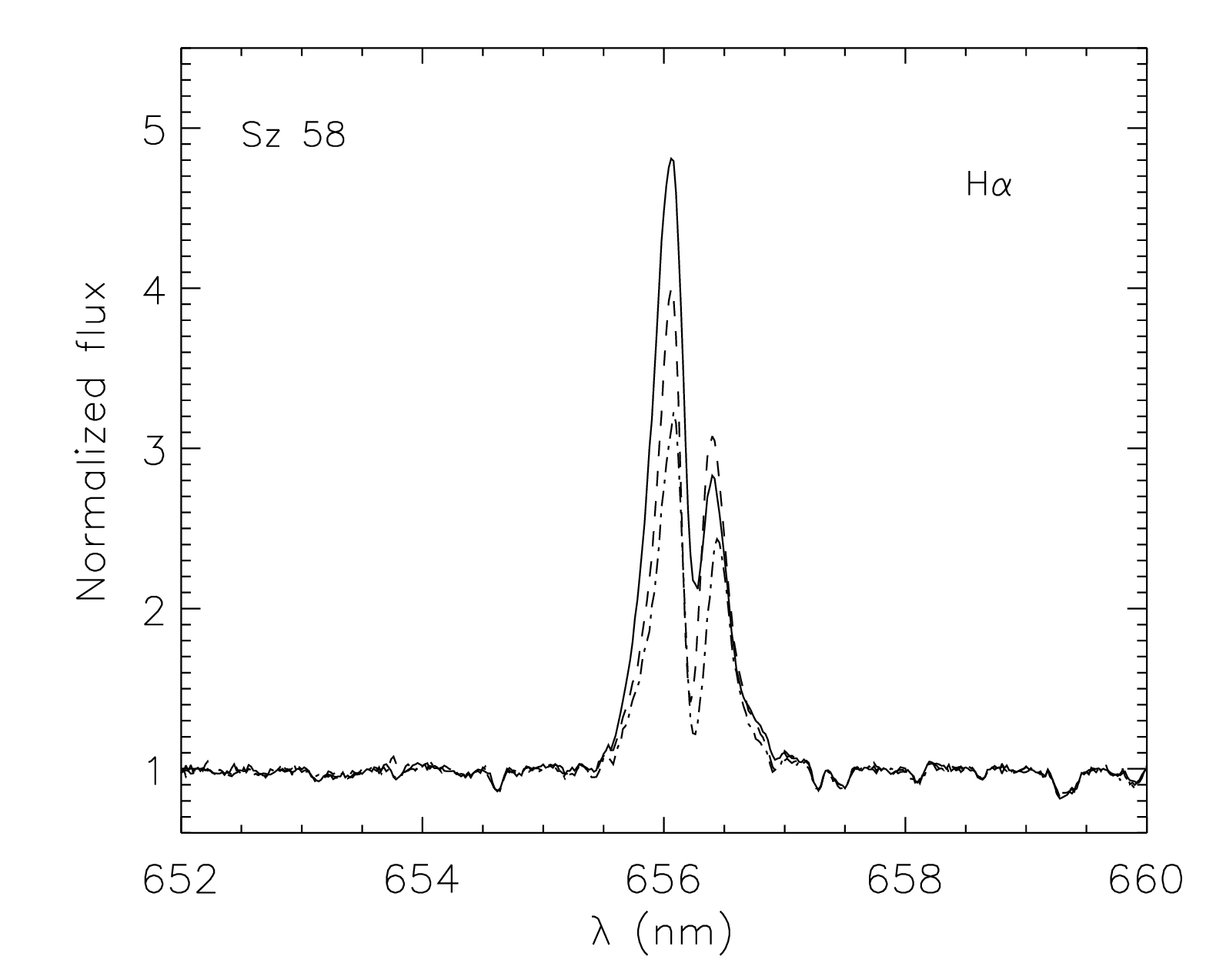}\\
\includegraphics[width=4.5cm]{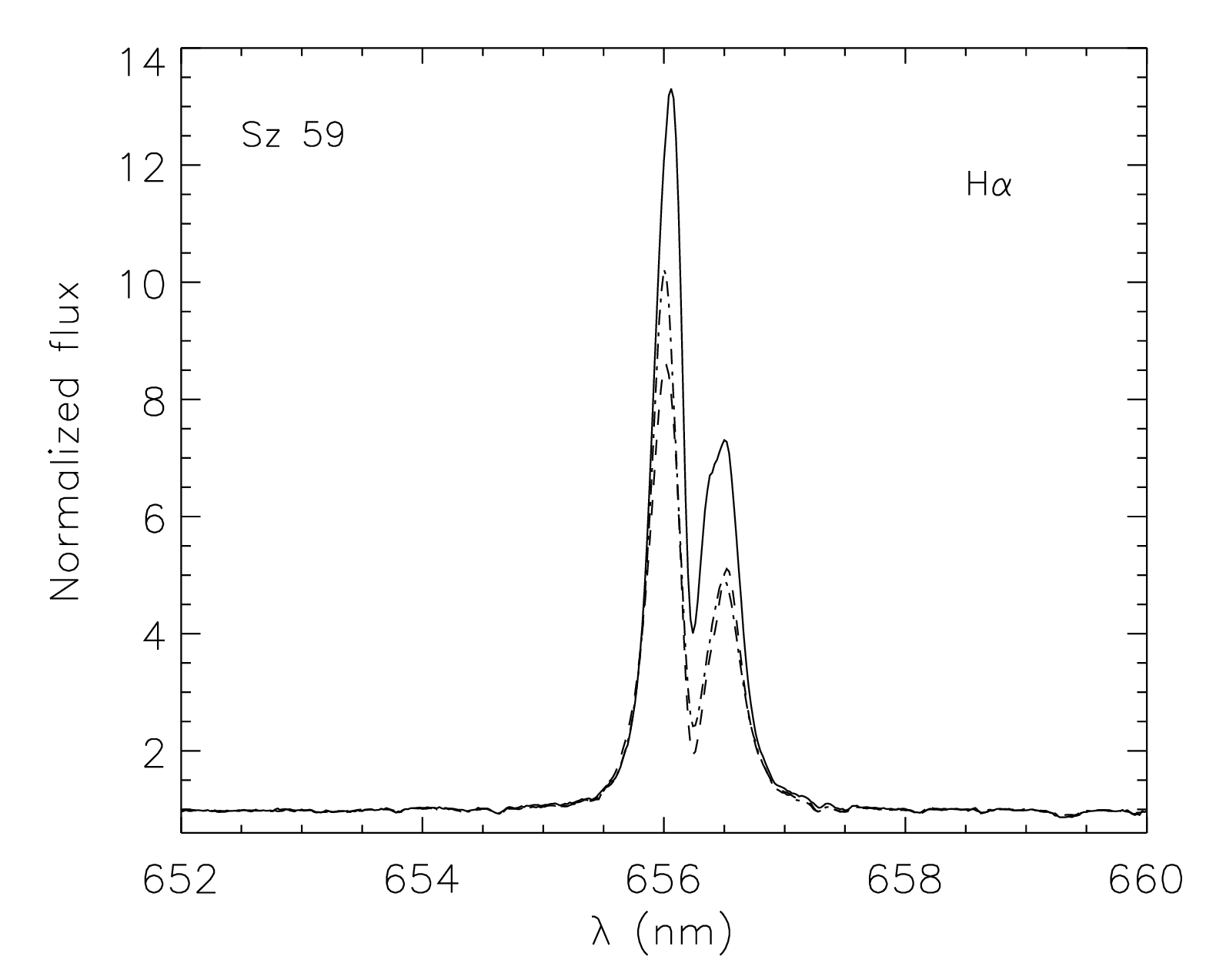}
\includegraphics[width=4.5cm]{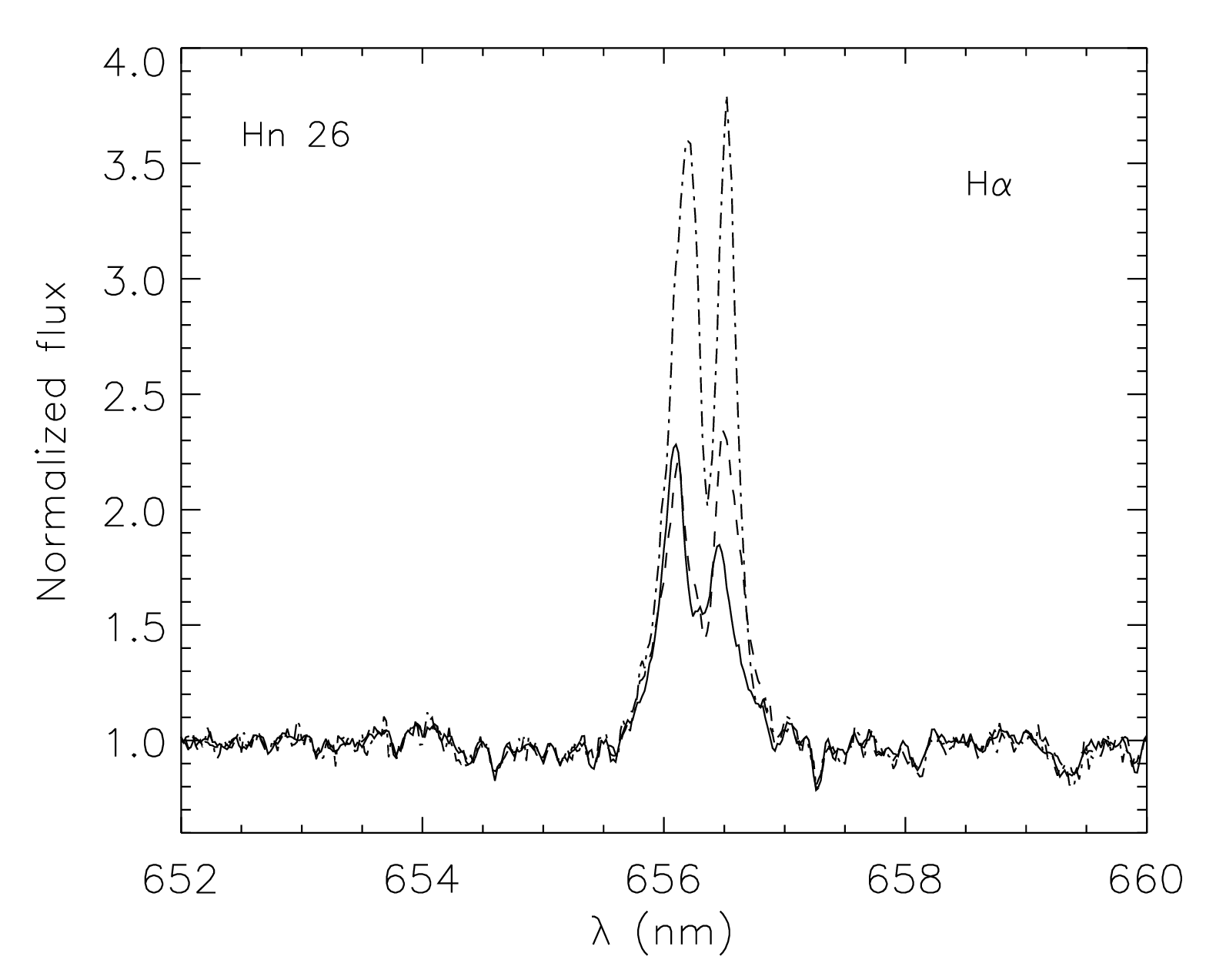}
\includegraphics[width=4.5cm]{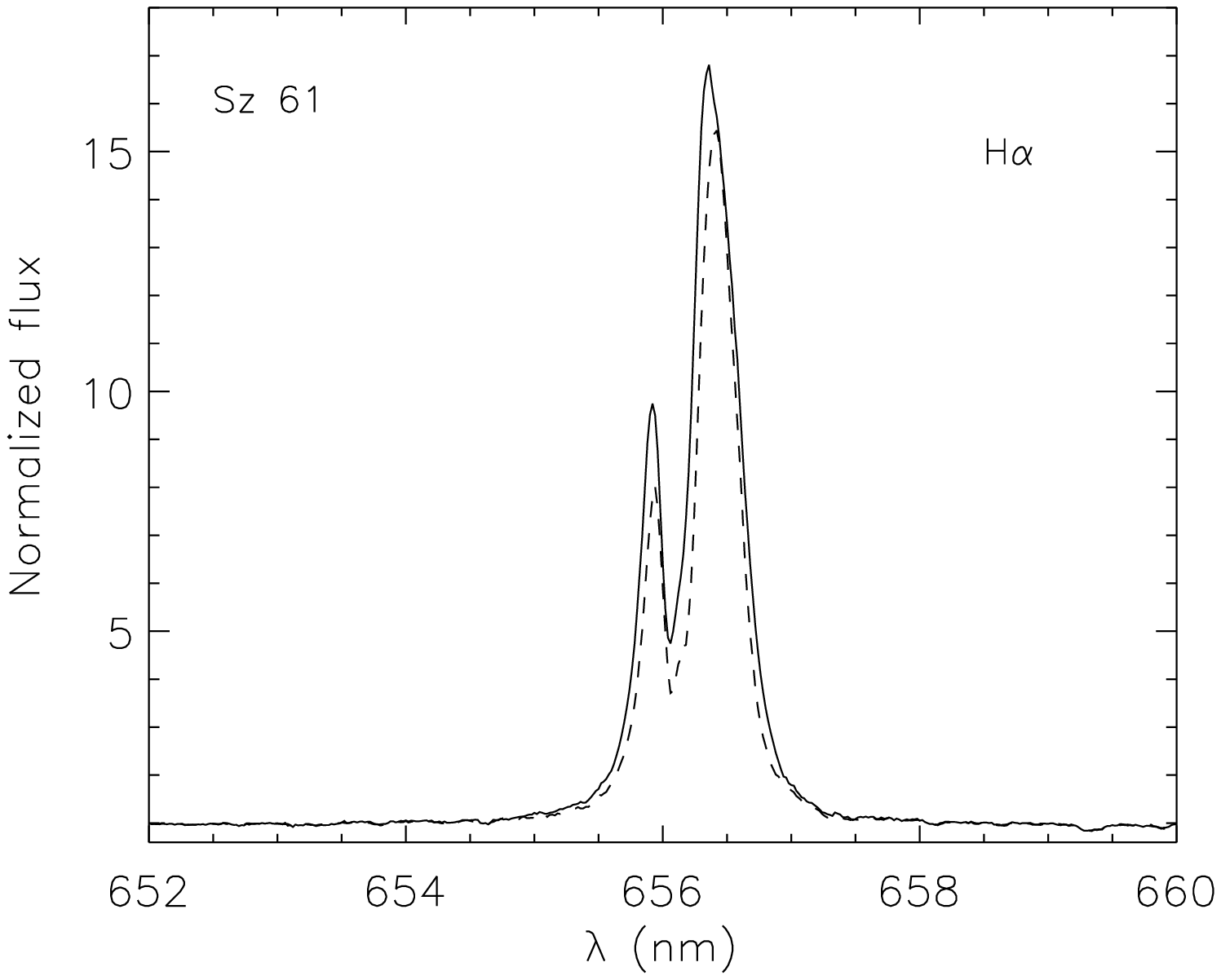}
\caption{Examples of H$\alpha$ emission line profile variations of nine stars. The fluxes are normalized to the continuum. 
The data refer to the FLAMES/GIRAFFE configuration. The solid/dashed/dash-dotted line represents the first/second/third observation, respectively. 
}
\label{fig:halpha_profiles_giraffe}
 \end{center}
\end{figure}

\begin{figure}[h!]
\begin{center}
\includegraphics[width=4.5cm]{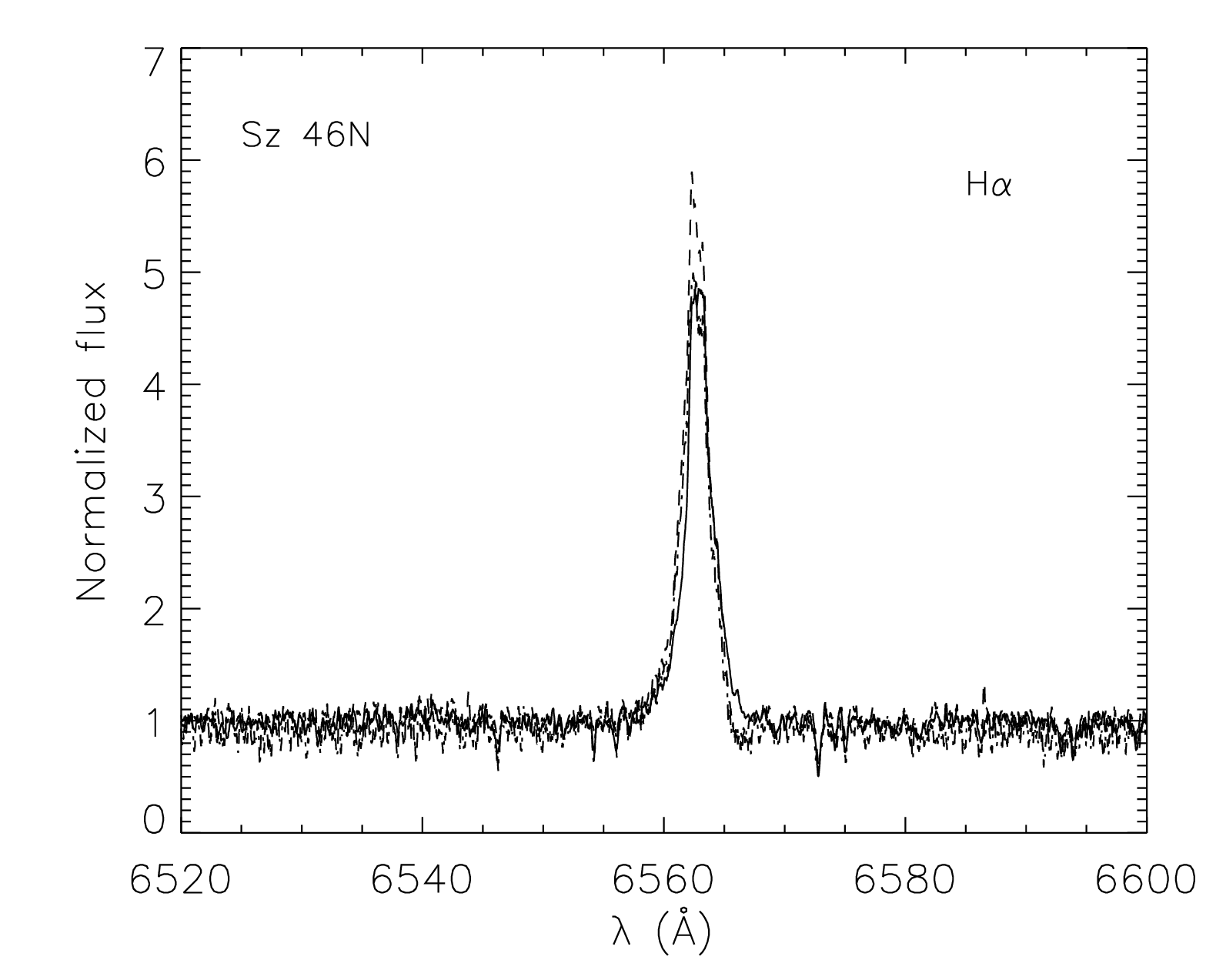}
\includegraphics[width=4.5cm]{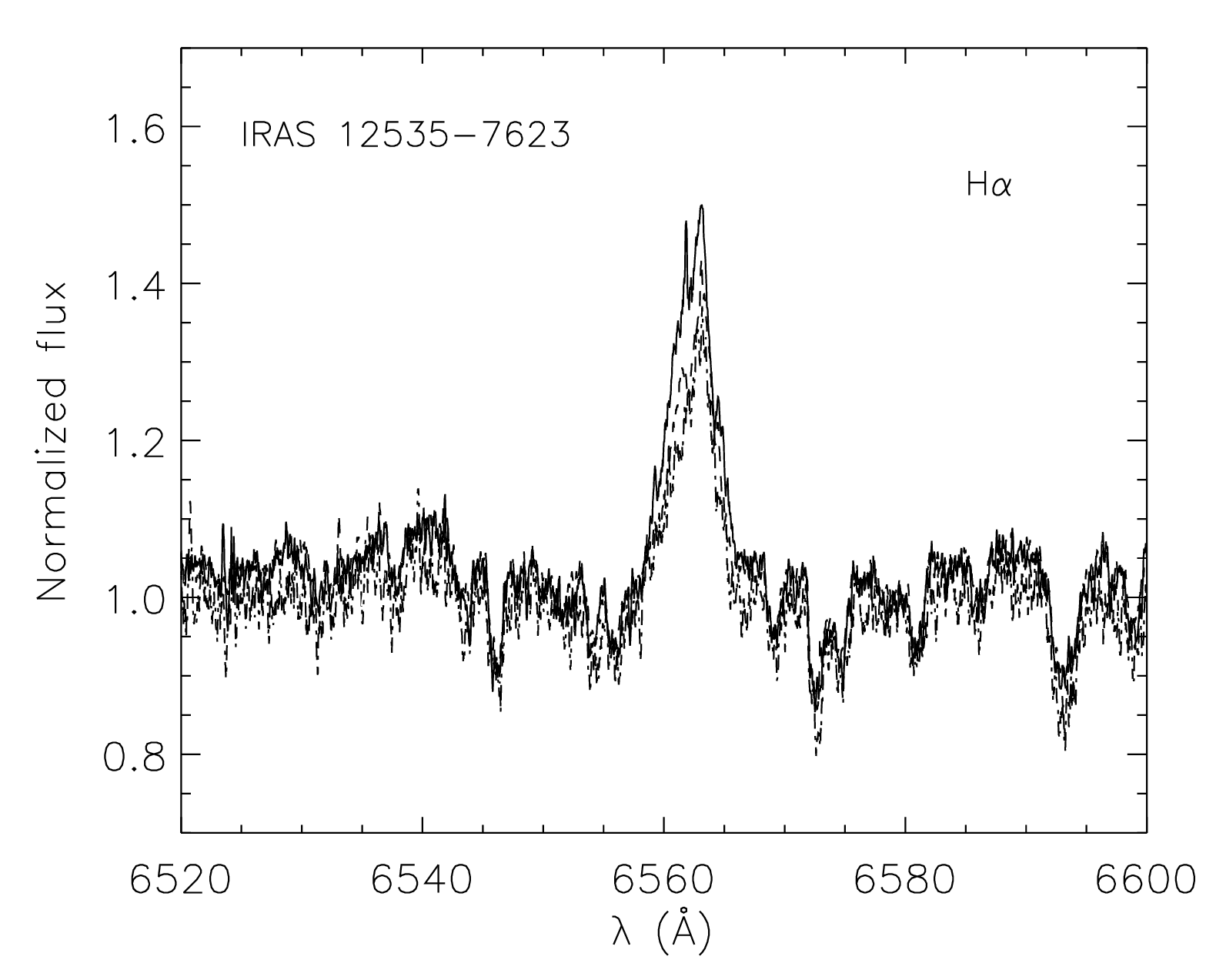}
\includegraphics[width=4.5cm]{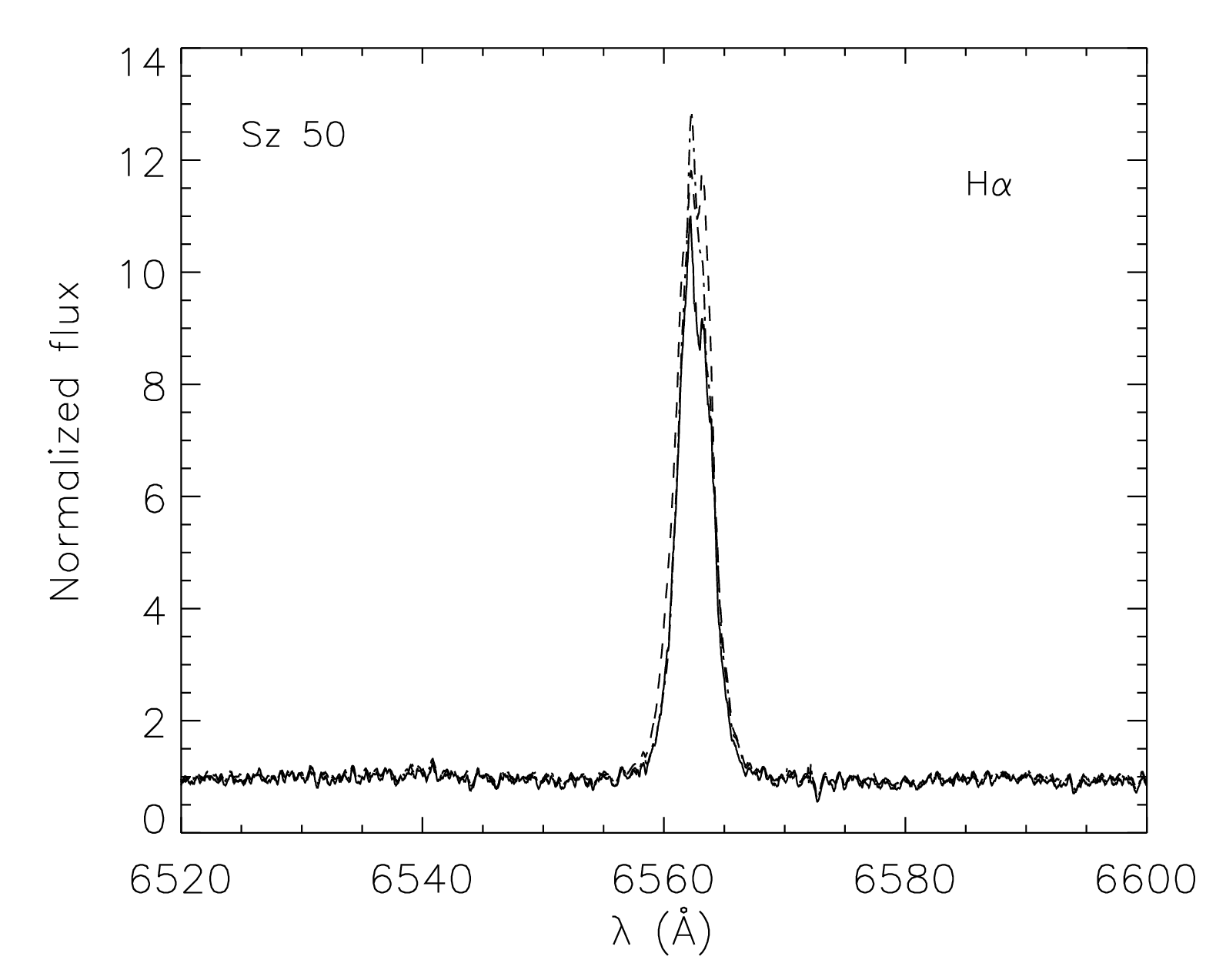}\\
\includegraphics[width=4.5cm]{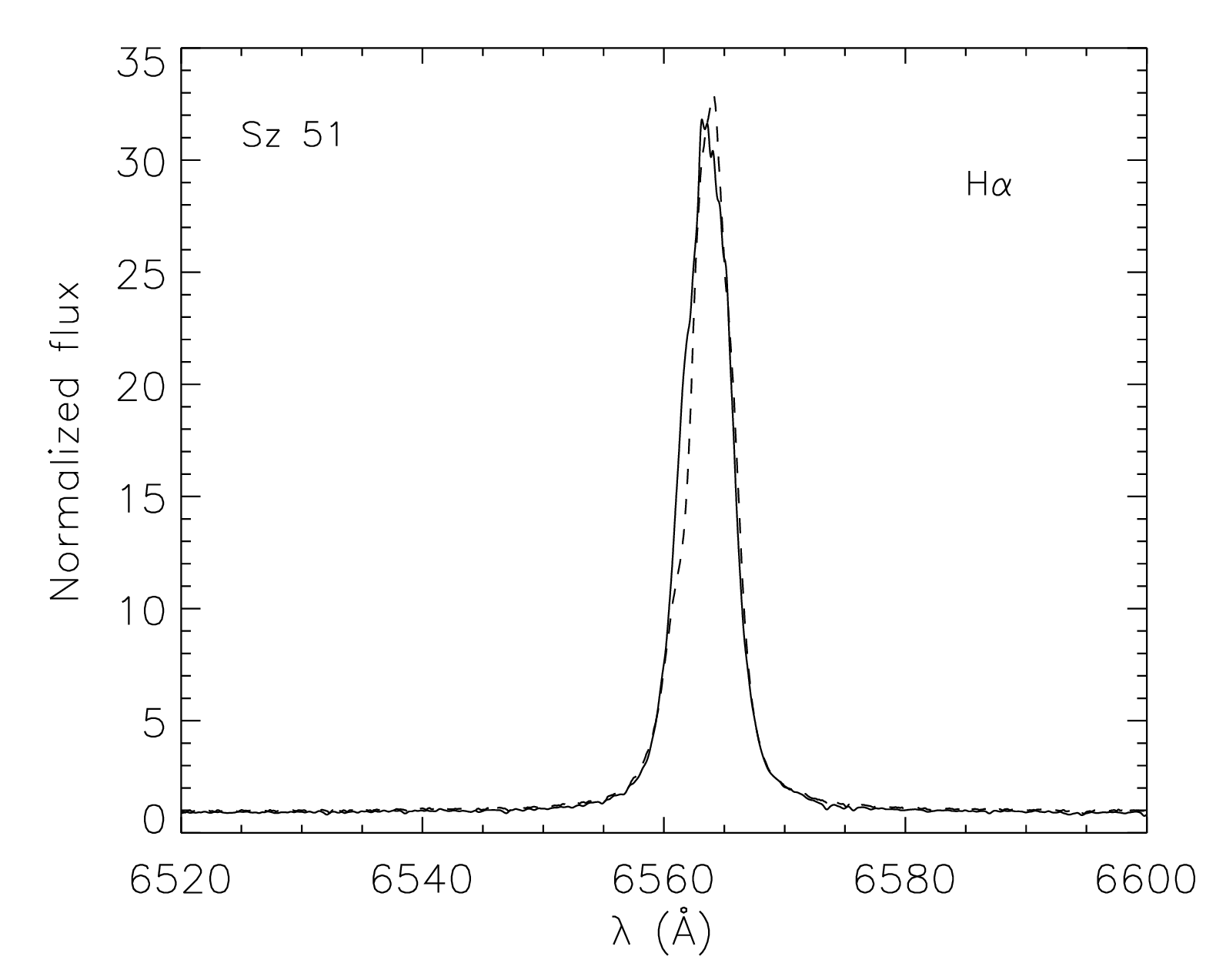}
\includegraphics[width=4.5cm]{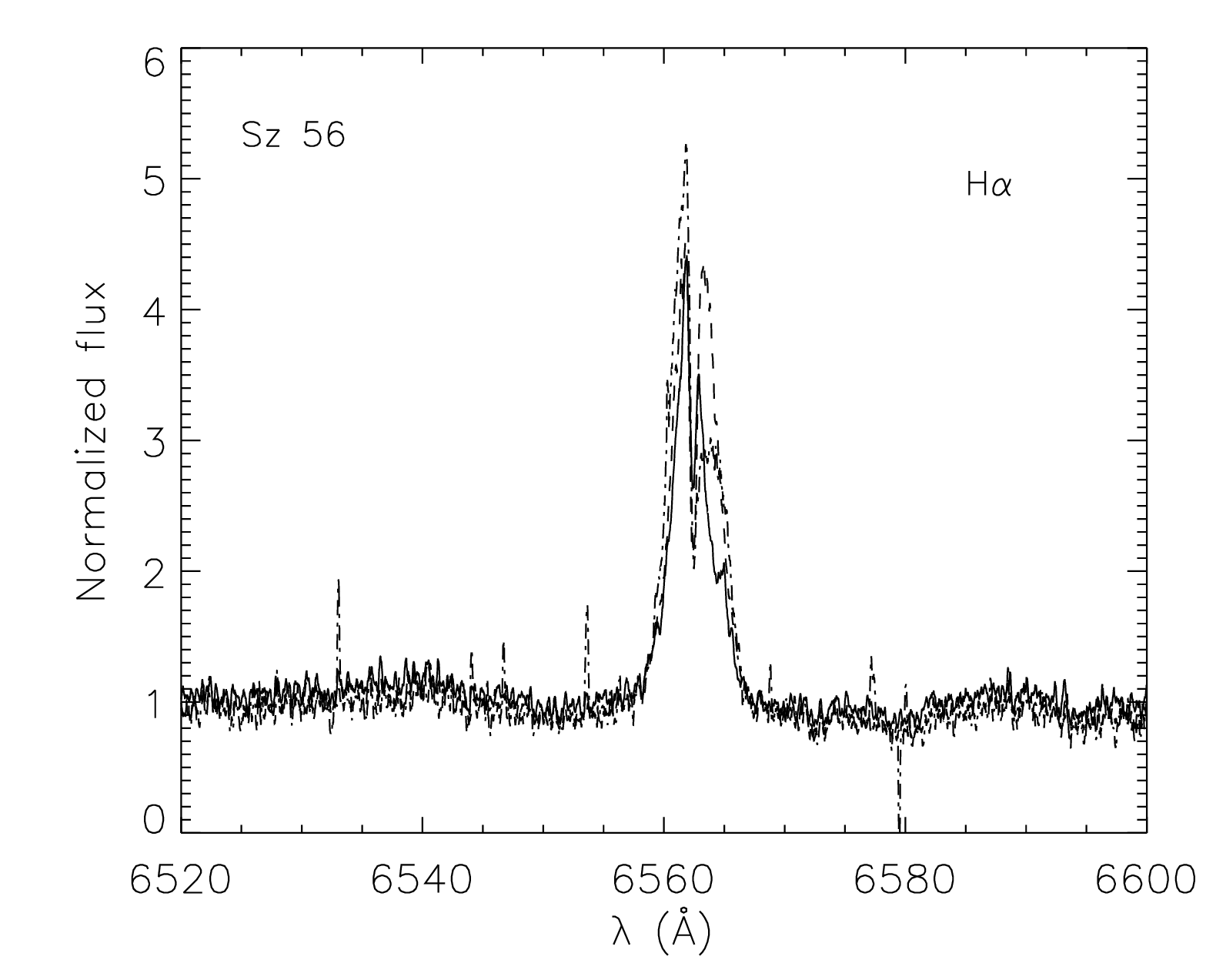}
\includegraphics[width=4.5cm]{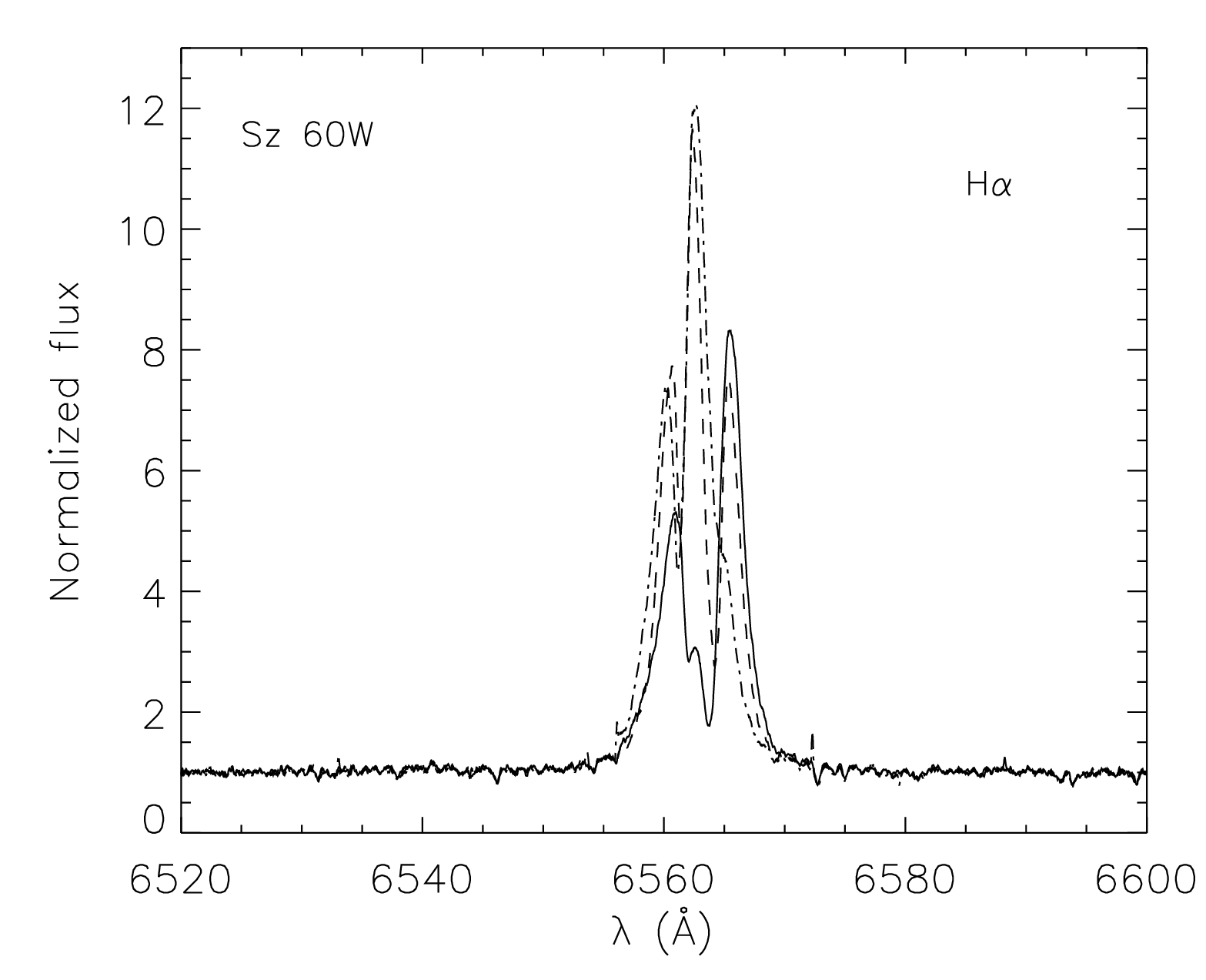}\\
\caption{Same as in Fig.~\ref{fig:halpha_profiles_giraffe}, but for the UVES data.
}
\label{fig:halpha_profiles_uves}
 \end{center}
\end{figure}

\begin{figure}[h!]
\begin{center}
\includegraphics[width=4.5cm]{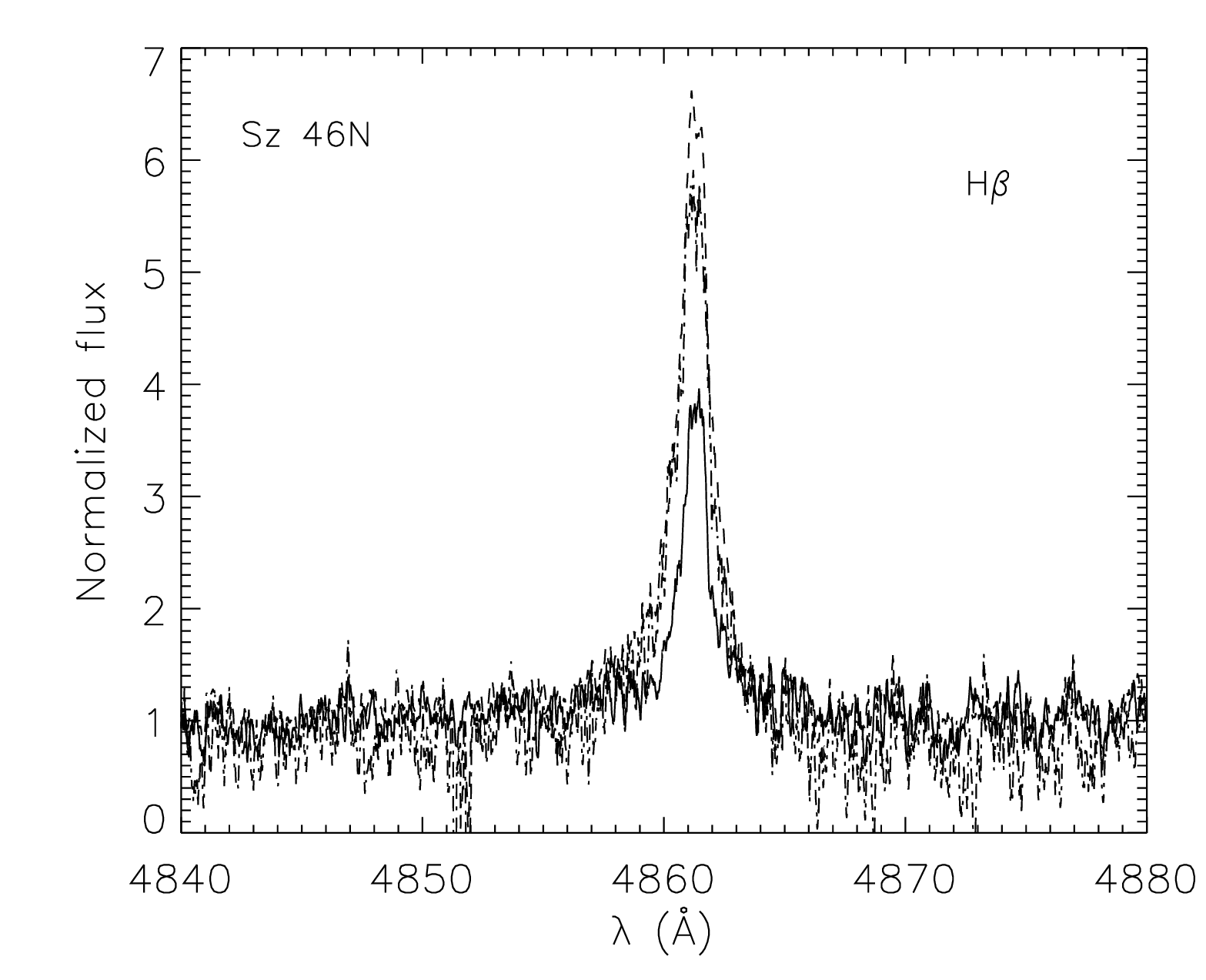}
\includegraphics[width=4.5cm]{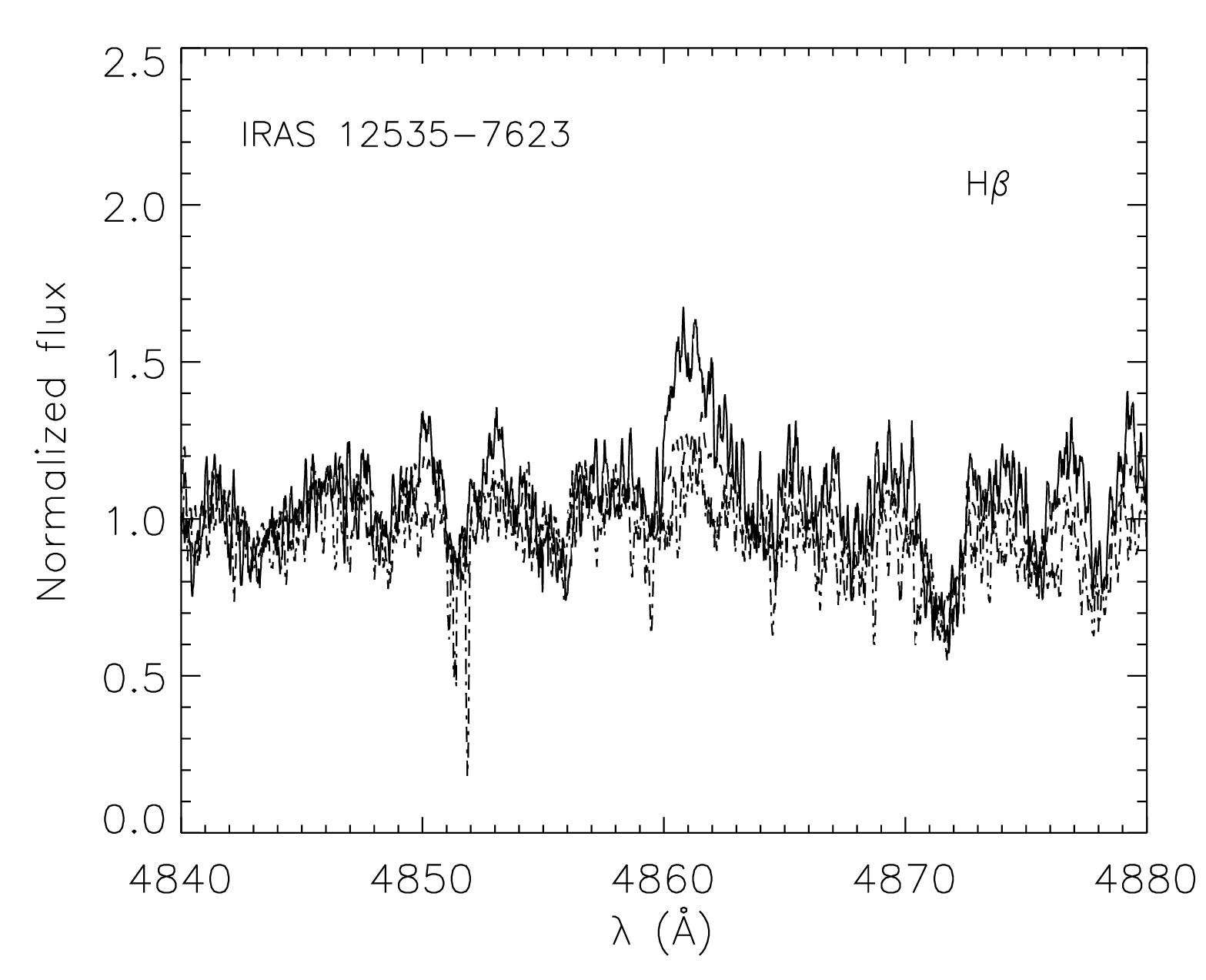}
\includegraphics[width=4.5cm]{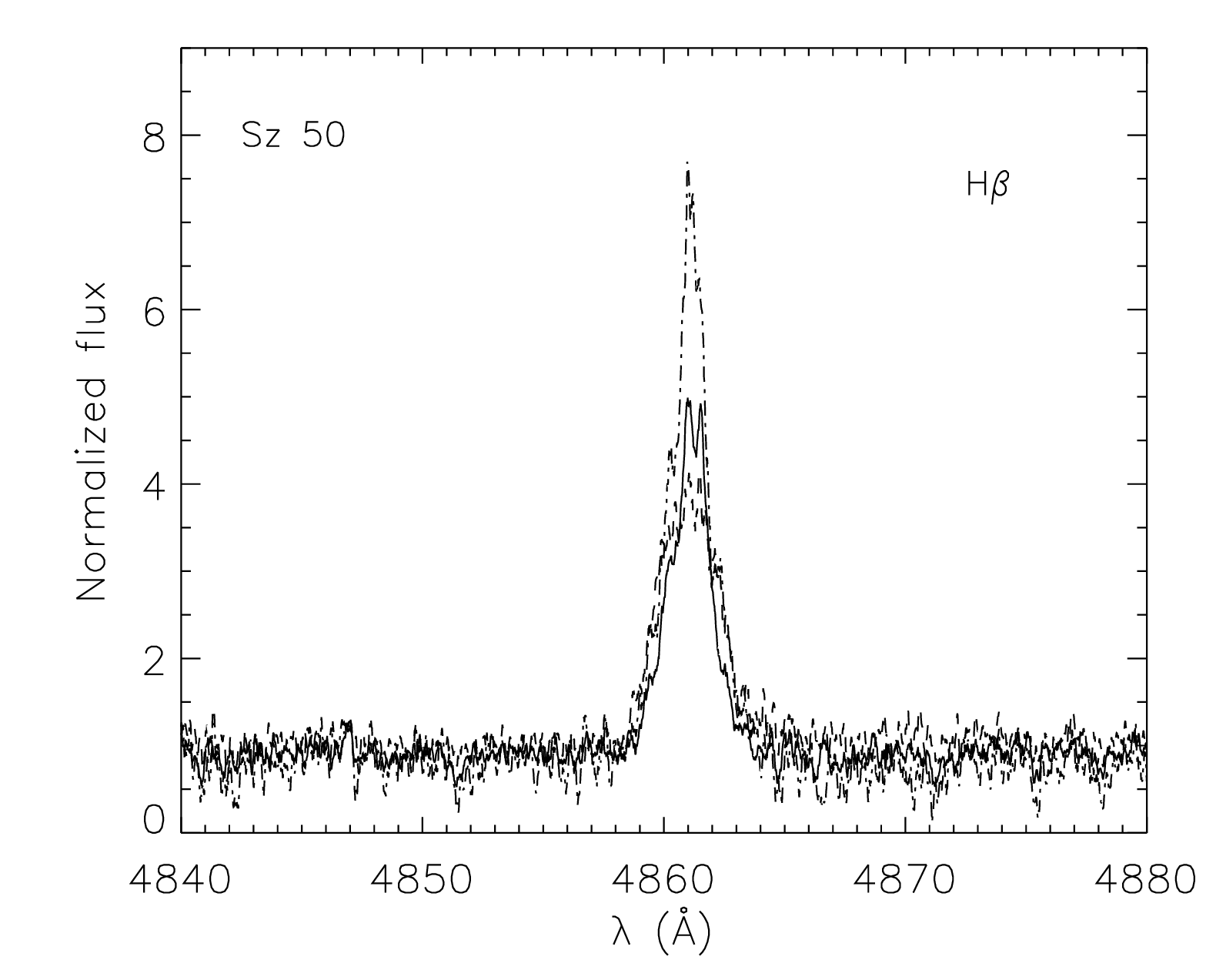}\\
\includegraphics[width=4.5cm]{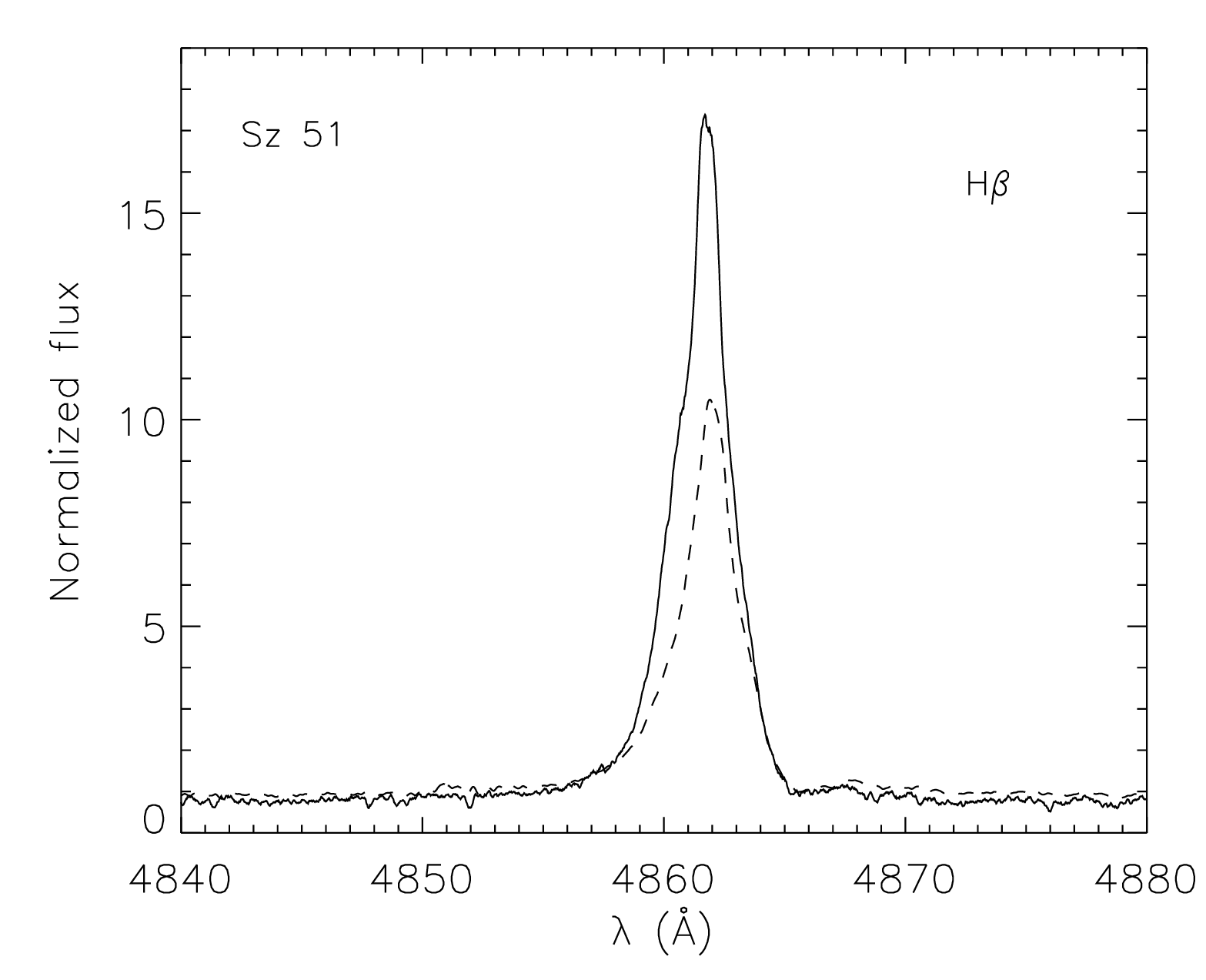}
\includegraphics[width=4.5cm]{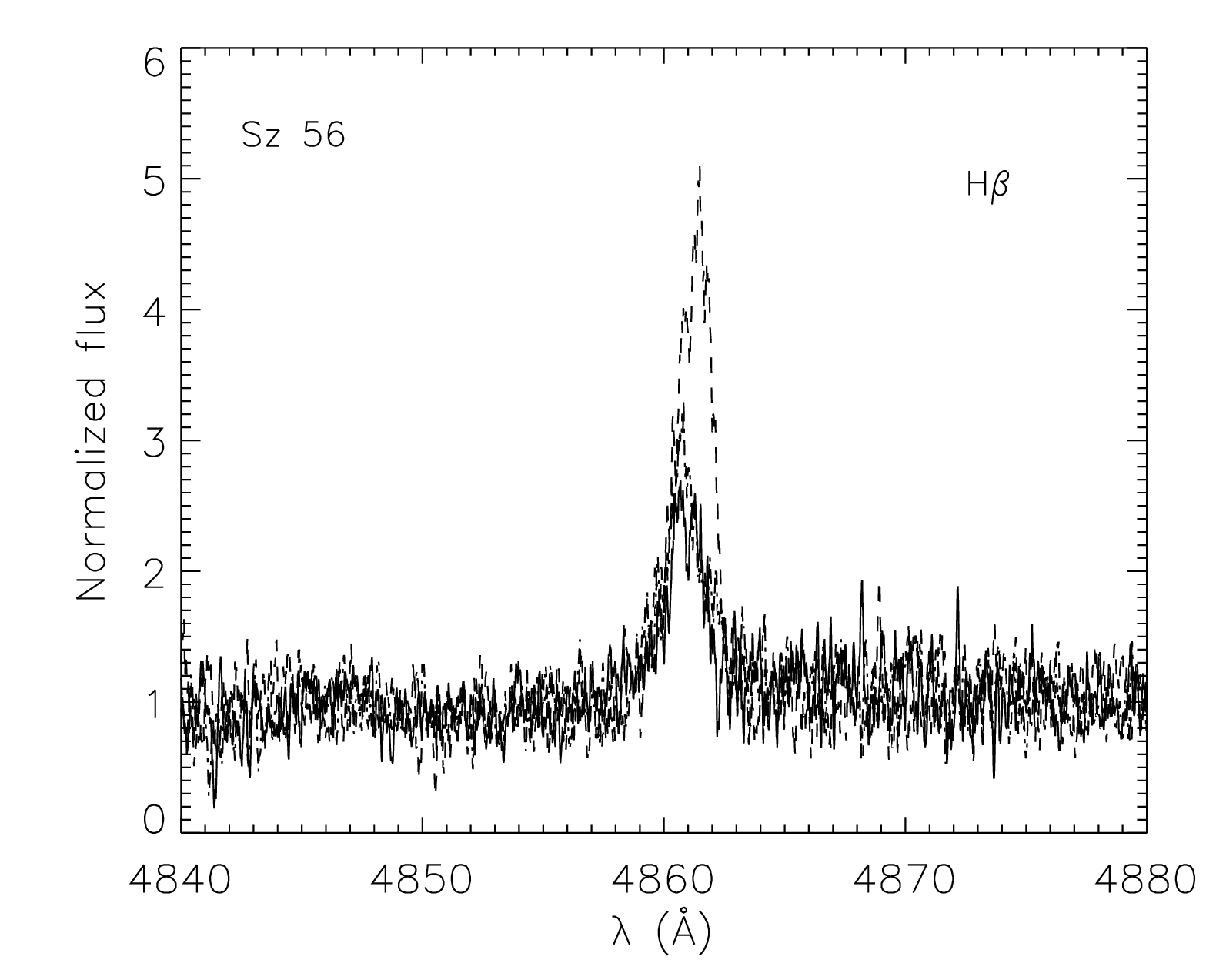}
\includegraphics[width=4.5cm]{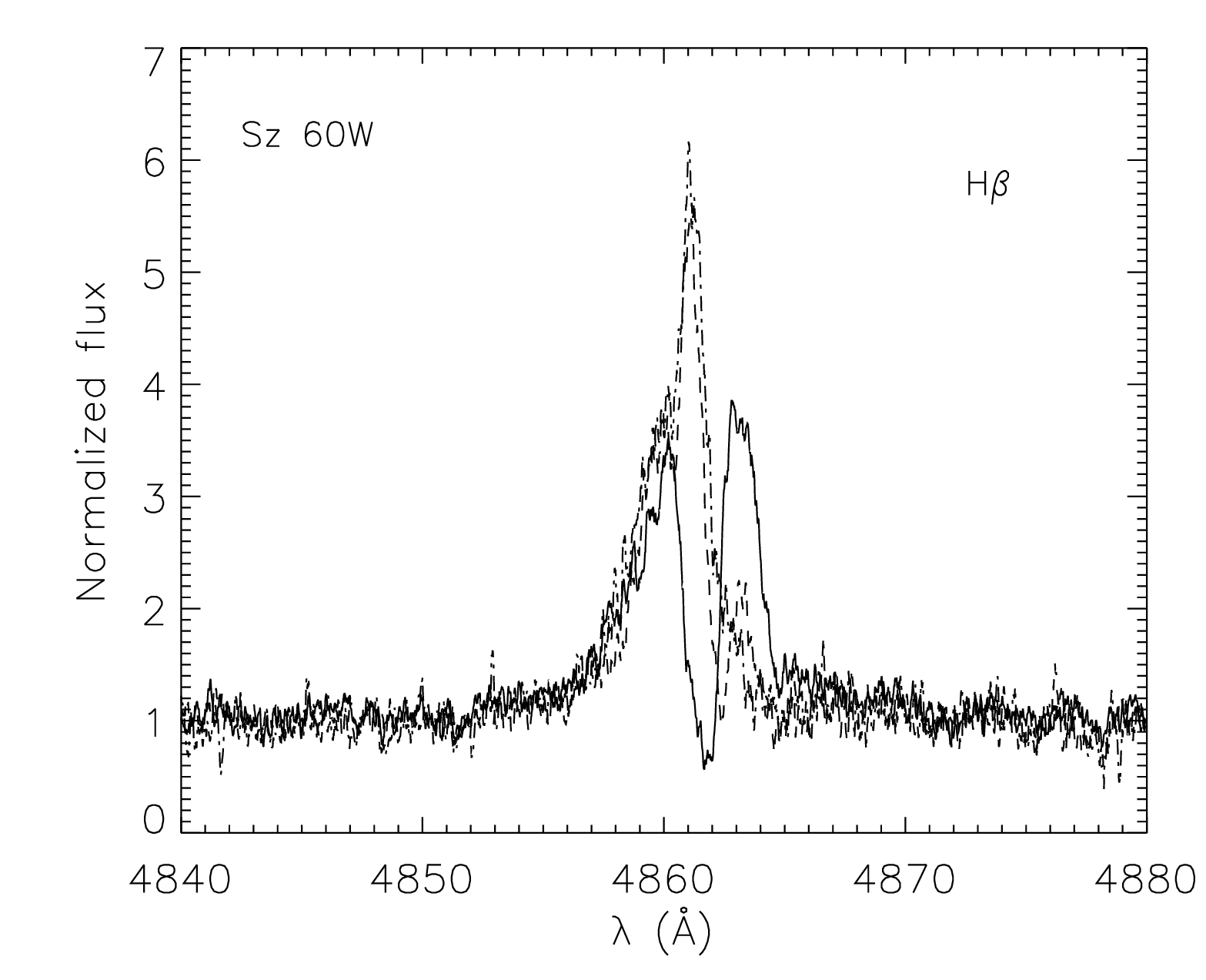}\\
\caption{Same as in Fig.~\ref{fig:halpha_profiles_giraffe}, but for the UVES data and the H$\beta$ profiles.
}
\label{fig:hbeta_profiles_uves}
 \end{center}
\end{figure}

Here, we display the H$\alpha$ (Figs.~\ref{fig:halpha_profiles_giraffe}, \ref{fig:halpha_profiles_uves}) and H$\beta$ 
(Figs.~\ref{fig:hbeta_profiles_uves}) profiles of the targets showing changes in their line shape and/or intensity. Below, we briefly 
comment on each object.

\noindent{{\it Sz~48SW}. During the first observing night, the spectrum showed a symmetric, relatively narrow H$\alpha$ emission profile with a peak 
close to the line center (type I profile, following the classification of \citealt{reipurthetal1996}). During the second night, the profile appeared 
slightly asymmetric with lower emission on the blue than on the red side (type IIIB), and then turned back to the type I profile 
on the third night.}

\noindent{{\it Hn~24}. The H$\alpha$ line profile changes from type IVR (an inverse P-Cygni-like profile) during the first two nights to type IIIR, 
where less emission in the red than in the blue is seen.}

\noindent{{\it Sz~53}. The H$\alpha$ line shows always a type IIB profile, with a central reversal at the line center and the 
blue-wing peak intensity lower than the red one. The line strength changes, being maximum during the first observation. 
The secondary peak always exceeds half the strength of the primary peak. This type of peak is probably due to the interplay of variable 
accretion and mass loss.}

\noindent{{\it Sz~55}. This star shows an H$\alpha$ line profile changing from IIB to IIIB (where the secondary peak is slightly less intense 
than half the strength of the primary peak) to IIB.}

\noindent{{\it Sz~57}. It always shows a narrow and symmetric H$\alpha$ emission type I profile with a peak close to the line center.}

\noindent{{\it Sz~58}. The H$\alpha$ profile changes from IIIR (first night) to IIR (second and third night).}

\noindent{{\it Sz~59}. IIR/IIR/IIIR profiles are observed for this star during the three observations.}

\noindent{{\it Hn~26}. This star shows IIR/IIB/IIB H$\alpha$ line profiles.}

\noindent{{\it Sz~61}. It shows an H$\alpha$ profile slightly changing from IIB type to IIIB.}

\noindent{{\it Sz~46N}. This star shows H$\alpha$ and H$\beta$ line profiles that are always symmetric (type I).}

\noindent{{\it IRAS~12535-7623}. The H$\alpha$ and H$\beta$ emission line profiles are always consistent with type I.}

\noindent{{\it Sz~50}. Its H$\alpha$ and H$\beta$ emission line profiles are always narrow and almost symmetric (type I).}

\noindent{{\it Sz~51}. H$\alpha$ and H$\beta$ are always of type I.}

\noindent{{\it Sz~56}. The H$\alpha$ profile changes from IIR (during the first and second night) to IIIR (third night), while the H$\beta$ 
profile is always of type I.}

\noindent{{\it Sz~60W}. The H$\alpha$ line has a complex profile, starting from IIB type during the first night; during the second night, it shows broad wings, 
red-shifted and blue-shifted absorption, and a narrow emission in the center (like a YY Orion H$\alpha$ profile, normally associated with 
high infall and outflow rates, consistent with the high value of $\dot M_{\rm acc}$; \citealt{walker1972}); during the third night, it turned back 
to a IIB type profile. The H$\beta$ line changes from YY Orionis H$\beta$-like profile to type IIIR to type I.}

\end{appendix}

\begin{appendix}
\section{Stellar variability}

\begin{figure*}[h!]
\begin{center}
 \begin{tabular}{c}
\includegraphics[width=9cm]{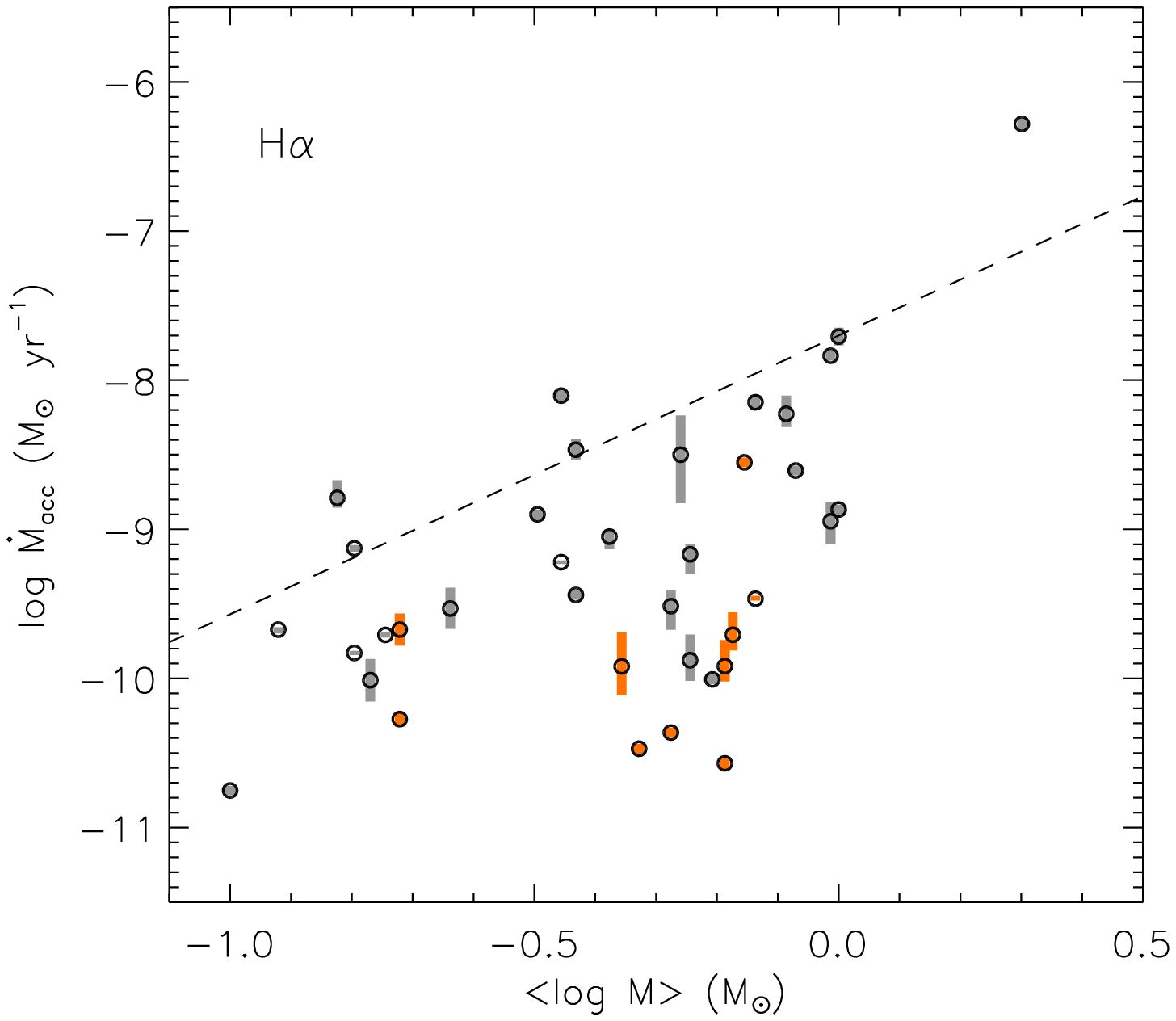}
\includegraphics[width=9cm]{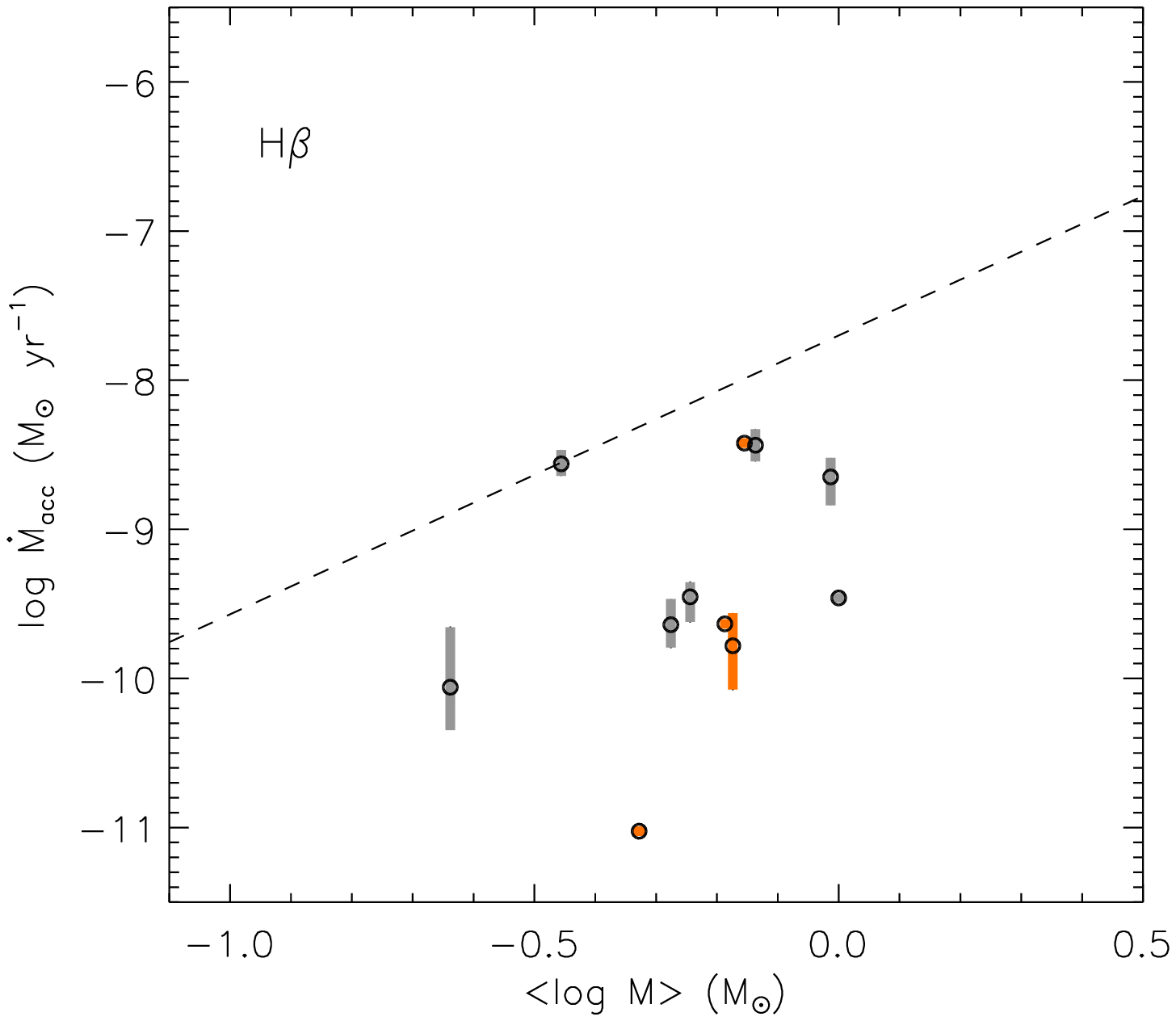}\\
\includegraphics[width=9cm]{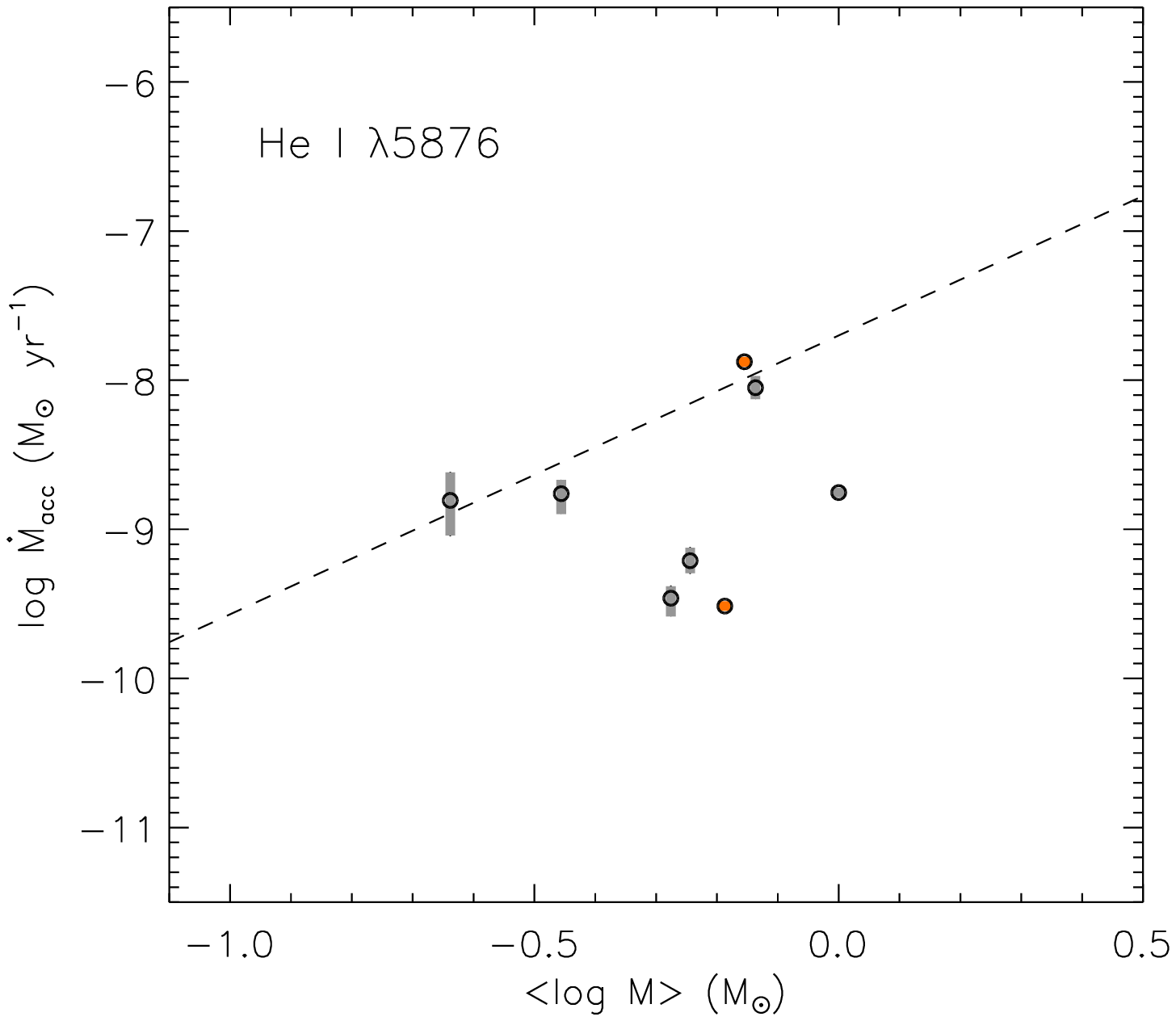}
\includegraphics[width=9cm]{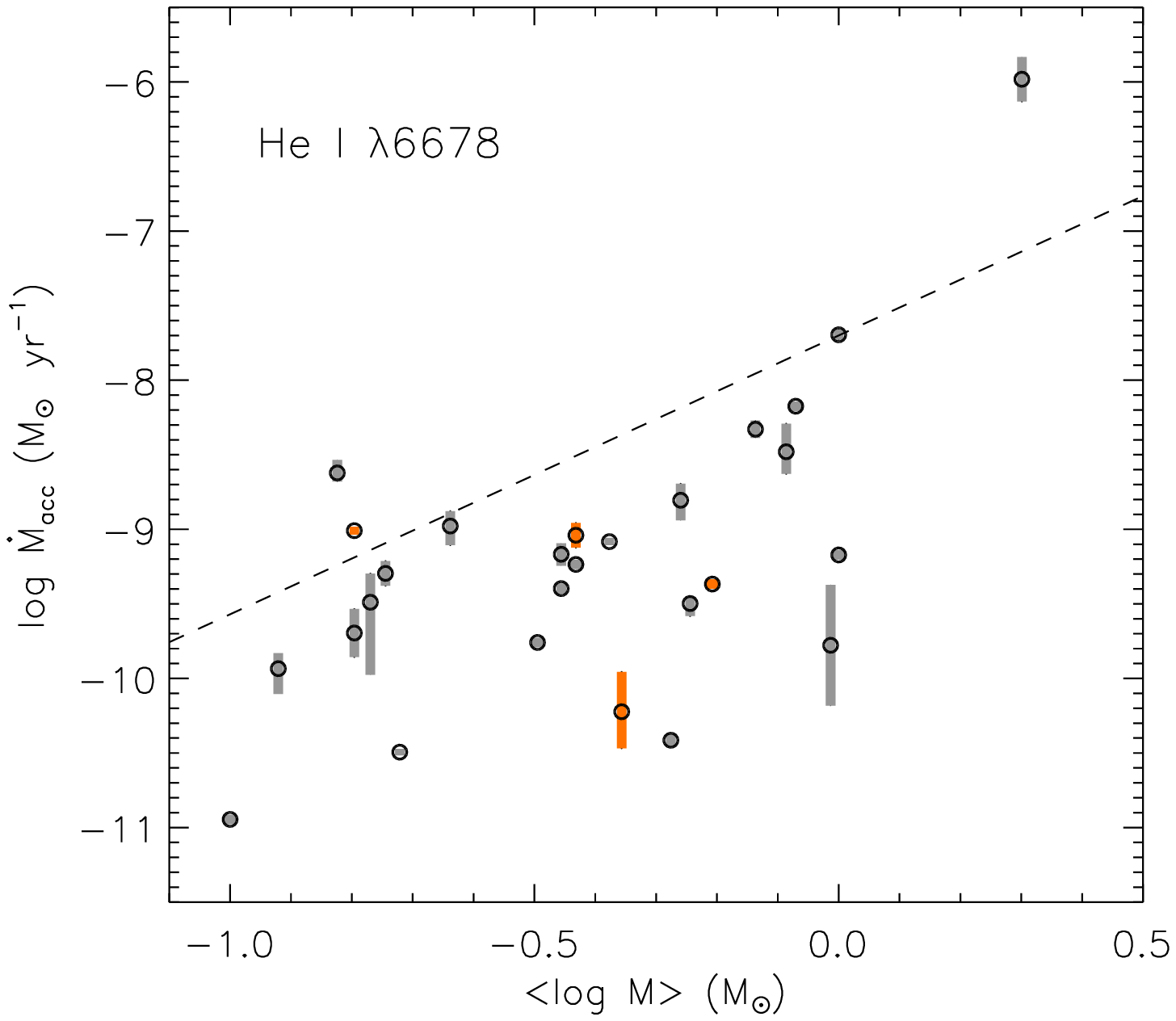}
\vspace{-.3cm}
 \end{tabular}
\caption{Mass accretion rates measured from the H$\alpha$ ({\it left top panel}), H$\beta$ ({\it right top panel}), 
\ion{He}{i} $\lambda$5876\,\AA\ ({\it left bottom panel}), and \ion{He}{i} $\lambda$6678\,\AA\ ({\it right bottom panel}) lines versus mass. 
Vertical bars connect the minimum and maximum values of $\dot M_{\rm acc}$ obtained per each star at different epochs. Stars are 
divided into high H$\alpha$ emitters ($EW_{\rm H\alpha}>10$\,\AA; grey bars) and low H$\alpha$ emitters ($EW_{\rm H\alpha}<10$\,\AA; orange bars), 
i.e., stars where the line emission cannot be unambiguously attributed to accretion activity, as it is most likely caused by chromospheric activity. 
The dashed line shows the relation $\dot M_{\rm acc} \propto M_\star^{1.87}$ obtained by \cite{herczeghillenbrand2008} for Taurus members. The highest 
$\dot M_{\rm acc}$ values in the first and fourth panels refer to the early-type star DK~Cha. The lower number of points in the H$\beta$ and 
\ion{He}{i} plots is due to two reasons: absence of the line in the given star or line out of the wavelength range (see Table~\ref{tab:all_param}). 
}
\label{fig:mass_accr_rate_mass}
 \end{center}
\end{figure*}
\end{appendix}

\end{document}